\documentclass[11pt,a4paper]{article}

\pdfoutput=1
\usepackage{jheppub}

\usepackage{amsmath, amsopn, amsthm}
\usepackage{mathtools} 
\usepackage{extarrows} 
\usepackage{verbatim}
\usepackage{amssymb}
\usepackage{inputenc, array}
\usepackage{textcomp}
\usepackage{bm}
\usepackage{physics}
\usepackage{simpler-wick}
\usepackage{appendix}
\usepackage{dsfont}

\def\be{\begin{equation}}
	
\def\ee{\end{equation}}

\def\bea{\begin{eqnarray}}
	
\def\eea{\end{eqnarray}}

\def\vk{\vect{k_1}}
\def\vkk{\vect{k_2}}
\def\vkkk{\vect{k_3}}
\def\vkfour{\vect{k_4}}
\def\half{\frac{1}{2}}

\newcommand{\cM}{\mathcal{M}}

\renewcommand{\Re}{{\rm Re}\,}

\renewcommand{\Im}{{\rm Im}\,}

\newcommand{\vect}[1]{{\bf{#1}}}

\def\Or[#1]{{\text{O}}\left({#1}\right)}
\def\dotl[#1,#2]{\left\langle #1,\, #2 \right\rangle}
\def\dotlb[#1,#2]{\left\langle #1,\, #2 \right\rangle}
\def\dotlm[#1,#2]{\left[ #1,\, #2 \right]}
\def\dotp[#1,#2]{(\vect{#1} \cdot\vect{#2})}

\pdfoutput=1
\usepackage{array}
\usepackage{multirow}
\usepackage{amsmath}
\usepackage{makecell}

\definecolor{green}{rgb}{0.1,0.8,0.2}

\usepackage{cancel}
\usepackage{graphicx} 

\usepackage{xcolor}
\usepackage{hyperref}
\usepackage{cleveref}
\usepackage{float}
\usepackage{subcaption}
\usepackage{physics}
\usepackage{slashed}
\usepackage{amsfonts}
\usepackage{caption}
\usepackage{mathrsfs}

\preprint{}
\title{Inflationary non-Gaussianities in alpha vacua and consistency with conformal symmetries}

\author[a]{Arhum Ansari}
\author[b, c]{, Pinak Banerjee}
\author[b]{, Prateksh Dhivakar}
\author[a]{, Sachin Jain}
\author[b]{, and Nilay Kundu}

\affiliation[a]{Indian Institute of Science Education and Research, Homi Bhabha Road, Pashan, Pune 411 008, India}
\affiliation[b]{Department of Physics, Indian Institute of Technology Kanpur, Kalyanpur, Kanpur 208016, India}
\affiliation[c]{Department of Theoretical Physics, Tata Institute of Fundamental Research, Colaba, Mumbai, India, 400005}

\emailAdd{ansari.arhum@students.iiserpune.ac.in}
\emailAdd{banerjee.pinak30@gmail.com}
\emailAdd{prateksh@iitk.ac.in}
\emailAdd{sachin.jain@iiserpune.ac.in}
\emailAdd{nilayhep@iitk.ac.in}

\preprint{}
\abstract{We study the conformal invariance of inflationary non-Gaussianities associated with scalar fluctuations in a non-Bunch-Davies initial state, known as the $\alpha$-vacuum, in single-field slow-roll inflation. The $\alpha$-vacuum is a one-parameter family of states, including the Bunch-Davies one, that preserves the conformal symmetry of inflationary dynamics in a nearly de-Sitter space-time. Working within the leading slow-roll approximation, we compute the four-point scalar correlator (the trispectrum) in $\alpha$-vacuum using the in-in formalism. We check that the conformal Ward identities are met between the three and four-point scalar $\alpha$-vacua correlators. Surprisingly, this contrasts the previously reported negative result of the Ward identities being violated between the two and the three-point correlators. We have also extended the wave-functional method, previously used for correlators with Bunch-Davies initial condition, to compute the three and four-point scalar correlators in $\alpha$-vacua. The results obtained from the wave-function method match the corresponding in-in results, adding further justification to our check of Ward identities with $\alpha$-vacua correlators.}


\begin{document}

\maketitle

\section{Introduction}\label{intro}

Inflation is a paradigm justifying why our universe is approximately homogeneous and isotropic at the cosmological scales, with tiny fluctuations seeding subsequent structure formation. Measurement of the correlation functions of these cosmological quantum fluctuations encoded in the CMB radiation reveals how our universe was immediately after the Big Bang. The CMB anisotropies follow a nearly Gaussian distribution, and future observations will put tighter bounds on the small non-Gaussianities \cite{Planck:2018jri, Planck:2019kim}. In inflation, the objects of primary interest are the correlation functions of the scalar and tensor perturbations produced during inflation at the late time slice when the modes crossed the horizon. The two-point correlator signifies the dominant Gaussian distribution, whereas the three and higher-point correlators reflect the non-Gaussianities in the CMB \cite{Chen:2006nt}.

Traditionally, symmetries have been very useful in understanding physical phenomena, especially when intricate details about such phenomena are lacking. In the context of inflation, a symmetry-based analysis will undoubtedly be advantageous compared to calculations done in specific models since the lessons learned would be robust and universal. The space-time geometry during inflation was \emph{nearly} de-Sitter ($dS_4$). Now, exact $dS_4$ has a symmetry group of $SO(4,1)$, same as that of a three-dimensional Euclidean conformal field theory (CFT). However, inflation slightly breaks the exact conformal symmetries, which is controlled perturbatively by the small slow-roll parameters. A vast literature studies inflation from this conformal symmetry perspective, but our approach is primarily aligned with the pioneering work in \cite{Maldacena:2002vr, Maldacena:2011nz}. Some of the other works with the same point of view are the following \cite{Antoniadis:1996dj, Larsen:2002et, McFadden:2010vh, Creminelli:2011mw, Bzowski:2011ab, Kehagias:2012td, Schalm:2012pi, Mata:2012bx, Kehagias:2015jha, Ghosh:2014kba, Kundu:2014gxa, Kundu:2015xta, Arkani-Hamed:2015bza}. Recently, ideas of bootstrap and Mellin space techniques have also been implemented in cosmological correlators, see \cite{Arkani-Hamed:2018kmz, Sleight:2019mgd, Sleight:2019hfp, Pajer:2020wxk, Green:2020ebl, Ghosh:2023agt} \footnote{Our list of references are certainly not an exhaustive one, see \cite{Baumann:2022jpr} and the references therein for more complete and detailed references.}. 

An insightful way in which the slightly broken conformal symmetries during inflation manifest themselves is through the Ward identities satisfied by the correlation functions of the cosmological perturbations \cite{Maldacena:2002vr, Mata:2012bx, Ghosh:2014kba, Kundu:2014gxa, Kundu:2015xta}. These Ward identities are a set of relations between an $(n+1)$-point function and the $n$-point function, in the limit when one of the external momenta in the $(n+1)$-point function becomes soft \cite{Maldacena:2002vr}. The Maldacena consistency condition \cite{Maldacena:2002vr} is the leading instance of this infinite set of Ward identities: a relation between the two-point and three-point correlators \footnote{Such consistency conditions have previously been discussed in the context of squeezed limit, for a partial list see\cite{Creminelli:2004yq, Cheung:2007sv, Leonardo2012, Maldacena:2011nz, Bartolo:2011wb, Creminelli:2012qr, Assassi:2012zq, Hinterbichler:2013dpa, Berezhiani:2013ewa, Sreenath:2014nca}.}. In this sense, the correlation functions and, therefore, the non-Gaussianities in the CMB are constrained by conformal symmetries. It was shown in \cite{Kundu:2015xta} that these Ward identities follow from more general spatial re-parametrization invariance and thus should hold even when conformal symmetries are broken. Even though the symmetry is not exact, we can use the Ward identities by treating the breaking of the exact $dS_4$ symmetries perturbatively in a conformal perturbation theory. Consequently, the Ward identities that would have worked with exact $dS_4$ symmetries get additional contributions proportional to the slow-roll parameters.   

Conventionally, the inflationary correlation functions are calculated using the Bunch-Davies initial condition for the perturbations. This is a natural choice since, at the beginning of inflation, the perturbative modes in the Bunch-Davies vacuum are essentially in the Minkowski vacuum. However, it is known in the literature that $dS_4$ admits a larger class of conformally invariant vacuum \cite{PhysRevD.32.3136, PhysRevD.35.3771, PhysRevD.31.754}. Ignoring a possible phase factor, these are a one-parameter family of vacuum states denoted by the real parameter $\alpha$ for a quantum field theory in $dS_4$ space-time \footnote{It should be noted that the $\alpha$ vacua states in \cite{PhysRevD.31.754} are constructed in Global $dS_4$. However, we will be working in the Poincar\'e patch, and $\alpha$ vacua can be considered a squeezed state in terms of the Bunch-Davies vacua described above.}. Expectedly, one recovers the standard Bunch-Davies vacuum for $\alpha=0$. Since the $\alpha$-vacuum maintains the isometries of $dS_4$, it becomes an exciting and intriguing question to ask if the correlation functions of the scalar and tensor perturbations in $\alpha$-vacua also satisfy the Ward identities that follow from the conformal symmetry. This is the primary motivation of our present paper. 

Although the $\alpha$-vacua as a state respects the conformal symmetries of $dS_4$, certain peculiarities raise concern. It has been reported that for field theories on $dS_4$ Green's functions in $\alpha$-vacuum suffer from non-local singularities between antipodal points \cite{Bousso:2001mw, Banks:2002nv, PhysRevD.31.754, Kaloper:2002cs, Kundu:2011sg, Kundu:2013gha}. However, there have been counterarguments \cite{Goldstein:2003ut} suggesting an appropriate choice of counter-terms can deal with such pathologies. In this paper, we simply assume $\alpha$-vacua is sensible and focus on the symmetry properties of correlators computed in such vacua without aiming to resolve any such debates mentioned above.

Correlators of scalar or tensor vacuum fluctuations in non-Bunch-Davies initial states (e.g., the $\alpha$-vacuum) of inflation have been extensively studied in the literature, see \cite{Chen:2006nt, Holman:2007na, Xue:2008mk, Meerburg:2009ys, Chen:2010xka, Agullo:2010ws, Ashoorioon:2010xg, Ganc:2011dy, LopezNacir:2011kk, Kundu:2011sg, Agarwal:2012mq, Gong:2013yvl, Flauger:2013hra, Aravind:2013lra, Ashoorioon:2014nta, Bahrami:2013isa, Kundu:2013gha, Shukla:2016bnu, Akama:2020jko, Ragavendra:2020vud, Kanno:2022mkx, Ghosh:2022cny, Gong:2023kpe, Akama:2023jsb} for a partial list of references. The primary reference of our present work is \cite{Shukla:2016bnu}, investigating the conformal invariance of the scalar three-point correlator in $\alpha$-vacuum in great detail. In \cite{Shukla:2016bnu}, both of the following situations were considered - a scalar field with cubic self-interaction in the probe limit on a rigid $dS_4$ space-time and the more non-trivial case of inflation where the scalar field fluctuations are coupled to the gravity fluctuations. The conformal Ward identities were shown to be met for the probe scalar field on rigid $dS_4$. However, for inflation, the Maldacena consistency condition was found to be violated in $\alpha$-vacua \footnote{Also see \cite{Agullo:2012cs, Kundu:2013gha} for related explorations in the past.}. They argued that this presumably is due to the diverging back-reaction of the quantum stress tensor in the $\alpha$-vacuum, signaling a possible breakdown of the classical background solution and invalidating the perturbative approximation of slow-roll inflation.

Our paper examines the inflationary Ward identities between the three-point and the four-point scalar correlators computed in $\alpha$-vacuum. Interestingly, we find that these Ward identities are satisfied between the three-point and the four-point scalar inflationary correlators. This is surprising since it is at odds with the negative result for the Ward identities between the two-point and the three-point $\alpha$-vacua correlators reported in \cite{Shukla:2016bnu}. The main result of our present paper is to report this apparently unexpected observation. We provide some speculative comments to justify this anomalous result: Why do the Ward identities work out between the three-point and the four-point correlators but fail for the two-point and the three-point correlators? However, we have been quite unsatisfactorily unable to pinpoint the exact reason. Nevertheless, our result provides enough motivation to revisit the case between the two-point and the three-point correlator worked out in \cite{Shukla:2016bnu}. 

To check the Ward identities mentioned above, we need to know the expressions of the three-point and the four-point scalar correlator in $\alpha$-vacua. In \cite{Shukla:2016bnu}, the three-point inflationary scalar correlator in $\alpha$-vacua was already computed using the in-in formalism \cite{Schwinger:1960qe, Keldysh:1964ud, Weinberg:2005vy} in the leading slow-roll approximation. In this paper, we calculate the four-point inflationary scalar correlator in $\alpha$-vacua, which was not known before, to the leading order in slow-roll parameters. This is one of our new results. 

Calculating the inflationary scalar correlators in $\alpha$-vacua using in-in methods involves a highly technical computation. We also use an alternate method to compute the correlators that further bolster our in-in result. In this method, pioneered in \cite{Maldacena:2002vr}, the correlators are calculated from the wave-function of the universe, evaluated as a function of the values of the fluctuations at the late time slice of $dS_4$. In \cite{Ghosh:2014kba}, the scalar four-point correlator with Bunch-Davies initial condition was obtained in this method, reproducing the expression obtained in \cite{Seery:2006vu, Seery:2008ax} from the in-in method. Additionally, it was noted that the constraints of the conformal symmetry were more clearly manifested in this wave-function method than in the in-in methods. Motivated by this, in \cite{Shukla:2016bnu}, the authors used the wave-function formalism for the scalar three-point function but again only for the Bunch-Davies vacuum. 

In this paper, we also extend the wave-functional method for computing scalar correlators in $\alpha$-vacua, previously restricted to Bunch-Davies initial conditions. Implementing this requires appropriate modifications in the techniques of evaluating Feynmann-Witten diagrams in AdS/CFT correspondence due to $\alpha$-vacua via analytic continuation. Unlike the standard practice in AdS/CFT calculations, we will need to impose non-regular boundary conditions for the modes that diverge in the deep radial interior of the Euclidean $AdS_4$ by hand. The three and four-point scalar correlator in $\alpha$-vacua, computed with this set of rules in the wave-functional method, reproduces the in-in method expressions. This matching provides another justification for the Ward identity being satisfied between three and four-point scalar correlators in $\alpha$-vacua.

We compute the scalar correlators in momentum space, which is common in studying cosmological correlators. Recently, in \cite{Jain:2022uja}, some of the authors of this paper looked at the possible CFT interpretations of three-point correlators of scalar and spinning operators in $\alpha$-vacua. By solving the Ward identities in momentum space, they found that the three-point function in $\alpha$-vacua has a specific structure entirely determined by the corresponding Bunch-Davies result. The results in \cite{Jain:2022uja} apply to any generic CFT. In this paper, we work from the gravity side with a particular example: slow-roll inflation. However, our final result for the three-point correlator in $\alpha$-vacua and a part of our $\alpha$-vacua four-point correlator is consistent with \cite{Jain:2022uja}. 

The sections in the rest of this paper are organized as follows: In \S\,\ref{sec:setup}, we explain the basic idea of our paper, reviewing some of the conventions, approximations, and relevant technical ingredients.
Following that, we get into our main calculations in \S\,\ref{sec:4pininalpha}, where we compute the four-point scalar correlator in $\alpha$-vacua using the in-in formalism. Next, in \S\,\ref{secWI:check}, we check the consistency of conformal Ward identities between the scalar three and four-point correlator computed in $\alpha$-vacua in inflation, which is one of our crucial results in this paper. In \S\,\ref{sec:4pwfalpha}, we compute the scalar correlators (three- and four-point ones) in $\alpha$-vacua using the alternative wave-function method. We summarise with comments and discussion in \S\,\ref{sec:disco}. Finally, many technical details supporting the analysis in the main text are presented in the Appendices.

\section{Basic ideas and setup} \label{sec:setup}

This section will recap the basic facts and technical ingredients needed to compute the three- and four-point inflationary scalar correlators in $\alpha$-vacuum and check the Ward identities. This will also be useful in setting up the conventions. We will be brief in our description, and for more details, the reader should consult \cite{Maldacena:2002vr, Ghosh:2014kba, Shukla:2016bnu}.

\subsection{Setting up notations and gauge fixing the inflationary fluctuations} \label{ssec:dsintro}

We will start with an action that describes a theory of gravity coupled to the inflaton, a scalar field that drives the process of inflation:
\be
\label{eq:action1}
S=\int d^4x\sqrt{-g}M_{Pl}^2  \bigg[{1\over 2} R-{1\over 2} (\nabla \phi)^2   -V(\phi) \bigg] \, ,
\ee
where the Planck mass is related to the Newton's constant $M_{Pl}^2= 1 /(8 \pi G_N)$. The background metric of de-Sitter  space-time is given by:
\begin{equation}\label{dsmetric}
	\begin{split}
		ds^2 =-  dt^2 + a^{2}(t) \, d\vec{x}^2 \, ,  \quad \text{with} \quad 
		a^2 (t) =  e^{2 Ht} \, , 
	\end{split}
\end{equation}
where the scale factor $a(t)$ describes exponential expansion of space-time. The Hubble constant $H$ is defined as $H = \dot{a}/a$. Another commonly used coordinate system to write de-Sitter space-time, involving the conformal time $\eta = - e^{-Ht}/H$ with $-\infty < \eta \le 0$, is 
\begin{equation} \label{dSmeteta}
	ds^2={1\over H^2\eta^2} \left(-d\eta^2 + d\vec{x}^2\right) \, .
\end{equation}

During inflation, the de-Sitter symmetries are broken, resulting in the Hubble parameter being a function of time. In the slow-roll scenario, inflation is modeled by a scalar field $\phi$ called the inflaton, slowly rolling down a slightly inclined potential. As a result, we get a small breaking of the exact de-Sitter geometry controlled by the slow roll parameters, as defined below 
\be
\label{releps3}
\epsilon=-{\dot{H}\over H^2} = {1\over 2}{\dot{\phi}^2\over H^2} \ll 1 \,, \quad \quad \eta=-{\ddot{H}\over 2H \dot{H}} \ll 1 \,  .
\ee
We use conventions where $\dot{a} = da/dt$ and $a'=da/d\eta$. 

In $(3+1)$-dimensions, the symmetry group of de-Sitter is $SO(4,1)$. Apart from the standard spatial translations and rotations, $dS_4$ is also symmetric under the scale transformations and special conformal transformations (SCT) given by 
\begin{equation}\label{eq:scaleSCTtrnsf}
\begin{split}
	\text{Scaling:}& \quad t \to t -(\log s)/H , \quad x^i \to s \, x^i \, , \\
	\text{SCT:}& \quad x^i  \to x^i - 2 (c_j x^j) x^i + c^i \big( (x_j x^j) - (e^{-2Ht}/H^2) \big) \, , \quad t  \to t + (2 c_j x^j/H)
\end{split}
\end{equation}
where $c_j$ ($j=1,2,3$) are the infinitesimal parameters. The Latin indices on the components $x^i,c^i,\dots$ are raised and lowered using the flat space metric $\delta_{ij}$ so that $x^i = x_i, c^i = c_i, \dots$. 

In this paper, we are interested in studying the conformal symmetry of the correlators of operators inserted at the late time boundary of $dS_4$ (i.e., $\eta=0$ in conformal time coordinate of eq.\eqref{dSmeteta}). The symmetries of $dS_4$ are the same as that of a $3$-dimensional Euclidean CFT. The generators of scaling and SCT in eq.\eqref{eq:scaleSCTtrnsf} at $\eta \to 0$ operate on the boundary values of quantum fields in the same way as the generators of $3$-dimensional Euclidean CFT, see \S\,(2.2.1) of \cite{Arkani-Hamed:2018kmz}.

\paragraph*{Gauge fixing for inflationary fluctuations:}
We must fix the gauge appropriately to compute correlation functions of the perturbations about the background solution produced during inflation. It is very convenient to work with the ADM decomposition of the metric
\begin{equation} \label{NNizero}
		ds^2 = - N^2 dt^2 + h_{ij}(dx^i + N^i \, dt)(dx^j + N^j \, dt) \, ,  
\end{equation}
where $N$ and $N^i$ are the lapse and shift functions. We make a choice of gauge to fix $N = 1 \, , \, N^i = 0$ \footnote{This is also known as the synchronous gauge.} such that the metric becomes 
\begin{equation}\label{eq:dSmetric1}
	ds^2 = - dt^2 + h_{ij} dx^i dx^j \, .
\end{equation}
It should be noted that exact $dS_4$ metric is with $h_{ij} = e^{2Ht} \delta_{ij}$ and we can now define the metric perturbations given by
\be
\label{gfversion1}
h_{i j}=e^{2 H t} g_{i j} \, , \quad \quad \text{with} \quad g_{ij}  =  \delta_{ij} + \gamma_{ij} \,  , \quad \quad 
\text{and} \quad \gamma_{ij}  =  2 \zeta \delta_{ij} + \widehat{\gamma}_{ij} \, . 
\ee
note that we have extracted out the trace part of $\gamma_{ij}$ as $\zeta$ and hence $\widehat{\gamma}_{ij}$ is traceless $\widehat{\gamma}_{ii}=0$. 

In addition to the metric perturbations, we also have inflaton perturbations about its background solution $\bar{\phi}(t)$ 
\begin{equation}\label{eq:scalperturb}
	\phi = \bar{\phi}(t) + \delta \phi \, .
\end{equation}

As detailed in \cite{Ghosh:2014kba}, the coordinate transformations that brought the metric to the form of eq.\eqref{eq:dSmetric1} do not completely fix the gauge of diffeomorphism invariance. It turns out one can further use the freedom of residual time and spatial reparametrization to set either of the two following gauges \footnote{Apart from deciding the scalar perturbation to be either denoted by either $\delta \phi$ or $\zeta$, in both the gauges, the graviton perturbation is traceless by construction and also transverse.} 
\begin{equation}\label{eq:gauge12}
	\begin{split}
		\text{Gauge 1:} \quad \delta \phi = 0, \,\, \zeta \neq 0 \quad  &\text{and} \quad  \partial_i \widehat{\gamma}^{ij} = 0 \, , \\
		\text{Gauge 2:} \quad \zeta = 0, \,\, \delta\phi \neq 0 \quad &\text{and} \quad  \partial_i \widehat{\gamma}^{ij} = 0 \, .
	\end{split} 
\end{equation}
These two gauges are related to each other by
\be
\label{infper}
\delta \phi= -(\dot{\bar{\phi}} / H)\, \zeta = - \sqrt{2 \epsilon} \, \zeta \, . 
\ee
where in the last step, we have used eq.\eqref{releps3}. The gauge invariant combination $ \zeta - (H/\dot{\bar{\phi}}) \delta \phi$ is actually the variable that freezes out (i.e., stays constant) once the modes cross the horizon \cite{Kundu:2015xta,Maldacena:2002vr} \footnote{The horizon crossing of any mode is understood as the situation when the physical momenta of the modes, denoted by $k/a$ with $a$ being the scale factor, becomes much smaller than the Hubble parameter $H$.}. We use the following prescription in our calculations: while the modes are inside the horizon, we use Gauge-$2$ such that $\zeta =0$ and $\delta\phi$ denote the scalar fluctuations and compute their correlation functions. Once the modes leave the horizon, we go over to gauge-$1$ using eq.\eqref{infper} such that now $\delta\phi=0$ and the correlator is expressed in terms of $\zeta$. 

The appearance of the slow-roll parameter as $\sqrt{\epsilon}$ on the RHS of the relation eq.\eqref{infper} is also crucial to note. For calculating correlators, we start in gauge-$2$ and compute $\langle \delta\phi\delta\phi\delta\phi\delta\phi \rangle$ in exact $dS_4$ ignoring the breaking due to slow-roll. Finally, we move to gauge-$1$, using eq.\eqref{infper} and obtain $\langle \zeta \zeta \zeta \zeta \rangle$ in the leading slow-roll approximation. More precisely, as we will see later, 
\begin{equation}
\langle \delta\phi\delta\phi\delta\phi\delta\phi \rangle \,  \sim \, \mathcal{O}(\epsilon^0) \quad \text{implying} \quad  \langle \zeta \zeta \zeta \zeta \rangle \, \sim \,  \mathcal{O}(\epsilon^{-2})
\end{equation} 
to the leading order in the slow-roll parameter. In our present paper, we will see that the same holds even for the same correlator in $\alpha$-vacuum.

\subsection{A generic interacting $\alpha$-vacua}
\label{ssec:alphavacua}

We will now briefly describe some defining properties of the $\alpha$-vacua for a massless free scalar field $\phi$ in $dS_4$ background with the following action
\begin{equation} \label{eq:phids2}
		S = -\dfrac{1}{2} \int d^4 x \, \sqrt{-g} \, g^{\mu\nu} \partial_{\mu} \phi \partial_{\nu} \phi \, = \dfrac{1}{2} \int d^3 \vect{x}\, d \eta \, \dfrac{1}{\eta^2 H^2} \left( (\partial_{\eta} \phi)^2 - (\partial_i \phi)^2 \right) \, ,
\end{equation}
where in the second step, we used the $dS_4$ metric in eq.\eqref{dSmeteta}. At this point, let us mention that we are following a convention similar to what was used in \cite{Ghosh:2014kba} where vectors are denoted by boldface letters $\vect{x},\vect{k}$ and the components are denoted by $k^i$. We thus have $\vect{k} \cdot \vect{x} = k_i x_i$ since Latin indices are raised and lowered by the flat Euclidean metric $\delta_{ij}$.

The equations of motion are given by
$\partial^2_{\eta} \phi - (2/\eta) \partial_{\eta} \phi -\partial^2_i \phi = 0$, which admits the general solutions $U_k^\alpha(\eta)$ as given by 
\begin{equation}\label{eq:modefunctions}
	U_k^{\alpha}(\eta) = \dfrac{H}{\sqrt{2k^3}} \left[ c_1(1-ik\eta) e^{ik \eta} + c_2 (1 + ik\eta) e^{-ik\eta} \right] \, .
\end{equation}
One can now write a mode expansion of the scalar field (now thought of as an operator) in terms of the creation and annihilation operators ($a^+_{\vect{k}}$ and $a^-_{\vect{k}}$) as 
\begin{equation}\label{eq:modeexpand}
	 \phi(\eta,\vect{x}) = \int \dfrac{d^3\vect{k} \, e^{i \vect{k} \cdot \vect{x}}}{(2\pi)^3} \left[ a^-_{\vect{k}} \,  U_k^{\alpha*}(\eta) + a^+_{-\vect{k}} \, U_k^{\alpha}(\eta) \right] \, .
\end{equation}
The normalization of $U_k^\alpha(\eta)$ can be maintained by writing the coefficients $c_1$ and $c_2$ as
\begin{equation}\label{eq:c1c2def}
	c_1 = \cosh(\alpha) \, , \quad \quad c_2 = -i\sinh(\alpha) \, .  
\end{equation}
Also, the standard commutation relations between $\phi$ and its momentum conjuagte $\pi$, as well as between $a^-_{\vect{k}}$ and $a^+_{\vect{k}}$ are imposed, \cite{PhysRevD.32.3136},  
\begin{equation}
\begin{split}
&[\phi(\eta,\vect{x}),\pi(\eta,\vect{x}')] = i \delta^3(\vect{x} - \vect{x}') \, , \\
	[a^-_{\vect{k}},a^+_{\vect{p}}] = &(2\pi)^3 \delta^3(\vect{k} - \vect{p}) \, , ~~ [a^-_{\vect{k}}, a^-_{\vect{p}}] = 0 \, , ~~  [a^+_{\vect{k}}, a^+_{\vect{p}}] = 0 \, .
	\end{split}
\end{equation}
We can now define the free $\alpha$-vacuum state as \footnote{Note that, for the mode functions $U_k^\alpha(\eta)$ to be normalized, we can always have a relative phase factor between $c_1$ and $c_2$ in eq.\eqref{eq:c1c2def}: $c_1 = \cosh(\alpha) \, , \quad c_2 = -i\, e^{-i\beta} \sinh(\alpha)$. We have specifically chosen to set $\beta =0$. 
Let us also emphasize that this choice is not just to simplify our calculations. In fact, we will obtain the expressions for the scalar three and four-point functions for arbitrary $\alpha$ and $\beta$. However, for further examinations of the conformal invariance of the correlators, we will focus on $\beta =0$. There are concerns regarding realizing the de-Sitter symmetries for $\beta \neq 0$. As noted in \cite{PhysRevD.32.3136} (see \S II.C), the Green's function is invariant under the full disconnected de-Sitter group only when $\beta=0$.} $a^-_{\vect{k}} | \alpha \rangle = 0 \, ,~ \forall ~ \vect{k}$.

The standard Bunch-Davies vacuum and the corresponding mode expansion can be obtained by setting $\alpha =0$ in the relations written above,
\begin{equation} \label{defUkBD}
\begin{split}
	\phi(\eta,\vect{x}) &= \int \dfrac{d^3\vect{k} \, e^{i \vect{k} \cdot \vect{x}}}{(2\pi)^3} \left[ b^-_{\vect{k}} U^{BD*}_k(\eta) + b^+_{-\vect{k}} U_k^{BD}(\eta) \right] \, ,\\
	\text{with}, \quad U_k^{BD}(\eta) &= \dfrac{H}{\sqrt{2k^3}}  (1 - ik\eta) e^{ik \eta} \, .
	\end{split}
\end{equation}
The mode function for the Bunch-Davies vacuum is defined to be the one that vanishes as we let $\eta \to -\infty(1-i\epsilon)$ with $\epsilon>0$. The creation and annihilation operators for the Bunch-Davies vacua (i.e., $b^-_{\vect{k}}$ and $b^+_{\vect{k}}$) are related to the corresponding ones in alpha vacua by Bogoliubov transformations of the form $a^-_{\vect{k}} = c_1\, b^-_{\vect{k}} - c_2 \,b^+_{-\vect{k}} \, , \, b^-_{\vect{k}} = c^*_1 \,a^-_{\vect{k}} + c_2\, a^+_{-\vect{k}}$. 

One can express the free vacuum $| \alpha \rangle$ as a squeezed state in terms of the free Bunch-Davies vacuum $| 0 \rangle$, defined as $b^-_{\vect{k}} |0 \rangle = 0 \, , \, \forall \, \vect{k}$. Similarly, once we introduce interactions, one can get an interacting $\alpha$-vacuum $| \alpha \rangle_{\text{int}}$ as a squeezed state in terms of an interacting Bunch-Davies vacuum, say $|\Omega\rangle$, as
\begin{equation} \label{alphaBDrel1}
	| \alpha \rangle_{\text{int}} = \dfrac{1}{\mathcal{N}} \exp\left( \dfrac{i}{2} \tanh(\alpha) \int \dfrac{d^3\vect{k}}{(2\pi)^3} b^-_{\vect{k}} b^+_{-\vect{k}} \right) |\Omega \rangle \, .
\end{equation}
In \cite{Shukla:2016bnu} \footnote{See Appendix-B in \cite{Shukla:2016bnu} for the details. It was extensively checked for $\phi^3$ interaction.}, by writing the generators of the conformal transformations in terms of the creation and annihilation operators and working out their action on the $\alpha$-vacuum, it was argued that the interacting $\alpha$-vacuum preserves conformal symmetry. This fact is the primary motivation for us to expect that the scalar correlators computed in a conformally invariant initial state, such as the $\alpha$-vacuum, should also be conformally invariant. 


\subsection{Techniques to compute inflationary scalar correlators in $\alpha$-vacuum}

Since our primary goal in this paper is to compute correlation functions of scalar perturbations produced during inflation, let us present a short overview of the techniques that we will use to compute the correlation functions. These are known as (a) the Schwinger-Keldysh or In-In formalism, and (b) a formalism involving the wave-function in $dS$. For the sake of brevity, for each of them we will just list out the operational steps that one follows in practice.

\subsubsection{Schwinger-Keldysh or In-In formalism}
\label{ssec:inin_review}

Cosmological correlators of inflationary fluctuations are traditionally calculated using the in-in formalism, alternatively referred to as the Schwinger-Keldysh formalism \footnote{See for example \cite{Chen:2010xka, Wang:2013zva, Baumann:2018muz} for reviews of the in-in formalism.}. This method uses a double line contour to perform a path-integral and obtain $\langle\mathcal{O}(\eta^*)\rangle =\bra{\Omega}\mathcal{O}(\eta^*)\ket{\Omega}$. Starting with the \emph{`in'} vacuum state (a ket-state $\ket{\Omega}$) at $\eta = -\infty$, one evolves along a future-directed contour to encounter the operator insertions at $\eta^* = 0$ and then evolve back to another \emph{`in'} state (a bra-state $\bra{\Omega}$) at $\eta = -\infty$. In the context relevant for inflationary correlators, $\eta^*$ is the late time slice at $\eta^* = 0$, and the operator in $\langle\mathcal{O}(\eta^*)\rangle$ is going to be the four scalar fluctuations $\langle \zeta \zeta\zeta\zeta\rangle $ where all $\zeta$'s are inserted at the late time slice. This double-line contour evolution can be formally written as 
\begin{equation}\label{eq:master_form_inin}
\begin{split}
\bra{\Omega}\mathcal{O}(\eta^*)\ket{\Omega} =\bra{0}\bar{T}\left[e^{i\int_{-\infty}^{\eta^*}d\eta'H^I_{\text{int}}(\eta')}\right]\, \mathcal{O}(\eta^*) \,\, T \left[e^{-i\int_{-\infty}^{\eta^*}d\eta'H^I_{\text{int}}(\eta')} \right] \ket{0} \, ,
	\end{split}
\end{equation}
where $\ket{\Omega}$ is the interacting vacuum of the full theory, $\ket{0}$ is the vacuum of the free theory, and $T$ and $\bar{T}$ represent time ordering and anti-time ordering of the operators on the future-directed and past-directed contour respectively. Also, $H^I_{\text{int}}$ is the interacting part of the Hamiltonian written in the interaction picture. Starting with eq.\eqref{eq:master_form_inin}, one has to perform a perturbative expansion in the coupling constant of the interaction term (contained within $H^I_{\text{int}}$) and keep terms up to quadratic order to obtain 
\begin{equation}\label{eq:form_inin1}
\begin{split}
\bra{\Omega}\mathcal{O}(\eta^*)\ket{\Omega} = & \,  i\int_{-\infty}^{\eta^*}d\eta' \, \bra{0}\left[H^I_{\text{int}}(\eta'),\mathcal{O}(\eta^*)\right] \ket{0} \\ & -\int_{-\infty}^{\eta^*} d\eta' \int_{-\infty}^{\eta'}d\eta''\, \bra{0}\left[H^I_{\text{int}}(\eta''), \, \left[H^I_{\text{int}}(\eta'),\mathcal{O}(\eta^*)\right]\right]\ket{0} + \cdots \, , 
\end{split}
\end{equation}
where on the RHS the $`\cdots'$ represents higher order terms in $H^I_{\text{int}}$ which we have neglected. This eq.\eqref{eq:form_inin1} is going to be the main formula that we will use while computing $\langle \zeta \zeta\zeta\zeta\rangle $ using in-in formalism in \S\,\ref{sec:4pininalpha}. 

The information of $\alpha$-vacuum will enter through the mode expansion of the $\delta\phi$ in eq.\eqref{eq:modeexpand}, which will contribute to $H^I_{\text{int}}$ relevant to calculate the scalar four-point correlator. To ensure the convergence of the $\eta$-integrations in eq.\eqref{eq:form_inin1} in the far past at $\eta \to - \infty$, we will follow the rules specified in \cite{Shukla:2016bnu}. The prescription is to deform the contour of the $\eta$-integration as $\eta \to \eta(1 \pm i \epsilon)$, with $\epsilon > 0$ as an infinitesimal parameter.

The expressions for the scalar two-point and three-point inflationary correlators in $\alpha$-vacuum are already worked out using this in-in formalism in the literature. For example, in \cite{Shukla:2016bnu}, the three-point scalar correlator was calculated, see eq.(5.17) in that paper, which upon taking $\alpha =0$ reduces to the answer in Bunch-Davies vacuum. Following the steps outlined in this sub-section in \S\,\ref{sec:4pininalpha} we will compute $\langle \alpha | \zeta(\vect{k_1}) \zeta(\vect{k_2}) \zeta(\vect{k_3}) \zeta(\vect{k_4})|\alpha\rangle$ in $\alpha$-vacuum.

\subsubsection{Wave-function at late times in $dS_4$ and analytic continuation to $EAdS_4$}
\label{sec:analyticcont}

An equivalent procedure to obtain the inflationary correlation functions is by computing \emph{the wave function of the universe} \cite{Maldacena:2002vr, Ghosh:2014kba}, called $\psi$, which is a functional of late time values (at $\eta =0$) of the scalar or tensor perturbations $\delta \phi$ and $\gamma_{ij}$. Formally, it is defined as the following path-integral 
\begin{equation}\label{eq:pathintwf}
	\psi[\varphi(\vect{x}, \, \eta =0)] = \int^{\varphi(\vect{x}, \, \eta =0)}_{\varphi(\vect{x}, \, \eta = -\infty)} \mathcal{D} \varphi \, e^{i \, S[\varphi]} \, ,
\end{equation}
where, $\varphi$ collectively denotes the perturbation fields $\delta \phi$ and $\gamma_{ij}$. Also, $S$ represents the effective action for the perturbations $\varphi$, obtained by substituting eq.\eqref{gfversion1} and eq.\eqref{eq:scalperturb} in eq.\eqref{eq:action1} and eliminating the background fields. The path-integral in eq.\eqref{eq:pathintwf} is performed with the field configurations at far past $\varphi(\vect{x}, \, \eta = -\infty)$ as an initial condition. In this paper this initial condition at $\eta = -\infty$ for the fluctuations $\varphi$ will be considered to be in the $\alpha$-vacuum, eq.\eqref{eq:modefunctions} \footnote{Note that the path-integration in eq.\eqref{eq:pathintwf} is made consistent through the gauge-fixing procedure that we discussed in \S\,\ref{ssec:dsintro}, specifically the transverse gauge condition for the graviton.}.

In the semiclassical limit $M_{Pl} \gg H$ \footnote{This approximation is valid since the action $S$ in eq.\eqref{eq:pathintwf} contains an overall factor $(M_{Pl}/ H)^2$.} we can implement the saddle point approximation to obtain 
\begin{equation} \label{wavefnsaddle}
\psi[\varphi(x)] = \exp\left[i \, S^{\text{dS}}_{\text{on-shell}}[\varphi(x)]\right] \, .
\end{equation}
To the leading order in slow-roll parameters, the on-shell action $S^{\text{dS}}_{\text{on-shell}}[\delta\phi(x)]$ for the scalar perturbations will essentially be that of a free scalar field in the background $dS_4$ geometry. The mode expansion discussed in \S\,\ref{ssec:alphavacua} can also be straightforwardly used for $\delta\phi(x)$.

It is known that the scalar and tensor perturbations produced during inflation follow a nearly Gaussian distribution statistic with small deviations denoted by the non-Gaussianities. Working in momentum space \footnote{For Fourier transform our convention is the following: $\varphi(\vect{x}) = \int (d^3\vect{k}/(2\pi)^3) \, \varphi(\vect{k}) e^{i \vect{k} \cdot \vect{x}}$.},  this feature can be made manifest by writing the wave function at late times in the following schematic form 
\begin{equation}\label{eq:wfgauge2}
	\begin{split}
		\psi[\delta \phi(\vect{k}),&\gamma^{ij}(\vect{k})] = \exp \left[ \dfrac{M^2_{Pl}}{H^2} \left( - \dfrac{1}{2} \int \dfrac{d^3 \vect{k}_1}{(2\pi)^3} \dfrac{d^3 \vect{k}_2}{(2\pi)^3} \,\delta \phi(\vect{k}_1) \delta \phi(\vect{k}_2) \,\langle O(-\vect{k}_1) O(-\vect{k}_2) \rangle\right. \right. \\
		& - \dfrac{1}{2} \int \dfrac{d^3 \vect{k}_1}{(2\pi)^3} \dfrac{d^3 \vect{k}_2}{(2\pi)^3} \gamma_{ij} (\vect{k}_1) \gamma_{kl} (\vect{k}_2) \,\langle T^{ij} (-\vect{k}_1) T^{kl}(-\vect{k}_2) \rangle  \\ 
		& - \dfrac{1}{4} \int \left[\prod_{J=1}^{3}  \dfrac{d^3 \vect{k}_J}{(2\pi)^3} \right] \delta \phi(\vect{k}_1) \delta \phi(\vect{k}_2) \gamma_{ij} (\vect{k}_3)\, \langle O(-\vect{k}_1) O(-\vect{k}_2) T^{ij} (-\vect{k}_3) \rangle \\
		& + \left. \left. \dfrac{1}{4!} \int \left[\prod_{J=1}^{4}  \dfrac{d^3 \vect{k}_J}{(2\pi)^3} \right] \left[\prod_{J=1}^{4} \delta \phi(\vect{k}_J) \right] \langle O(-\vect{k}_1) O(-\vect{k}_2) O(-\vect{k}_3) O(-\vect{k}_4) \rangle \right) \right] \, ,
	\end{split}
\end{equation}
where the objects $\langle O(-\vect{k}_1) O(-\vect{k}_2) \rangle$, $\langle T^{ij} (-\vect{k}_1) T^{kl}(-\vect{k}_2) \rangle$, $\langle O(-\vect{k}_1) O(-\vect{k}_2) T^{ij} (-\vect{k}_3) \rangle$, etc. are for now just coefficients of the wave-function. However, they are written in a suggestive manner such that they will become correlation functions in a three-dimensional Euclidean CFT \footnote{In gauge-$1$ the graviton is transverse and traceless we may write the two polarisations of it defined as $\gamma_{ij}(\vect{k}) = \sum_{s=1}^{2} \gamma_s(\vect{k}) \epsilon^s_{ij}(\vect{k})$, where $\epsilon^s_{ij}(k)$ are the transverse and traceless polarization tensors}. In eq.\eqref{eq:wfgauge2}, we have explicitly written down only the scalar and graviton two-point correlator and the graviton-scalar-scalar three-point correlator, relevant for the four-point scalar correlator. 

The wave function can be interpreted as the probability amplitude and its modulus square as the probability distribution for the scalar and tensor fluctuations at the $\eta=0$ surface at the end of the inflation. The correlation functions of the scalar or tensor fluctuations can now be straightforwardly obtained by inserting them in the path integration, weighted by the modulus square of the wave function. The scalar $n$-point correlator will be computed as
\begin{equation} \label{def_cor_fn}
\begin{split}
\langle  \delta \phi(\vect{k}_1) \delta \phi(\vect{k}_2) \cdots \delta \phi(\vect{k}_n) \rangle =& \dfrac{1}{\mathcal{C}_{norm}} \int \mathcal{D} \delta \phi \, \mathcal{D} \gamma_{ij} \left(\prod_{i=1}^n \delta \phi(\vect{k}_i) \right)\, \bigg|\psi[\delta \phi(\vect{k}),\gamma_{ij}(\vect{k})]\bigg|^2 \, ,\\
\text{with} \quad \mathcal{C}_{norm} = & \int \mathcal{D} \delta\phi \, \mathcal{D} \gamma_{ij} \, \bigg|\psi[\delta \phi(\vect{k}),\gamma_{ij}(\vect{k})]\bigg|^2 \, .
\end{split}
\end{equation}
It is straightforward now to perform the transformation to gauge-$1$ and write the final expression in terms of $\zeta$.   

\paragraph*{Analytic continuation to $EAdS_4$:}
Following \cite{Ghosh:2014kba}, we will perform the calculations by analytically continuing to Euclidean anti-de-Sitter space ($EAdS$). The coefficients functions are boundary CFT correlation functions, which can be computed from the Witten diagrams in $EAdS$. We will precisely use this approach to calculate the four-point and three-point scalar correlators in alpha vacua in the section \ref{sec:4pwfalpha}. In this procedure, we do all the calculations first in $EAdS_4$ and then analytically continue back them to $dS_4$. 

After performing the analytic continuations 
\be
\label{anaz}
\eta = i \, z\, , \quad \quad \text{and} \quad \quad H= i\,/ \text{R}_{\text{AdS}} \, ,
\ee
the $dS_4$ metric in eq.\eqref{dSmeteta} goes over to the Poincare $EAdS_4$ metric
\be
\label{eads}
ds^2 = \frac{\text{R}_{\text{AdS}}^{\,2}}{z^2} \, \bigg(dz^2 + \sum_{i=1}^3 dx_i \, dx^i \bigg)\, ,
\ee
where $\text{R}_{\text{AdS}}$ is the $EAdS$ radius, and $z \in [0,\infty)$ is the radial coordinate with $z = 0$ being the boundary. Also, the wave function in de Sitter space becomes the partition function in $EAdS_4$ space via the same analytic continuation, implying 
\be
\label{eq:dsads}
Z_{\text{EAdS}}[\varphi(x)] \equiv \text{e}^{-S^{\text{EAdS}}_{on-shell}[\varphi(x)]} \xLeftrightarrow[\text{R}_{\text{AdS}} = i/H]{z = -\,i\eta} \Psi[\varphi(x)] \equiv \text{e}^{i \, S^{\text{dS}}_{on-shell}[\varphi(x)]} \, .
\ee
The $EAdS_4$ partition function $Z_{\text{EAdS}}$ is a functional of the boundary value of the fields. For our interest in $\alpha$-vacuum of $dS_4$, as $z \to 0$ is approached in $EAdS_4$, we get
\begin{equation}
	\varphi(\vect{x}) = \int \dfrac{d^3\vect{k} \, e^{i \vect{k} \cdot \vect{x}}}{(2\pi)^3} \, \varphi(\vect{k}) \, (c_1 + c_2) \, ,
\end{equation}

\paragraph*{Equivalent boundary condition in $EAdS_4$ for $\alpha$-vacuum in $dS_4$:}
The path integral in eq.\eqref{eq:pathintwf} is evaluated by choosing appropriate boundary conditions in the far past at $\eta = - \infty$. Imposing the Bunch-Davies boundary condition corresponds to demanding that the mode functions are effectively in the Minkowski background. After performing analytic continuation, this corresponds to requiring regular ingoing boundary conditions at $z=\infty$ for the fields that are now in $EAdS_4$,
\begin{equation}
U_k^{BD}(\eta) = \dfrac{H}{\sqrt{2k^3}}  (1-ik\eta) e^{ik \eta} \quad \Leftrightarrow \quad U_k^{reg}(z) = \dfrac{ (1+kz) e^{-kz}}{\sqrt{2k^3} \, \text{R}_{\text{AdS}}}  \, .
\end{equation} 
On the other hand, the $\alpha$-vacuum corresponds to an excited state of the mode functions in the far past in $dS_4$. Hence, after the analytic continuation to $EAdS_4$, this should correspond to a deviation from imposing the regular boundary condition at $z=\infty$.

From eq.\eqref{eq:modeexpand}, eq.\eqref{eq:modefunctions} we know that for a scalar mode $\varphi = f_{\vect{k}}(\eta) e^{i \vect{k} \cdot \vect{x} }$ in the $\alpha$-vacuum one should obtain 
\begin{equation}\label{eq:modefkalpha}
\begin{split}
	f_{\vect{k}} &=  \varphi(\vect{k}) \left[ \cosh(\alpha) \, (1-ik\eta) \, e^{ik\eta} -i \sinh(\alpha) \, (1+ik\eta)e^{-ik\eta} \right] \, \, .
	\end{split}
\end{equation}
Similarly, a generic solution to the Klein-Gordon equation in $EAdS_4$ can be obtained after using the analytic continuation eq.\eqref{anaz} in eq.\eqref{eq:modefkalpha} as
\begin{equation}\label{eq:solscalads}
	\varphi(z,\vect{x}) = \int \dfrac{d^3 \vect{k} \, e^{i\vect{k} \cdot \vect{x}}}{(2\pi)^3}\,  \varphi(\vect{k}) \, \left(c_1 \, (1+kz)e^{-kz} + c_2 \, (1-kz)e^{kz}\right) \, .
\end{equation}
Due to the presence of the $e^{kz}$ piece on the RHS in eq.\eqref{eq:solscalads}, blowing up at $z \to \infty$, this boundary condition is not regular. Our approach would be to ignore the possible implications of this non-standard choice of boundary condition. We would instead take a viewpoint that we impose this by hand; it is forced upon us by the analytic continuation from $dS_4$, and as we will see later, it produces the expected result for us.

In this paper, in \S\,\ref{sec:4pwfalpha} we will compute first the $4$-point $\langle  \zeta \zeta \zeta \zeta \rangle_\alpha$ in $\alpha$-vacuum by using the wave-function method. We will then compute the $3$-point $\langle  \zeta \zeta \zeta \rangle_\alpha $ using the same formalism. To compute $\langle  \zeta \zeta \zeta \zeta \rangle_\alpha$ we will need to know the wave-function coefficients $\langle OOT^{ij}\rangle$ and $\langle OOOO\rangle$ for the $\alpha$-vacuum. We will perform that task in \S\,\ref{sec:4pwfalpha} using the basic setup discussed in this sub-section. 

\subsection{Conformal symmetry and Ward identities for inflationary correlators}

Checking conformal symmetry of the $\alpha$-vacua correlators is implemented by verifying if the correlators satisfy the Ward identities of conformal invariance   \footnote{It was argued in \cite{Kundu:2015xta} that the Ward identities follow from spatial and time reparametrization invariance which are residual freedom remaining after one has chosen gauge-$1$ or gauge-$2$. Hence, since diffeomorphism invariance implies the Ward identities, they are expected to hold in even in situations where the conformal symmetry is badly broken (like DBI inflation \cite{Alishahiha:2004eh}).}. Since the interacting $\alpha$-vacuum defined in \S\,\ref{ssec:alphavacua} is conformally invariant \cite{Shukla:2016bnu}, we are going to demand that the correlators computed in this vacuum should naturally maintain conformal symmetry. 

The wave-function method is particularly insightful when checking conformal invariance compared to the in-in method. The wave function, defined in eq.\eqref{eq:pathintwf}, contains all the information about the scalar and tensor fluctuations imprinted on the late time slice at $\eta=0$ when inflation ends. Therefore, we demand that the wave function should be conformally invariant, inheriting the symmetry of the background geometry. This requirement imposes certain constraints on the structure of the correlation functions, expressed in the form of Ward identities satisfied by the correlation function of metric and inflaton perturbations \cite{Mata:2012bx, Kundu:2014gxa, Kundu:2015xta, Shukla:2016bnu}. The most general form of these Ward identities are written in eq.(3.25) and eq.(3.38) in \cite{Kundu:2015xta}. 

We write down the Ward identities between the following two pairs,  the two-point with three-point and the three-point with the four-point correlator, which are relevant to our discussions in this paper. For scaling transformations, we get the following Ward identities
\begin{equation}
\Bigg(3+\sum_{a=1}^{2}\vect{k}_a\cdot\frac{\partial}{\partial\vect{k}_a}\Bigg)\langle\zeta(\vect{k}_1)\zeta(\vect{k}_{2})\rangle'  =-\frac{\langle\zeta(\vect{k}_1)\zeta(\vect{k}_{2})\zeta(\vect{k}_{3})\rangle'}{\langle\zeta(\vect{k}_{3})\zeta(-\vect{k}_{3})\rangle'}\Bigg|_{\vect{k}_{3}\rightarrow 0} \, , \label{eq:scal_ward1}
\end{equation}
and 
\begin{equation}
\Bigg(6+\sum_{a=1}^{3}\vect{k}_a\cdot\frac{\partial}{\partial\vect{k}_a}\Bigg)\langle\zeta(\vect{k}_1)\zeta(\vect{k}_{2})\zeta(\vect{k}_{3})\rangle' =-\frac{\langle\zeta(\vect{k}_1)\zeta(\vect{k}_{2})\zeta(\vect{k}_{3})\zeta(\vect{k}_{4})\rangle'}{\langle\zeta(\vect{k}_{4})\zeta(-\vect{k}_{4})\rangle'}\Bigg|_{\vect{k}_{4}\rightarrow 0} \,  \label{eq:scal_ward2}
\end{equation}
Here, we are using the convention that a prime in $\langle\zeta(\vect{k}_1)\zeta(\vect{k}_{2})...\zeta(\vect{k}_{n})\rangle'$ means that the momentum conserving delta function has been stripped off from the correlator
\begin{equation}
\langle\zeta(\vect{k}_1)\zeta(\vect{k}_{2})...\zeta(\vect{k}_{n})\rangle = (2\pi)^3 \delta^3 \left(\vect{k_1}+\vect{k_2}+...+\vect{k_n} \right)\,  \langle\zeta(\vect{k}_1)\zeta(\vect{k}_{2})...\zeta(\vect{k}_{n})\rangle'\, .
\end{equation}
In the literature, the relation eq.\eqref{eq:scal_ward1} is known as the Maldacena consistency condition, which is traditionally written as
\begin{equation}\label{maldcon}
\begin{split}
\lim_{\vect{k}_{3}\rightarrow 0}\, \langle\zeta(\vect{k}_1)\zeta(\vect{k}_{2})\zeta(\vect{k}_{3})\rangle' = -n_s \, \langle\zeta(\vect{k}_{1})\zeta(-\vect{k}_{1})\rangle'\, \langle\zeta(\vect{k}_{3})\zeta(-\vect{k}_{3})\rangle' \, ,
\end{split}
\end{equation}
where $n_s$ is called the scalar tilt \footnote{It can be argued that the tilt $n_s$ is of order the slow-roll parameter, i.e., $n_s \sim \mathcal{O}(\epsilon)$.}, defined through the two-point function 
\begin{equation} \label{2point_titl}
\langle\zeta(\vect{k}_1)\zeta(\vect{k}_{2})\rangle = (2\pi)^3 \delta^3 \left(\vect{k_1}+\vect{k_2}\right)\,  \dfrac{H^2}{M_{Pl}^2} \dfrac{1}{4\epsilon} k_1^{(-3+n_s)}\, .
\end{equation}
Similarly, for special conformal transformation (SCT) with a vector parameter $\vect{b}$, we have the following Ward identities,
\begin{equation}
\Bigg(\sum_{a=1}^{2}\hat{\mathcal{L}}^{\vect{b}}_{\vect{k}_a}\Bigg)\langle\zeta(\vect{k}_1)\zeta(\vect{k}_{2})\rangle =  -2\Bigg[\vect{b}\cdot\frac{\partial}{\partial\vect{k}_{3}}\Bigg]\frac{\langle\zeta(\vect{k}_1)\zeta(\vect{k}_{2})\zeta(\vect{k}_{3})\rangle}{\langle\zeta(\vect{k}_{3})\zeta(-\vect{k}_{3})\rangle'}\Bigg|_{\vect{k}_{3}\rightarrow 0} \, , \label{eq:sct_ward_1}
\end{equation}
and 
\begin{equation}
\Bigg(\sum_{a=1}^{3}\hat{\mathcal{L}}^{\vect{b}}_{\vect{k}_a}\Bigg)\langle\zeta(\vect{k}_1)\zeta(\vect{k}_{2})\zeta(\vect{k}_{3})\rangle =-2\Bigg[\vect{b}\cdot\frac{\partial}{\partial\vect{k}_{4}}\Bigg]\frac{\langle\zeta(\vect{k}_1)\zeta(\vect{k}_{2})\zeta(\vect{k}_{3})\zeta(\vect{k}_{4})\rangle}{\langle\zeta(\vect{k}_{4})\zeta(-\vect{k}_{4})\rangle'}\Bigg|_{\vect{k}_{4}\rightarrow 0} \, ,\label{eq:sct_ward_2}
\end{equation}
where $\hat{\mathcal{L}}^{\vect{b}}_{\vect{k}}$ is given as,
\begin{equation} \label{defLbk}
    \hat{\mathcal{L}}^{\vect{b}}_{\vect{k}}=2\Bigg(\vect{k}\cdot\frac{\partial}{\partial\vect{k}}\Bigg)\Bigg(\vect{b}\cdot\frac{\partial}{\partial\vect{k}}\Bigg)-(\vect{b}\cdot\vect{k})\Bigg(\frac{\partial}{\partial\vect{k}}\cdot\frac{\partial}{\partial\vect{k}}\Bigg)+6\Bigg(\vect{b}\cdot\frac{\partial}{\partial\vect{k}}\Bigg) \, .
\end{equation}
It is worth highlighting that the breaking of exact $dS_4$ symmetries has been incorporated in the Ward identities written above. This is, for example, obvious from the appearance of scalar tilt $n_s$ in the Maldacena consistency condition written in eq.\eqref{maldcon}. 

Using the wave function in eq.\eqref{eq:wfgauge2}, one can re-write these identities involving the coefficient functions, the correlators in a three-dimensional Euclidean CFT. For example, the scaling Ward identity becomes
\begin{equation}\label{WI_O3_O4}
\Bigg(\sum_{a=1}^{3}\vect{k}_a\cdot\frac{\partial}{\partial\vect{k}_a}\Bigg)\langle O(\vect{k}_1)O(\vect{k}_{2})O(\vect{k}_{3})\rangle =\dfrac{\dot{\bar{\phi}}}{H} \, \langle O(\vect{k}_1)O(\vect{k}_{2})O(\vect{k}_{3})O(\vect{k}_{4})\rangle \bigg|_{\vect{k}_{4}\rightarrow 0} \, .
\end{equation}
With exact conformal symmetry, the RHS above would be zero. However, the RHS can be computed using a conformal perturbation theory in inflation with the small breaking of that symmetry incorporated. Hence, the four-point function is calculated in an exact CFT (or in an exact $dS_4$) to the leading order in the breaking captured by $\dot{\bar{\phi}}/H = \sqrt{2\epsilon}$.

Finally, we end this section by mentioning that these Ward identities, in eq.\eqref{eq:scal_ward1}, \eqref{eq:scal_ward2}, \eqref{eq:sct_ward_1} and eq.\eqref{eq:sct_ward_2}, were shown to be satisfied for scalar correlators computed in the Bunch-Davies vacuum \cite{Ghosh:2014kba}, \cite{Kundu:2014gxa}, \cite{Kundu:2015xta}. Interestingly, for scalar two-point and three-point correlators computed in the $\alpha$-vacuum, the Maldacena consistency condition eq.\eqref{maldcon} was not satisfied \cite{Shukla:2016bnu}. Our primary goal in this paper is to check the Ward identity between scalar three-point and four-point correlators computed in the $\alpha$-vacuum. To our surprise, we will find them consistent with the Ward identity.


\section{Four point scalar correlator in $\alpha$-vacua: using in-in formalism}\label{sec:4pininalpha}

In this section, we compute the four point correlation function of scalar perturbations in inflation (also known as the inflationary scalar trispectrum) for $\alpha$-vacua employing the in-in formalism. We will closely follow \cite{Seery:2006vu, Seery:2008ax} where they have computed the same correlator for the standard Bunch-Davies vacuum. Following the set up in \S\,\ref{ssec:inin_review}, we identify the interacting Hamiltonian $H^I_{\text{int}}$ ($I$ denotes the Interaction picture) for the scalar four-point function and use eq.\eqref{eq:form_inin1},  

We will work in the gauge-$2$ in eq.\eqref{eq:gauge12}, and hence at the late time slice $\eta^*$ will consider
\begin{align}
	\mathcal{O}(\eta^*)=\delta\phi(\eta^*,\vect{k}_1) \,\delta\phi(\eta^*,\vect{k}_2) \,\delta\phi(\eta^*,\vect{k}_3) \,\delta\phi(\eta^*,\vect{k}_4) \, .
\end{align}
We only need to focus on a subset of terms in the full interacting Hamiltonian which is relevant to compute the scalar four-point function given in  \cite{Seery:2006vu, Seery:2008ax}. The only difference in dealing with $\alpha$-vacuum will be to change the mode expansion for the fluctuations $\varphi = \delta\phi$ or $\gamma_{ij}$ as in eq.\eqref{eq:modeexpand}, eq.\eqref{eq:modefunctions}, eq.\eqref{eq:c1c2def}. The rest, including the $H^I_{\text{int}}$, will be the same as it was for the case of Bunch-Davies vacuum in \cite{Seery:2006vu, Seery:2008ax}. 

Two types of interactions in $H^I_{\text{int}}$ contribute to $\langle \delta\phi\delta\phi\delta\phi\delta\phi \rangle$. The first one comes due to a four point contact interaction, which we denote with a subscript ``CI". The other one involves  three point scalar-scalar-graviton interactions and a graviton exchange between two such three point vertices contributes to four point scalar correlator. We denote this term with a suffix ``GE". For the Bunch-Davies case, CI term was calculated in\,\cite{Seery:2006vu}, and GE term was calculated in\,\cite{Seery:2008ax}.

From the RHS of eq.\eqref{eq:form_inin1} we can see there are two terms at different orders of $H^I_{\text{int}}$. Up to $1^{st}$ order in $H^I_{\text{int}}$, we get
\begin{equation}\label{eq:1orderininexp}
	\langle\mathcal{O}(\eta^*)\rangle^{\text{$1^{st}$ order}}=i \, \int_{-\infty}^{\eta^*}d\eta'\bra{0}[H^I_{\text{int}}(\eta')\, , \, \mathcal{O}(\eta^*)]\ket{0} \,
\end{equation}
and we need to pick the four point scalar contact interaction vertex for $H^I_{\text{int}}$. We will use eq.\eqref{eq:1orderininexp} to calculate the contact interaction (``CI") contribution to the four-point function in \S \,\ref{ssec:CI_cont}.

Similarly, working at the $2^{nd}$ order in $H^I_{\text{int}}$ contribution we need to compute
\begin{equation}\label{eq:2orderininexp}
	\langle\mathcal{O}(\eta^*)\rangle^{\text{$2^{nd}$ order}}=- \int_{-\infty}^{\eta^*}d\eta'\int_{-\infty}^{\eta'}d\eta'' \, \bra{0}[H^I_{\text{int}}(\eta''),[H^I_{\text{int}}(\eta')\, , \, \mathcal{O}(\eta^*)]]\ket{0}  \, ,
\end{equation} 
with $H^I_{\text{int}}$ as the scalar-scalar-graviton three point vertex. We will use eq.\eqref{eq:2orderininexp} to compute the graviton exchange (``GE") contribution to the scalar four-point function in \S \,\ref{ssec:GE_cont}.


\subsection{Contribution due to the four-point contact interaction}\label{ssec:CI_cont}
The form of the action relevant for the computation of contact interaction contribution is given by the following (see eq.(37) of \cite{Seery:2006vu}), involving quartic order terms in $\delta\phi$, 
\begin{equation}\label{eq:actionfourth_mb}
	S_4 \simeq \int dt \, d^3x \, \left( - \dfrac{1}{4 a} \beta_{2j} \,\partial^2 \beta_{2j} + a\, \vartheta_2 \,\Sigma + \dfrac{3}{4} a^3\, \partial^{-2} \Sigma \, \partial^{-2}\Sigma - a \,\delta \dot{\phi} \,\beta_{2j}\, \partial_j \delta \phi \right) \, ,
\end{equation}
where the metric is used in coordinates of eq.\eqref{dsmetric}, and, hence, $a$ is the scale factor \footnote{The $\partial^{-2}$ should be understood the following way: if $\partial^{2} A = B$, then $A=\partial^{-2}B$.}. Also, each of $\beta_{2j}$, $\vartheta_2$ and $\Sigma$ involve quadratic factors of $\delta\phi$ and their explicit forms are given in eqs.\eqref{eq:beta2j},\eqref{eq:theta2} and \eqref{eq:sigma2} respectively. We can read off the interaction Hamiltonian from eq.\eqref{eq:actionfourth_mb}. Further using the mode expansion of $\delta\phi$ eq.\eqref{eq:modeexpand} in eq.\eqref{eq:1orderininexp} we now compute the contact interaction (``CI") contribution to the four-point scalar correlator. The calculations of eq.\eqref{eq:1orderininexp} are straightforward following the Bunch-Davies computations in \cite{Seery:2006vu}, but complicated. We encountered integrations of the following kind 
\begin{equation}\label{eq:wijci_0}
	\begin{split}
	\tilde{W}_{ij} &= i\, \tilde{k}_t \int^0_{-\infty} d \eta \, (1 \pm i\,k_i \eta)(1 \pm  i\,k_j \eta) e^{i \tilde{k}_t \eta} \, , \\
	\tilde{W}_{lmn} &= i\, \tilde{k}_t \int^0_{-\infty} d\eta \, (1 \pm i\,k_l \eta)(1 \pm i\,k_m \eta) (1\pm i\,k_n \eta) e^{i\tilde{k}_t \eta} \, , \\
	\tilde{W}_i &= -i \,\tilde{k}_t \int_{-\infty}^{0} d\eta \, \eta^2 \, (1 \pm i\, k_i \eta) e^{i \tilde{k}_t \eta} \, .
	\end{split}
\end{equation}
with $\tilde{k}_t = \sum_{i=1}^4 s_i \, k_i$, with $s_i$ being either $+1$ or $-1$, giving a total of $16$ possibilities for $\tilde{k}_t$. 

The extra complication for the $\alpha$-vacua appears in terms of the huge proliferation of the number of integrations to be performed. This can be traced back to the mode expansion of $\delta\phi$ expressed in terms of $U^\alpha_k(\eta)$ in eq.\eqref{eq:modefunctions}. In contrast to the Bunch-Davies answer ($c_1=1, \, c_2=0$), one now has to keep both $e^{\pm i k \eta}$ with coefficients $c_1$ and $c_2$. However, from eq.\eqref{eq:modefunctions} we must note that $U^*_{k}=U_{-k}$. Using this it should be obvious that the $\alpha$-vacua mode function, when expanded can be written in terms of the Bunch-Davies mode function. This realization simplifies our computations considerably. In summary, we can express the contact interaction part of the full four-point function in $\alpha$-vacuum in terms of the corresponding expression in Bunch-Davies case available in \cite{Seery:2006vu}. We will now explain how this can be achieved. 

The Bunch-Davies expression for `CI" contribution is given as \cite{Seery:2006vu},
\begin{equation}\label{eq:citerm}
	\begin{split}
		&\langle \delta \phi(\vect{k}_1) \delta \phi(\vect{k}_2) \delta \phi(\vect{k}_3) \delta \phi(\vect{k}_4) \rangle^{\text{CI}}_{\text{\tiny BD}} \\ &= (2\pi)^3 \delta \left( \sum_{i=1}^4 \vect{k}_i \right) \dfrac{H^6}{\prod_{a=1}^4 (2 k^3_a) } \sum_{\text{perms}} \mathcal{M}_4(\vect{k}_1,\vect{k}_2,\vect{k}_3,\vect{k}_4) \, .
	\end{split}
\end{equation}
The sum is over all $4!$ permutations of $k_i$. The CI label is used to signify the contact interactions. The form factor $\mathcal{M}_4(\vect{k}_1,\vect{k}_2,\vect{k}_3,\vect{k}_4)$ is given by
\begin{equation}\label{eq:formfactorinin}
	\begin{split}
		\mathcal{M}_4(\vect{k}_1,\vect{k}_2,\vect{k}_3,\vect{k}_4) &= -2 \dfrac{k^2_1 k^2_3}{k^2_{12} k^2_{34}}\dfrac{W_{24}}{k_t} \left( \dfrac{\mathbf{Z}_{12} \cdot \mathbf{Z}_{34}}{k^2_{34}} + 2 \vect{k}_2 \cdot \mathbf{Z}_{34} + \dfrac{3}{4} \sigma_{12} \sigma_{34} \right) \\
		& ~~~ - \dfrac{k^2_3}{k^2_{34}}\sigma_{34} \left( \dfrac{\vect{k}_1 \cdot \vect{k}_2}{k_t} W_{124} + \dfrac{k^2_1 k^2_2}{k_t}W_4 \right) \, ,
	\end{split}
\end{equation}
where the various quantities are given by
\begin{equation}\label{eq:quantininlong}
	\begin{split}
		\sigma_{ab} &= \vect{k}_a \cdot \vect{k}_b + k^2_b \, , \quad  
		\mathbf{Z}_{ab} = \sigma_{ab} \vect{k}_a - \sigma_{ba} \vect{k}_b \, ,  \quad W_{ab} = 1 + \dfrac{k_a + k_b}{k_t} + \dfrac{2k_a k_b}{k^2_t} \, , \\
		W_{abc} &= 1 + \dfrac{k_a+k_b+k_c}{k_t} + \dfrac{2(k_a k_b + k_b k_c + k_a k_c)}{k^2_t} + \dfrac{6k_a k_b k_c}{k^3_t} \, , \\
		W_a &= \dfrac{2}{k^2_t} + 6 \dfrac{k_i}{k^3_t} \, , \quad k_t = k_1 + k_2 + k_3 + k_4 \, ,
	\end{split}
\end{equation}

It is also useful to know the integral representations for these quantities when trying to arrive at a generalized result for $\alpha$-vacua
\begin{equation}\label{eq:wijci}
	\begin{split}
	W_{ij} &= i k_t \int^0_{-\infty} d \eta \, (1-ik_i \eta)(1 - ik_j \eta) e^{i k_t \eta} = 1 + \dfrac{k_i + k_j}{k_t} + \dfrac{2 k_i k_j}{k^2_t} \, , \\
	W_{lmn} &= i k_t \int^0_{-\infty} d\eta \, (1-ik_l \eta)(1 - ik_m \eta) (1-ik_n \eta) e^{ik_t \eta} \\
		&= 1 + \dfrac{k_l+k_m+k_n}{k_t} + \dfrac{2(k_l k_m + k_l k_n + k_m k_n)}{k^2_t} + \dfrac{6 k_l k_m k_n}{k^3_t} \, , \\
	W_i &= -i k_t \int_{-\infty}^{0} d\eta \, \eta^2 \, (1- i k_i \eta) e^{i k_t \eta} = \dfrac{2}{k^2_t} + 6 \dfrac{k_i}{k^3_t} \, .
	\end{split}
\end{equation}
These integrals are the Bunch-Davies counterpart of eq.\eqref{eq:wijci_0} and for $\alpha$-vacuum case they should be added to their complex conjugates appropriately according to eq.\eqref{eq:1orderininexp}. 

Finally, we can write the scalar four-point function in $\alpha$-vacua as, 
\begin{equation}\label{eq:alpha_CI}
	\begin{split}
		\langle \delta &\phi(\vect{k}_1) \delta \phi(\vect{k}_2) \delta \phi(\vect{k}_3) \delta \phi(\vect{k}_4) \rangle^{\text{CI}}_\alpha \\ &= (2\pi)^3 \delta \left( \sum_{i=1}^4 \vect{k}_i \right) \dfrac{H^6}{\prod_{a=1}^4 (2 k^3_a) } \sum_{\text{perms}} \mathcal{M}^{\text{new}}_4(\vect{k}_1,\vect{k}_2,\vect{k}_3,\vect{k}_4) \, .
	\end{split}
\end{equation}
where the form factor now gets determined in terms of the Bunch-Davies form factor as follows
\begin{equation}\label{eq:alphaci2}
\begin{split}
		\cM^{\text{new}}_4&(k_1,k_2,k_3,k_4) = \text{Re}\bigg\{ (c^*_1+ c^*_2)^4 \bigg[ (c^4_1 - c^4_2) \, \cM_4(k_1,k_2,k_3,k_4)\\
			&+(c^3_1 c_2 - c_1 c^3_2) \bigg(\cM_4(-k_1,k_2,k_3,k_4) + \cM_4(k_1,-k_2,k_3,k_4)\\
			&+  \cM_4(k_1,k_2,-k_3,k_4) + \cM_4(k_1,k_2,k_3,-k_4) \bigg) \bigg] \bigg\} \, .
\end{split}
\end{equation}
The details of eq.\eqref{eq:alphaci2} are given in eq.\eqref{eq:newformfactor}. Putting in the values of $c_1$ and $c_2$ from eq.\eqref{eq:c1c2def} we get,
\begin{equation}\label{eq:finalformfactorci}
\begin{split}
   \cM&^{\text{new}}_4(k_1,k_2,k_3,k_4)  =\bigg\{\frac{1}{4} (5 \cosh (2 \alpha)-\cosh (6 \alpha))\bigg\}\cM_4(k_1,k_2,k_3,k_4)\\
   &+\bigg\{\sinh ^2(2 \alpha) \cosh (2 \alpha)\bigg\}\bigg(\cM_4(-k_1,k_2,k_3,k_4) + \cM_4(k_1,-k_2,k_3,k_4)\\
			&+  \cM_4(k_1,k_2,-k_3,k_4) + \cM_4(k_1,k_2,k_3,-k_4) \bigg) \, .
\end{split}
\end{equation}
Since the form factor of the Bunch-Davies $\mathcal{M}_4(k_1,k_2,k_3,k_4)$ vacuum is real, the reality condition only acts on the coefficients to ensure that odd powers of $c_2$ vanish from the final expression of the four-point correlator. The final result for the contact interaction contribution for the four-point function in $\alpha$-vacuum is therefore given by eq.\eqref{eq:alpha_CI} with the $\mathcal{M}^{\text{new}}_4(\vect{k}_1,\vect{k}_2,\vect{k}_3,\vect{k}_4)$ written in eq.\eqref{eq:finalformfactorci}.  

We learn that $\langle \delta \phi \delta \phi\delta \phi\delta \phi\rangle^{\text{CI}}_\alpha$ is determined by the corresponding expression in Bunch-Davies vacuum following the algorithm outlined in eq.\eqref{eq:finalformfactorci}: flip the sign and permute the absolute value of the external momentum in $\mathcal{M}_4$ to obtain $\mathcal{M}^{\text{new}}_4$. 
Let us also highlight that the prescription of eq.\eqref{eq:finalformfactorci} was previously reported in \cite{Jain:2022uja} where general solutions to the CFT Ward identities in momentum space was studied. However, that study involved only three point functions. Here we have observed the same behavior for a part of the inflationary scalar four point function, but from a bulk analysis (i.e., calculations done in the gravity side in $dS_4$). 

\subsection{Contribution due to the graviton exchange}\label{ssec:GE_cont}

For the scalar four point function coming from the graviton exchange we need to focus on the following relevant vertex containing a scalar-scalar-graviton interaction. Following \cite{Seery:2008ax} (from eq.(2.9) there), we write down the following cubic order part of the action 
\begin{equation}\label{eq:intterm}
	S_3 = \dfrac{1}{2} \int d\eta \, d^3x \, a^2 \gamma^{ij} \partial_i \delta \phi \, \partial_j \delta \phi \, . 
\end{equation}
From this we can easily read off the interaction Hamiltonian and calculate the graviton exchange contribution to the scalar four-point correlator using eq.\eqref{eq:2orderininexp}.
We also need to know the mode expansions of the scalar perturbation and the graviton in $\alpha$-vacuum. For $\delta \phi$ field, we use eq.\eqref{eq:modefunctions}. The free graviton field, which we are encountering for the first time, satisfies the free equations of motion and therefore have the mode expansion given as,
\begin{equation}
	\gamma_{ij}(\eta,\vect{x}) = \sum_{s=1,2} \int \dfrac{d^3\vect{k}\, e^{i \vect{k}\cdot \vect{x}}}{(2\pi)^3} \left[ a^{-}_{\vect{k},s} \, \epsilon^s_{ij}(\vect{k}) \, \gamma^{s,\alpha}_k(\eta) + a^{+}_{-\vect{k},s} \, \epsilon^{s*}_{ij}(-\vect{k}) \, {(\gamma^{s,\alpha}_k(\eta))}^* \right]  \, ,
\end{equation}
where the mode function $\gamma^{s,\alpha}_k(\eta)$ for the case of $\alpha$-vacua can be written as \footnote{Here $U_k(\eta) = U^{BD}_{k}(\eta)$ which is the Bunch-Davies mode function. We will subsequently drop the superscript $BD$ for Bunch-Davies modes to avoid cluttering of expressions. However, for $\alpha$-vacua modes, we will keep the superscript $\alpha$.},
\begin{equation}\label{eq:gammamode}
	\gamma^{s,\alpha}_k(\eta) = \dfrac{H}{\sqrt{k^3}}\left[ c_1(1-ik\eta) e^{ik \eta} + c_2 (1+ ik\eta) e^{-ik\eta} \right] = c_1 \, U^{BD}_k(\eta)+ c_2 \, {U^{BD}_k}^*(\eta) \, ,
\end{equation}
where $s$ labels the two polarization states of the spin $2$ graviton. We have the same mode function eq.\eqref{eq:gammamode} for either polarization states and $\epsilon^s_{ij}(\vect{k})$ denotes the corresponding polarization tensors, and $c_1$ and $c_2$ have been defined earlier in eq.\eqref{eq:c1c2def} \footnote{The polarization tensors are transverse and traceless: $\epsilon^s_{ii}(\vect{k})=k^i \epsilon^s_{ij}(\vect{k}) =0$. They are also chosen to satisfy the completeness relation $\epsilon^s_{ij}(\vect{k}) \epsilon^{s'*}_{ij}(\vect{k}) = 2 \delta_{ss'}$.}. 

Following the prescription of in-in formalism we need to define four different propagators for fields lying on the time ordered and the anti-time ordered path, see \cite{Seery:2008ax}. For the case of $\alpha$-vauca, the scalar propagators are given as,
\begin{equation}
	\begin{split}
		G^{++}_k(\eta,\eta') &= G^>_k(\eta,\eta')\Theta(\eta-\eta') + G^<_k(\eta,\eta')\Theta(\eta'-\eta) \, , \\
		G^{--}_k(\eta,\eta') &=  G^>_k(\eta,\eta')\Theta(\eta'-\eta) + G^<_k(\eta,\eta')\Theta(\eta-\eta') \, , \\
		G^{-+}_k(\eta,\eta') &= G^>_k(\eta,\eta') \, , \quad G^{+-}_k(\eta,\eta') = G^<_k(\eta,\eta') \, ,
	\end{split}
\end{equation}
where 
\begin{equation}\label{eq:scalretprop}
		G^>_k(\eta,\eta') = U_k^\alpha(\eta) \,  U^{*\alpha}_k(\eta') \, , \quad \quad 
		G^<_k(\eta,\eta') = U^{*\alpha}_k(\eta)  \, U_k^\alpha(\eta') \, ,
\end{equation}
with $U_k^\alpha(\eta)$ given by eq.\eqref{eq:modefunctions}. We have the exact same structure for the graviton propagators except that
\begin{equation}\label{eq:gravretprop}
	\begin{split}
		F^{s>}_k(\eta,\eta') = \gamma^{s,\alpha}_{k}(\eta) \, \gamma^{s,\alpha*}_k(\eta') \, , \quad \quad 
		F^{s<}_k(\eta,\eta') = \gamma^{s,\alpha*}_{k}(\eta) \, \gamma^{s,\alpha}_k(\eta') \, .
	\end{split}
\end{equation}

Now using eq.\eqref{eq:2orderininexp} and wick contracting the fields appropriately, we can write the scalar correlator in terms of these propagators as follows,
\begin{equation}\label{eq:ininbegin}
	\begin{split}
		\langle &\delta \phi(\eta^*,\vect{k}_1) \delta \phi(\eta^*,\vect{k}_2) \delta \phi(\eta^*,\vect{k}_3) \delta \phi(\eta^*,\vect{k}_4) \rangle^{\text{GE}} \\ = &-\dfrac{1}{4}(2\pi)^3 \delta^3\left( \sum_{i=1}^4 \vect{k}_i \right) \sum_s \epsilon^s_{ij}(\vect{k}_{12}) k^i_1 k^j_2 \epsilon^s_{lm}(\vect{k}_{34}) k^l_3 k^m_4   \times \\
		&\int^{\eta_*}_{-\infty} \dfrac{d\eta}{\eta^2} \int^{\eta}_{-\infty} \dfrac{d\eta'}{\eta'^{2}}\left[ G^{++}_{k_1}(\eta_*,\eta) G^{++}_{k_2}(\eta_*,\eta') G^{++}_{k_3}(\eta_*,\eta') G^{++}_{k_4}(\eta_*,\eta') F^{s++}_{k_{12}}(\eta,\eta') \right. \\
		&~~~ + G^{+-}_{k_1}(\eta_*,\eta) G^{+-}_{k_2}(\eta_*,\eta') G^{+-}_{k_3}(\eta_*,\eta') G^{+-}_{k_4}(\eta_*,\eta') F^{s--}_{k_{12}}(\eta,\eta') \\
		&~~~ - G^{++}_{k_1}(\eta_*,\eta) G^{++}_{k_2}(\eta_*,\eta') G^{+-}_{k_3}(\eta_*,\eta') G^{+-}_{k_4}(\eta_*,\eta') F^{s+-}_{k_{12}}(\eta,\eta') \\
		& ~~~ - \left. G^{+-}_{k_1}(\eta_*,\eta) G^{+-}_{k_2}(\eta_*,\eta') G^{++}_{k_3}(\eta_*,\eta') G^{++}_{k_4}(\eta_*,\eta') F^{s-+}_{k_{12}}(\eta,\eta') \right] \\
		& ~~~ + 23 ~\text{permutations} \, .
	\end{split}
\end{equation}
We now use $G^{++}_k(\eta_*,\eta) = G^>(\eta_*,\eta)$ for $\eta_* > \eta$, $F^{s++}_k(\eta,\eta') = F^{s>}_k(\eta,\eta')$ for $\eta>\eta'$, $G^> = G^{<*}$ and other relations to simplify the above expression as
\begin{equation}\label{eq:ininfourptfirst}
	\begin{split}
		\langle \delta \phi&(\eta^*,\vect{k}_1) \delta \phi(\eta^*,\vect{k}_2) \delta \phi(\eta^*,\vect{k}_3) \delta \phi(\eta^*,\vect{k}_4) \rangle^{\text{GE}}_\alpha \\ 
		= &(2\pi)^3 \delta^3\left( \sum_{i=1}^4 \vect{k}_i \right) \sum_s \epsilon^s_{ij}(\vect{k}_{12})\, k^i_1 k^j_2\, \epsilon^s_{lm}(\vect{k}_{34})\, k^l_3 k^m_4 \,\int^{\eta_*}_{-\infty} \dfrac{d\eta}{\eta^2} \int^{\eta}_{-\infty} \dfrac{d\eta'}{\eta'^{2}} \times\\
		& \text{Im}\left[G^>_{k_1}(\eta_*,\eta)\, G^>_{k_2}(\eta_*,\eta) \right] \times \text{Im}\left[G^>_{k_3}(\eta_*,\eta')\, G^>_{k_4}(\eta_*,\eta')\, F^{s>}_{k_{12}}(\eta,\eta') \right] \\
		&+ 23 ~\text{permutations} \, .
	\end{split}
\end{equation}
Upto this point, we have basically followed the steps of \cite{Seery:2008ax}. We now use the retarded form of the propagators eq.\eqref{eq:scalretprop} and eq.\eqref{eq:gravretprop} with the $\alpha$-vacua generalized modes as given in eq.\eqref{eq:modefunctions}. 
Substituting them in eq.\eqref{eq:ininfourptfirst}, we will end up with $1024$ integrals to solve \footnote{For instance because of eq.\eqref{eq:GG}, we have $2^{10}$ terms in each permutation of eq.\eqref{eq:ininfourptfirst}.}. 
 
All such integrals arise from one common integral with momenta labels appropriately dressed by ``helicity factors" $\varepsilon_i$ that take values $\pm 1$. Let us evaluate a protypical integral with coefficient $c^{10}_1$ 
\begin{equation}\label{eq:intc18}
	\begin{split}
		&I_{1234}(c^{10}_1) = \int_{-\infty}^{\eta_*} \dfrac{d\eta}{\eta^2} \int_{-\infty}^{\eta'} \dfrac{d\eta'}{\eta'^2} \text{Im}[ (1-ik_1 \eta_*)(1-ik_2\eta_*)(1+ik_1\eta) \\ &~~\times (1+ik_2\eta)e^{i(\eta_*-\eta)(k_1+k_2)}]
		\times \text{Im}[(1-ik_3 \eta_*)(1-ik_4\eta_*)(1+ik_3\eta') \\ &~~\times (1+ik_4\eta')(1-ik_{12}\eta)(1+ik_{12}\eta')e^{i(\eta_*-\eta')(k_3+k_4)}e^{i(\eta-\eta')k_{12}}] \, .
	\end{split}
\end{equation}
This should result in the standard Bunch-Davies answer of \cite{Seery:2008ax}. The other integrals of eq.\eqref{eq:ininfourptfirst} take the folllowing typical form
\begin{equation}\label{eq:genalphaint}
	\begin{split}
		I_{1234}(c^k_1 & c^{10-k}_2) = \int_{-\infty}^{\eta_*} \dfrac{d\eta}{\eta^2} \int_{-\infty}^{\eta'} \dfrac{d\eta'}{\eta'^2}\times \text{Im}\bigg[ (1+i \varepsilon_1 k_1 \eta_*)(1+i \varepsilon_2 k_2\eta_*)\\ 
		& \quad \quad \quad \quad (1+i \varepsilon_3 k_1\eta)(1+i \varepsilon_4 k_2\eta)e^{i(\eta_*( - \varepsilon_1 k_1 - \varepsilon_2 k_2) - \eta( \varepsilon_3 k_1 + \varepsilon_4 k_2 )}\bigg]\\ &
		\times \text{Im}\bigg[(1+i \varepsilon_6 k_3 \eta_*)(1+i \varepsilon_7 k_4\eta_*)(1+i \varepsilon_8 k_3\eta') (1+i \varepsilon_9 k_4\eta')(1+i \varepsilon_5 k_{12}\eta)\\ &
		\quad \quad\quad \quad \times (1+ i \varepsilon_{10} k_{12}\eta')  e^{i(\eta_*(-\varepsilon_6 k_3 - \varepsilon_7 k_4) - \eta'(\varepsilon_8 k_3 + \varepsilon_9 k_4)}e^{-i(\varepsilon_5 \eta+ \varepsilon_{10} \eta')k_{12}}\bigg] \, ,
	\end{split}
\end{equation}
where $\varepsilon_i = \pm 1$ are the helicity factors.
Upon evaluating eq.\eqref{eq:genalphaint} followed by summing over the various permutations in eq.\eqref{eq:ininfourptfirst} with the limit $\eta^*\rightarrow 0$, we finally get the expression for the scalar correlator as, 
\begin{equation}\label{eq:ininfourptfinal}
	\begin{split}
		\langle& \delta \phi(\vect{k}_1)  \delta \phi(\vect{k}_2)\delta \phi(\vect{k}_3) \delta \phi(\vect{k}_4) \rangle^{\text{GE}}_\alpha  = (2\pi)^3 \delta^3\left( \sum_{i=1}^4 \vect{k}_i \right) \dfrac{4 H^6}{\prod_a (2 k^3_a)} \times \\
		   \sum_s & \left[\dfrac{1}{k^3_{12}}\epsilon^s_{ij}(\vect{k}_{12}) \epsilon^s_{lm}(\vect{k}_{34}) \, k^i_1 k^j_2 k^l_3 k^m_4 \Bigg(\mathcal{I}_{1234}^\alpha+\mathcal{I}_{3412}^\alpha\Bigg) + \{ k_2 \leftrightarrow k_3\} + \{ k_2 \leftrightarrow k_4\} \right]  \, ,
	\end{split}
\end{equation}
where we have put the expression similar in structure as given in\,\cite{Seery:2008ax} for convenience. The explicit calculation leading to the form of $\mathcal{I}_{1234}^\alpha$ is given in Appendix\,\ref{ap:I1234_calc}. The final expression for $\mathcal{I}_{1234}^\alpha$ is given by eq.\eqref{eq:ialpha1234full}. It is clear from eq.\eqref{eq:ialpha1234full} and eq.\eqref{eq:i121234} that setting $\alpha=0$ one recovers the Bunch-Davies result  for the graviton exachange part of \cite{Seery:2008ax}.

It is interesting to note that, unlike the CI part in eq.\eqref{eq:finalformfactorci}, the structure of the final answer for the GE part eq.\eqref{eq:ininfourptfinal} cannot be obtained from the Bunch-Davies answer by flipping the signs of the magnitude of the momenta as predicted by a Ward identity analysis \cite{Jain:2022uja}. This is precisely due to the graviton exchange in the diagram via eq.\eqref{grav_cont} which has cross terms of the form $\cosh\alpha \sinh\alpha$. Also, as we will see in \S\ref{secWI:check}, this ties up nicely with the fact that in the Ward identities between three-point and four-point scalar correlator, only $\langle \delta \phi \delta \phi\delta \phi\delta \phi\rangle^{\text{CI}}_\alpha$ contributes, and $\langle \delta \phi \delta \phi\delta \phi\delta \phi\rangle^{\text{GE}}_\alpha$ is not sensitive to the Ward identities.

Finally, we can combine the ``CI" (eq.\eqref{eq:alpha_CI}) and the ``GE" (eq.\eqref{eq:ininfourptfinal}) contributions and write the full scalar four-point correlator, in $\alpha$-vacua, formally as,
\begin{equation}\label{eq:scalarfourpfinal_inin}
	\begin{split}
    \langle \delta \phi(\vect{k}_1) \delta \phi(\vect{k}_2) \delta \phi(\vect{k}_3) \delta \phi(\vect{k}_4)\rangle_\alpha=& \langle  \delta \phi(\vect{k}_1) \delta \phi(\vect{k}_2)\delta \phi(\vect{k}_3) \delta \phi(\vect{k}_4) \rangle^{\text{CI}}_\alpha\\
    &+\langle \delta \phi(\vect{k}_1) \delta \phi(\vect{k}_2)\delta \phi(\vect{k}_3) \delta \phi(\vect{k}_4) \rangle^{\text{GE}}_\alpha
	\end{split}
\end{equation}
So far, we have been working with the gauge-$2$, eq.\eqref{eq:gauge12}, where the scalar perturbations are denoted by $\delta \phi$. We can now perform the transformation to the gauge-$1$, using eq.\eqref{infper}, so that the scalar fluctuations are now expressed through $\zeta$, and we get the scalar four-point function in the leading order of slow-roll parameter as 
\begin{equation}\label{eq:zeta_phi_rel}
	\begin{split}
\langle\zeta(\vect{k}_1)\zeta(\vect{k}_2)\zeta(\vect{k}_3)\zeta(\vect{k}_4)\rangle_{\alpha}=\frac{H^4}{\dot{\bar{\phi}}^4}\langle \delta \phi(\vect{k}_1) \delta \phi(\vect{k}_2) \delta \phi(\vect{k}_3) \delta \phi(\vect{k}_4)\rangle_\alpha \\
= \frac{1}{\epsilon^2} \, \langle \delta \phi(\vect{k}_1) \delta \phi(\vect{k}_2) \delta \phi(\vect{k}_3) \delta \phi(\vect{k}_4)\rangle_\alpha\, .
	\end{split}
\end{equation}
This is one of the main results of our present work. It is also straightforward to see that in the limit of $\alpha = 0$ the corresponding answer for the Bunch-Davies case, \cite{Seery:2006vu}, \cite{Ghosh:2014kba}, is recovered from eq.\eqref{eq:zeta_phi_rel}. This assures the consistency of our result for the $\alpha$-vacuum. 
\section{Consistency check of conformal Ward Identities for $\alpha$-vacua correlators} \label{secWI:check}

As we have discussed before, in \cite{Shukla:2016bnu} it was shown that the ward identities for scale and special conformal transformation between two-point and three-point scalar correlators, eq.\eqref{eq:scal_ward1} and eq.\eqref{eq:sct_ward_1} are violated. However,in this section, we will check that the ward identities, eq.\eqref{eq:scal_ward2} and eq.\eqref{eq:sct_ward_2} are satisfied by the $3$-point and $4$-point scalar correlation functions in $\alpha$-vacuum. 

Let us rewrite those Ward identities again here for convenience, the scaling Ward identity is 
\begin{equation} \label{eq:scal_ward_3} 
\Bigg[6+\sum_{a=1}^{3}\vect{k}_a\cdot\frac{\partial}{\partial\vect{k}_a}\Bigg]\langle\zeta(\vect{k}_1)\zeta(\vect{k}_{2})\zeta(\vect{k}_{3})\rangle'_\alpha =-\frac{\langle\zeta(\vect{k}_1)\zeta(\vect{k}_{2})\zeta(\vect{k}_{3})\zeta(\vect{k}_{4})\rangle'_\alpha}{\langle\zeta(\vect{k}_{4})\zeta(-\vect{k}_{4})\rangle'_\alpha}\Bigg|_{\vect{k}_{4}\rightarrow 0} \, ,  
\end{equation}
and the special conformal (SCT) one as 
\begin{equation} \label{eq:sct_ward_3}
\Bigg[\sum_{a=1}^{3}\hat{\mathcal{L}}^{\vect{b}}_{\vect{k}_a}\Bigg]\langle\zeta(\vect{k}_1)\zeta(\vect{k}_{2})\zeta(\vect{k}_{3})\rangle_\alpha =-2\Bigg[\vect{b}\cdot\frac{\partial}{\partial\vect{k}_{4}}\Bigg]\frac{\langle\zeta(\vect{k}_1)\zeta(\vect{k}_{2})\zeta(\vect{k}_{3})\zeta(\vect{k}_{4})\rangle_\alpha}{\langle\zeta(\vect{k}_{4})\zeta(-\vect{k}_{4})\rangle'_\alpha}\Bigg|_{\vect{k}_{4}\rightarrow 0} \, .
\end{equation}
with $\hat{\mathcal{L}}^{\vect{b}}_{\vect{k}_a}$ given in eq.\eqref{defLbk}. 
The expression for $\langle\zeta(\vect{k}_1)\zeta(\vect{k}_{2})\zeta(\vect{k}_{3})\rangle_\alpha$ was already obtained in eq.(5.17) of \cite{Shukla:2016bnu} \footnote{We will derive this result from the wave-function formalism in \S\,\ref{sec:3ptwfalpha} with the final result in eq.\eqref{eq:threepointalpha}.}. In the previous section \S\,\ref{sec:4pininalpha}, we have also worked out $\langle\zeta(\vect{k}_1)\zeta(\vect{k}_{2})\zeta(\vect{k}_{3})\zeta(\vect{k}_{4})\rangle_\alpha$. 

For the scaling Ward identity in eq.\eqref{eq:scal_ward_3} we have checked that both the LHS and RHS vanish independently, and that is how this identity is satisfied. That the LHS of eq.\eqref{eq:scal_ward_3} vanishes can be understood easily. The logic is the following: firstly, while checking this Ward identity for Bunch-Davies correlators it has already been noted that the LHS of eq.\eqref{eq:scal_ward_3} vanishes. Now, $\langle\zeta \zeta \zeta \rangle_\alpha$ is determined in terms of the corresponding Bunch-Davies result $\langle \zeta \zeta \zeta \rangle_{\alpha=0}$ as follows
\begin{equation} \label{zeta3pt_form}
\begin{split}
\langle\zeta(k_1)\zeta(k_{2})\zeta(k_{3})\rangle_\alpha  \sim &  \, d_1(\alpha) \, \langle\zeta(k_1)\zeta(k_{2})\zeta(k_{3})\rangle_{\alpha=0}  + d_2(\alpha) \, \bigg( \langle\zeta(-k_1)\zeta(k_{2})\zeta(k_{3})\rangle_{\alpha=0} \\
& +\langle\zeta(k_1)\zeta(-k_{2})\zeta(k_{3})\rangle_{\alpha=0}+\langle\zeta(k_1)\zeta(k_{2})\zeta(-k_{3})\rangle_{\alpha=0}\bigg) \, , 
\end{split}
\end{equation}
where $d_1(\alpha)$ and $d_2(\alpha)$ are some $\alpha$ dependent coefficients \footnote{We note that the eq.\eqref{zeta3pt_form} is valid in a strict sense upto contact terms that satisfy the homogeneous part of the CFT Ward identities, compare eq.\eqref{zeta3pt_form} with eq.\eqref{eq:threepointalpha}. These contact terms are generated as we go from $\delta\phi$ to $\zeta$. Therefore, in terms of 
$\langle \delta \phi \delta \phi \delta \phi \rangle$ eq.\eqref{zeta3pt_form} is strictly valid.}. We must also impose that upon substituting $\alpha=0$ on the RHS of eq.\eqref{zeta3pt_form} one recovers the Bunch-Davies result , i.e., $d_1(\alpha =0) =1$, and $d_2(\alpha =0) =0$.  

Hence, it is expected that the LHS of eq.\eqref{eq:scal_ward_3} should vanish even for $\alpha$-vacuum $\langle \zeta \zeta \zeta \rangle_\alpha$. Alternatively, from eq.\eqref{eq:threepointalpha}, it is clear that $\langle \zeta \zeta \zeta \rangle'_\alpha$ (after stripping off the momentum conserving delta function) behaves as $\sim k^{-6}$, and hence the LHS of eq.\eqref{eq:scal_ward_3} should vanish when $(6+ \sum_{a=1}^{3} \vect{k}_a \cdot \partial_{\vect{k}_a})$ acts on it. 

Similarly, for the RHS of eq.\eqref{eq:scal_ward_3}, we need to work out the $\vect{k}_{4}\rightarrow 0$ limit of the four point function $\langle\zeta \zeta \zeta \zeta \rangle'_\alpha$. It is known that for the Bunch-Davies case this vanishes linearly in $k_4$. We have explicitly checked using Mathematica that $\langle\zeta \zeta \zeta \zeta \rangle'_\alpha$ vanishes in the $\vect{k}_4 \rightarrow 0$ limit; thus arguing the scaling Ward identity \footnote{The Mathematica file is available here: \url{https://github.com/arhumansari007/alpha_vacua_ward_identity}.}. 

For the SCT ward identity we have observed that in the limit $\vect{k}_{4}\rightarrow 0$, only the contact interaction part $\langle\zeta \zeta \zeta \zeta \rangle^{\text{CI}}_\alpha$ contributes to the RHS of eq.\eqref{eq:sct_ward_3} while the graviton exchange part $\langle\zeta \zeta \zeta \zeta \rangle^{\text{GE}}_\alpha$ vanishes. We have explicitly verified this using Mathematica \footnote{
The Mathematica files with the calculations supporting the check of SCT ward identity are available here: \url{https://github.com/arhumansari007/alpha_vacua_ward_identity}. We have employed the \emph{Triple K} package developed in \cite{Bzowski:2020lip} for our checks.}.

At this point, we have obtained one of the most important results in this paper, which is to explicitly check that the $3$-point and $4$-point scalar correlators computed in $\alpha$-vacuum are consistent with the scaling and special conformal Ward identities. Crucially, this result is at odds with the negative result reported in \cite{Shukla:2016bnu} involving the $2$-point and $3$-point scalar correlators in $\alpha$-vacuum. In this paper, we have not been able to explain this anomaly in the validity of Ward identities for $\alpha$-vacuum. We will return to this point in the discussions, \S\,\ref{sec:disco}, with some speculative comments. 

Prior to concluding this section, it's worth noting that our verification of the scaling Ward identity between the $4$-point and $3$-point scalar correlators, in eq.\eqref{eq:scal_ward_3}, differs from how it was done for the $3$-point and $2$-point scalar correlators in eq.\eqref{eq:scal_ward1}. The main difference lies in the level of precision in the perturbative expansion in the slow roll parameter.

As seen in eq.\eqref{maldcon}, also known as the Maldacena consistency condition, the scaling Ward identity is checked more rigorously by comparing two non-zero values on both sides of the Ward identity. This is possible because the correction due to slow-roll on the de-Sitter symmetries is easily incorporated in the $2$-point function as the scalar tilt $n_s$, see eq.\eqref{2point_titl}. However, as demonstrated earlier, the scaling Ward identity between the $4$-point and $3$-point scalar correlator in eq.\eqref{eq:scal_ward_3} is verified by establishing that both sides of this equation vanish to the leading order in the slow-roll parameter. Ideally, we should have compared both sides to the next sub-leading order in slow-roll, thereby comparing two non-vanishing quantities. However, we cannot perform that test with the expressions of the $4$-point and $3$-point scalar correlators available to us.
To check the scaling Ward identity for the $4$-point and $3$-point case at the next non-trivial order in the slow roll, we would first need to compute the correlators themselves up to the next sub-leading order in the slow roll expansion. 

Additionally, it's worth noting that this is how the scaling ward identity for the $4$-point and $3$-point correlator has been verified for the Bunch-Davies case as well (see \cite{Kundu:2014gxa}), and is not particularly a limitation of our analysis here for the alpha vacua. Nevertheless, it would be interesting to verify the scaling ward identity for the $4$-point and $3$-point to the next sub-leading order in the slow roll, particularly in view of the result in \cite{Kundu:2015xta} that the conformal Ward identities should be valid to all orders in the slow roll parameter. Once this is confirmed for the Bunch-Davies case, our analysis in this paper suggests the same will hold true for the $\alpha$-vacua case. 

Finally, we should note that, unlike the scaling ward identity, the SCT ward identity for the $4$-point and $3$-point correlators works out rather non-trivially. As we discussed above, both sides of eq.\eqref{eq:sct_ward_3} match precisely in the leading order of the slow roll expansion by comparing two non-vanishing quantities. This is similar to how the same ward identity was checked for the $3$-point and $2$-point scalar correlators (as in eq.\eqref{eq:sct_ward_1}), \cite{Kundu:2015xta}.


\section{Scalar correlators in $\alpha$-vacua: using wave-function formalism} \label{sec:4pwfalpha}

In this section, we compute the inflationary scalar correlators in $\alpha$-vacua using the wavefunctional formalism, reviewed in \S\,\ref{sec:analyticcont}, by generalizing the Bunch-Davies vacuum calculations of the four-point scalar correlator in \cite{Ghosh:2014kba}. We will give the main steps here and relegate all the details to Appendix-\ref{ap:OOOO_calc}.
 
Using the wave function $\Psi[\delta\phi,\gamma_{ij}]$ in eq.\eqref{eq:wfgauge2}, and using eq.\eqref{def_cor_fn}, one can express $\langle \delta \phi \delta \phi \delta \phi  \delta \phi \rangle$ in terms of the wave-function coefficients \footnote{Following \S\,5 of \cite{Ghosh:2014kba} we obtained eq.\eqref{eq:sc_corr_wf_coff_rel} by integrating out the graviton fluctuations in the path integral of eq.\eqref{def_cor_fn}. There is a factor of $\cosh 2\alpha$ that appears in eq.\eqref{eq:sc_corr_wf_coff_rel} when compared to the Bunch-Davies answer in eq.(5.3) of \cite{Ghosh:2014kba}. As we will see, this extra factor will be crucial for matching with the in-in formalism results.}, as written below
\begin{equation}\label{eq:sc_corr_wf_coff_rel}
	\begin{split}
		&\langle \delta \phi(\vect{k}_1) \delta \phi(\vect{k}_2)\delta \phi(\vect{k}_3) \delta \phi(\vect{k}_4) \rangle_{\alpha} = \dfrac{1}{\mathcal{C}_{norm}}\int \mathcal{D}[\delta \phi] \prod_{i=1}^{4} \delta \phi(\vect{k}_i) \\
		& \times \exp \left[ \dfrac{M^2_{pl}}{H^2} \left( - \int \dfrac{d^3k_1}{(2\pi)^3}\dfrac{d^3k_2}{(2\pi)^3} \delta \phi(\vect{k}_1) \delta \phi(\vect{k}_2) \langle O(-\vect{k}_1) O(-\vect{k}_2) \rangle_{\alpha} \right. \right. \\
		& + \int \prod_{I=1}^{4} \left\{ \dfrac{d^3k_I}{(2\pi)^3} \delta \phi(\vect{k}_I) \right\} \left\{ \dfrac{1}{12} \langle O(-\vect{k}_1) O(-\vect{k}_2) O(-\vect{k}_3) O(-\vect{k}_4) \rangle_{\alpha} \right. \\
		& + \dfrac{1}{8} \langle O(-\vect{k}_1) O(-\vect{k}_2) T_{ij}(\vect{k}_1+\vect{k}_2) \rangle'_{\alpha} \langle O(-\vect{k}_3) O(-\vect{k}_4) T_{kl}(\vect{k}_3+\vect{k}_4) \rangle'_{\alpha} \\
		& \left. \left. \left. (2\pi)^3 \delta^3 \left( \sum_{I=1}^{4} \vect{k}_I \right) \hat{P}_{ijkl}(\vect{k}_1 + \vect{k}_2) \dfrac{\cosh2\alpha}{|\vect{k}_1+\vect{k}_2|^3} \right\} \right) \right] \, ,
		\end{split}
	\end{equation}
where 
\begin{equation}\label{eq:tranproj}
	\hat{P}_{ijkl}(\vect{k}) = \left(\pi^{\vect{k}}_{l j} \pi^{\vect{k}}_{k i} + \pi^{\vect{k}}_{l i} \pi^{\vect{k}}_{k j} - 
	\pi^{\vect{k}}_{l k} \pi^{\vect{k}}_{j i} \right) \, , ~~~ \text{where} ~~ \pi^{\vect{k}}_{ij} = \delta_{ij} - \dfrac{k_i k_j}{k^2} \, .
\end{equation}
It is clear from eq.\eqref{eq:sc_corr_wf_coff_rel} that to compute $\langle \delta \phi \delta \phi \delta \phi  \delta \phi \rangle_{\alpha}$ we need to know the following wave-function coefficients: $\langle O(\vect{k}_1)O(\vect{k}_2)O(\vect{k}_3)O(\vect{k}_4)\rangle_{\alpha}$ and $\langle O(\vect{k}_1)O(\vect{k}_2)T_{ij}(\vect{k}_s)\rangle_{\alpha} $ \footnote{It should be noted that in eq.\eqref{eq:sc_corr_wf_coff_rel}, $\langle \delta \phi \delta \phi \delta \phi  \delta \phi \rangle$ is obtained by performing a path integral over the graviton mode weighted by the modulus square of the wave function, i.e., $|\psi|^2$, as given in eq.\eqref{def_cor_fn}. Consequently, $\langle \delta \phi \delta \phi \delta \phi  \delta \phi \rangle$ as written in eq.\eqref{eq:sc_corr_wf_coff_rel} actually depends on the modulus values of the coefficient functions $\langle OOOO\rangle$ and $\langle OOT_{ij}\rangle$, see eq.\eqref{eq:scal_corr_CF}, eq.\eqref{eq:scal_corr_ET}, eq.\eqref{eq:ghatgg}, and eq.\eqref{eq:gnew}. We are not very explicit about it here in eq.\eqref{eq:sc_corr_wf_coff_rel} to avoid notational clutter. We refer to appendix A.3 of \cite{Goon:2018fyu} for a review. \label{footnote26}}. Now, $\langle O(\vect{k}_1)O(\vect{k}_2)T_{ij}(\vect{k}_s)\rangle_{\alpha}$ is already known to us from \cite{Jain:2022uja} for $\alpha$-vacuum. 
In \cite{Ghosh:2014kba}, the other coefficient $\langle O(\vect{k}_1)O(\vect{k}_2)O(\vect{k}_3)O(\vect{k}_4)\rangle_{\alpha=0}$ was obtained by computing the partition function as a functional of the sources at the boundary ($z \to 0$) of $EAdS_4$. After carefully incorporating the modifications due to $\alpha$-vacuum, we will get $\langle O(\vect{k}_1)O(\vect{k}_2)O(\vect{k}_3)O(\vect{k}_4)\rangle_\alpha$ by taking appropriate derivatives of this partition function with respect to the sources. 

The effective action for the perturbations is obtained by starting from 
\begin{equation}\label{eq:eadsaction}
	S={M_{Pl}^2 \over 2}\int d^4 x \sqrt{g} \bigg[R-2 \Lambda - (\nabla \delta \phi)^2\bigg] \, ,
\end{equation}
where $\Lambda$ is the cosmological constant and $g_{\mu\nu}$ is the $EAdS_4$ metric eq.\eqref{eads}. Defining the metric perturbations as $g_{\mu\nu} = \bar{g}_{\mu\nu} + \delta g_{\mu\nu}$ \footnote{Appropriate gauge fixing is of course needed, we are going to choose $\delta g_{zz} = \delta g_{zi} = 0$.}, and then expanding the action in eq.\eqref{eq:eadsaction} about the background metric, we get 
\be
\label{eq:expact}
\begin{split}
	S= & S_0 - {M_{Pl}^2 \over 2} \int d^4 x \sqrt{\bar{g}} \bar{g}^{\mu\nu} \partial_\mu (\delta \phi) \partial_\nu (\delta \phi) + S_{\text{grav}}^{(2)} + S_{\text{\text{int}}} \, ,
\end{split}
\ee
where $S_0$ is the unperturbed action with Einstein gravity plus negative cosmological constant, leading to the background $EAdS_4$ space. The second term on the RHS of eq.\eqref{eq:expact} signifies the free part of the action of $\delta \phi$ on the background $EAdS_4$ with metric $\bar{g}^{\mu\nu}$. The third term, $S_{\text{grav}}^{(2)}$ is the quadratic action for $\delta g_{\mu\nu}$, see eq.(4.6) in \cite{Ghosh:2014kba}. Lastly, $S_{\text{int}}$, denoting the coupling of $\delta \phi$ and $\delta g_{\mu\nu}$, is taken to be 
\be
\label{eq:defint}
\begin{split}
	S_{\text{int}}=&{M_{Pl}^2 \over 2} \int d^4 x {\sqrt{\bar{g}}\over 2} \, \delta g^{\mu\nu} T_{\mu\nu} + \mathcal{O} \left(g_{\mu\nu}^2 \right)\, , \\ \text{with} \quad T_{\mu\nu}=& 2\partial_\mu (\delta \phi) \, \partial_\nu (\delta \phi)  -\bar{g}_{\mu\nu}\, 
\bar{g}^{\alpha\beta} \partial_\alpha (\delta \phi)\, \partial_\beta (\delta \phi) \, , 
\end{split}
\ee
where, $\delta \phi$, which is a massless scalar field in $EAdS_4$, acts like a source to $\delta g_{\mu\nu}$, through its stress energy tensor $T_{\mu\nu}$. 

To know the scalar four-point function, only a specific term in the on-shell action will be relevant to us. In the leading order of slow-roll approximation, it will be obtained from a Feynmann-Witten diagram involving two scalar-scalar-graviton three-point vertices (coming from $S_{\text{int}}$ in eq.\eqref{eq:defint}) properly Wick contracted via a graviton exchange in the bulk. This was explained in \cite{Ghosh:2014kba} and for $\alpha$-vacua correlator we also compute the same diagram \footnote{In eq.\eqref{eq:defint} we have just kept the linear order term in $\delta g_{\mu\nu}$ since this is sufficient for calculating the scalar $4$-point function to the desired leading order.}. The scalar bulk-to-boundary and the graviton bulk-to-bulk propagators are straightforwardly obtained from the corresponding quadratic terms in eq.\eqref{eq:expact}. 

The graviton exchanged diagram will give us a term in the on-shell action proportional to $\delta \phi^4(z\to 0)$, with precisely four factors of the scalar modes present in the boundary. Each of these four $\delta \phi$'s corresponds to four external momenta. Also, the coefficient of $\prod_{i=1}^4 \delta \phi(z\to 0, \vect{k}_i)$ in the $EAdS_4$ on-shell action will readily give us the coefficient function $\langle O(\vect{k}_1)O(\vect{k}_2)O(\vect{k}_3)O(\vect{k}_4) \rangle_{\alpha}$ to be used in eq.\eqref{eq:sc_corr_wf_coff_rel} \footnote{Upon analytical continuation, the onshell action is related to the wave-function in eq.\eqref{eq:wfgauge2}.}. 

The discussion above can be summarized into the following relevant expression for the on-shell action, see eq.(4.15) in \cite{Ghosh:2014kba}, 
\begin{equation}\label{eq:onshellact1}
	\begin{split}
		S^{\text{AdS}}_{\text{on-shell}} = {M_{\text{pl}}^2 R_{\text{\text{AdS}}}^2 \over 8} \int & {d z_1 \over z_1^4} {d z_2 \over z_2^4}  d^3 x_1 d^3 x_2  \, \bar{g}^{k_1 k_2} \bar{g}^{l_1 l_2}\bar{g}^{i_1 i_2} \bar{g}^{j_1 j_2}  \times \\ 
		& T_{k_1 l_1}(x_1, z_1)\, G^{{\text{grav}}}_{l_2 k_2, j_2 i_2} (x_1, z_1, x_2, z_2)\, T_{i_1 j_1} (x_2, z_2) \, ,
	\end{split}
\end{equation}
where $G^{{\text{grav}}}_{l_2 k_2, j_2 i_2} (x_1, z_1, x_2, z_2)$ is bulk-to-bulk propagator for the graviton $g_{\mu\nu}$. The four factors of $\delta\phi$ corresponding to the sources at the boundary come from the two factors of $T_{k l}$,  each contributing a pair of $\delta\phi$. For our purpose in this paper, we need to be careful about how $\alpha$-vacuum will affect the evaluation of eq.\eqref{eq:onshellact1}.

\subsection{Computing on-shell action in $EAdS_4$: new ingredients due to $\alpha$-vacua} \label{new_alpha_wfn}

We now focus on the elements that will modify the $EAdS_4$ on-shell action computation in \eqref{eq:onshellact1} for $\alpha$-vacua. Firstly, we take note of the non-regular boundary condition for $\delta \phi$ as $z\to \infty$. Additionally, the graviton propagator must also be modified to reflect the appropriate boundary condition due to $\alpha$-vacuum.  

\subsubsection*{On-shell action with non-regular boundary condition: }

To the leading order (ignoring the interactions), the field $\delta \phi$ behaves like a free massless scalar field \eqref{eq:expact}, satisfying the Klein-Gordon equation in $EAdS_4$ given by $\nabla^2 \delta \phi = 0$. The general solution for $\delta \phi$ to this equation will be given by eq.\eqref{eq:solscalads}. Next, we prepare $\delta \phi$ as a linear combination of the four solutions
\begin{equation}\label{eq:supsoleads}
	\delta \phi = \sum_{i=1}^{4} \phi(\vect{k}_i) \, \left(c_1\, (1+ k_i z) e^{-k_i z} + c_2\, (1-k_i z)e^{k_i z} \right) \, e^{i \vect{k}_i \cdot \vect{x}} \, ,
\end{equation}
with $c_1$ and $c_2$ given in eq.\eqref{eq:c1c2def}. Upon taking the boundary limit $z \to 0$ of eq.\eqref{eq:supsoleads}, we get
\begin{equation}\label{eq:scalsource}
	\delta \phi(z = 0) = \sum_{i=1}^{4} \phi(\vect{k}_i) (c_1 + c_2 ) e^{i \vect{k}_i \cdot \vect{x}} \, .
\end{equation}
In the on-shell action, this $ \phi (\vect{k}_i)$, with $i = 1, \cdots, 4$, are going to show up as the four scalar sources present at the boundary $z=0$ with four given external momenta. The signature of $\alpha$-vacuum is reflected as we retain the non-regular behaviour with exponentially growing (as $z \to \infty$) mode $e^{kz}$ in eq.\eqref{eq:supsoleads}. This is the first instance where the Bunch-Davies result (with $c_2 = 0$) in \cite{Ghosh:2014kba} will get modified. 

As $\delta\phi$ (from eq.\eqref{eq:scalsource}) gets substituted in $S^{\text{AdS}}_{\text{on-shell}}$ (in eq.\eqref{eq:onshellact1} through the $T_{kl}$), the non-regular part of eq.\eqref{eq:supsoleads} ($e^{k_i z}$ terms) should be treated carefully. They will make the integration over the radial coordinate $z$ in the on-shell action, eq.\eqref{eq:onshellact1}, divergent as $z \to \infty$. We extract the finite part of the integrals involving $e^{k_i z}$ terms, ignoring the divergent contribution. This essentially corresponds to the following: we first evaluate the integrals involving the finite pieces from $e^{-k_i z}$ terms and replace $k_i \to -k_i$ to obtain the answer of the integrals involving $e^{k_i z}$ terms. 

In other words, we are using the following prescription to isolate the finite pieces in the divergent integrals. Let us consider a function $A(z)$ \footnote{We have functions $A(z)$ that typically have poles at $z=0$. These will result in divergences as $z \to 0$. These manifest as divergences in the late time slice $\eta \to 0$. After introducing a suitable cut-off, one can show that all these divergences combine to cancel out amongst themselves when computing correlators. Thus, we only focus on the divergences as $z \to \infty$.}, 
and also assume that we are doing the following integration
\begin{equation}\label{eq:divintfinal}
	\int_{0}^{\infty} dz \, A(z)\,  e^{k z} = B(-k) + \text{Divergent terms as $z \to \infty$} \, . 
\end{equation}
Thus, apart from the finite piece $B(-k)$, we should, in principle, expect the divergences. But we will choose to ignore the divergent terms and only keep the finite piece as 
\begin{equation}\label{eq:divint}
	\int_{0}^{\infty} dz \, A(z) \,  e^{k z} = B(-k) \, . 
\end{equation}
Moreover, we note that $B(-k)$ can be obtained from $B(k)$ by performing the following convergent integration  
\begin{equation}
\int_{0}^{\infty} dz \, A(z) \, e^{-kz} = B(k)  \, .
\end{equation}
We have confidence in this prescription because, as we will see later, the four-point scalar correlator computed with this exactly matches the in-in result of eq.\eqref{eq:scalarfourpfinal_inin} \footnote{Actually, this divergent behavior at $z \to \infty$ in $EAdS_4$ has similar counterparts in $dS_4$ related via the analytic continuation eq.\eqref{anaz}, see comments following eq.\eqref{eq:solscalads}. We emphasize that this analytic continuation to go to $EAdS_4$ is just a trick for us to compute correlation functions. We have calculated the correlator from the in-in formalism in section \ref{sec:4pininalpha}. There, we use the prescription $\eta \to \eta(1 \pm i \epsilon)$ with $\epsilon \to 0$ to ensure the convergence of integrals in far past $\eta \to - \infty$. This ``$i \epsilon$'' prescription in dS amounts to choosing the finite part of the integrals like eq.\eqref{eq:divintfinal} in EAdS.}.  

\subsubsection*{A modified Graviton propagator due to $\alpha$-vacua:}
We now discuss the modifications in the graviton bulk-to-bulk propagator due to the change in initial conditions from Bunch-Davies to $\alpha$-vacuum. The four-point function is given by a Witten diagram with four scalar modes in the external legs ending at the boundary $z=0$. In the interior, the scalar modes pair-wise interact through a graviton exchange process arising in two three-point vertices $S_{int}$ of eq.\eqref{eq:expact}. The information of the $\alpha$-vacua is encoded in the boundary conditions of the $\delta\phi$ solutions eq.\eqref{eq:expact}. Thus, it is natural to expect those boundary conditions to reflect in the Green's function or the propagator.

The bulk-to-bulk propagators originate from the quadratic terms in eq.\eqref{eq:expact}. Let us first compute Green's function for the simpler case of a scalar field in $dS_4$ with metric eq.\eqref{dSmeteta} and then add the information of the polarization to obtain the graviton Green's function
\begin{equation}\label{eq:gravGreenScal}
	\begin{split}
		\widetilde{G}_{l k, j i}({\vect{k}}, \eta,\eta') = {1 \over 2}  \left(\pi^{\vect{k}}_{l j} \pi^{\vect{k}}_{k i} + \pi^{\vect{k}}_{l i} \pi^{\vect{k}}_{k j} - 
		\pi^{\vect{k}}_{l k} \pi^{\vect{k}}_{j i} \right) \, G_k(\eta,\eta') \, ,
	\end{split}
\end{equation} 
where $\pi^{\vect{k}}_{ij}$ was defined in eq.\eqref{eq:tranproj}.  
In momentum space, the scalar propagator solves the equation
\begin{equation}\label{eq:propseqns}
	\nabla^2 G_k(\eta,\eta') = i\delta(\eta-\eta') \, .
\end{equation}
However, most importantly, we should specify the boundary conditions to obtain the Green's function. This is where our propagator in $\alpha$-vacua will differ from the propagator with Bunch-Davies vacuum studied in \cite{Ghosh:2014kba}. 

Let us first look at the boundary condition for the Bunch-Davies case. Following \cite{Goodhew:2020hob, Goodhew:2021oqg, Bonifacio:2022vwa}, the appropriate boundary conditions for the bulk-to-bulk Green's function can be written as
\begin{equation}\label{eq:bdycondsBD}
	\lim_{\eta \to \eta_0} G_k(\eta,\eta') = 0 \, ,  \quad \text{and} \quad \lim_{\eta \to -\infty(1- i \epsilon)} G_k(\eta,\eta') = 0 \, .
\end{equation}
Here, $\eta_0$ is the late time slice, which eventually can be set to $\eta_0 = 0$. The first boundary condition of eq.\eqref{eq:bdycondsBD} indicates that the Green's function must vanish when both the points are taken to the boundary at $\eta_0$. The second boundary condition at far past can be interpreted as the choice of Bunch-Davies vacua, i.e., the mode asymptotes to the positive frequency Minkowski mode. With this boundary condition, we get the following general solution to eq.\eqref{eq:propseqns} for massless scalars \footnote{This matches with eq.(2.23) of \cite{Bonifacio:2022vwa}. Due to a choice of conventions, we have defined Green's function with an overall factor of $i$. See footnote-$6$ of \cite{Goodhew:2021oqg} for a discussion on this point.}:
\begin{align}\label{bb_bb_prop_def}
	G_k(\eta,\eta')=i\theta(\eta-\eta')f^*_k(\eta)f_k(\eta')+i\theta(\eta'-\eta)f^*_k(\eta')f_k(\eta)-\frac{if^*_k(\eta_0)}{f_k(\eta_0)}f_k(\eta)f_k(\eta') \, ,
\end{align}
where $f_k(\eta)$ is the solution to the free equation of motion. For Bunch-Davies vacuum, this is given by,
$f^{BD}_k(\eta)=(H / \sqrt{2k^3})(1-ik\eta)\,e^{ik\eta}$. Also, the bulk-to-boundary propagator for massless scalars with Bunch-Davies vacuum becomes $\mathcal{K}^{BD}_k(\eta) \equiv f^{BD}_k(\eta)/f^{BD}_k(\eta_0)$, satisfying the boundary condition $\text{lim}_{\eta \to\eta_0}\mathcal{K}^{BD}_k(\eta)=1$. 

We want to generalize this to $\alpha$-vacua. Thus, although the first condition in eq.\eqref{eq:bdycondsBD} will remain the same, the second one for $\eta \to -\infty$ must be modified. We know the most general solution to the Klein-Gordon equation for a massless scalar in $dS_4$ as
\begin{equation}\label{f_def_c1_c2}
	f_k(\eta)=\frac{H}{\sqrt{2k^3}}\left( c_1 \, U^{BD}_k(\eta)+ c_2 \, U^{BD*}_k(\eta)\right) \, ,
\end{equation}
where $U^{BD}_k(\eta)=(1-ik\eta)e^{ik\eta}$, see eq.\eqref{defUkBD}, eq.\eqref{eq:modefunctions}. 
Substituting eq.\eqref{f_def_c1_c2} in eq.\eqref{bb_bb_prop_def} the bulk-to-bulk propagator can be obtained as 
\begin{equation}\label{eq:gravpropstep}
\begin{split}
	G_k(\eta,\eta')&=\frac{i H^2}{2k^3}\Bigg[\theta(\eta-\eta')\{c_1^*U^{BD*}_k(\eta)+c_2^*U^{BD}_k(\eta)\}\{c_1U^{BD}_k(\eta')+c_2U^{BD*}_k(\eta')\} \\
	&\hspace{1.2cm}+\theta(\eta'-\eta)\{c_1^*U^{BD*}_k(\eta')+c_2^*U^{BD}_k(\eta')\}\{c_1U^{BD}_k(\eta)+c_2U^{BD*}_k(\eta)\} \\
	&\hspace{1.2cm}-\frac{c_1^*+c_2^*}{c_1+c_2}\{c_1 U^{BD}_k(\eta)+ c_2 U^{BD*}_k(\eta)\}\{c_1 U^{BD}_k(\eta')+ c_2 U^{BD*}_k(\eta')\}\Bigg] \, .
\end{split}
\end{equation}
For the case of $\alpha$-vacua, we have to put $ c_1=\cosh{\alpha}, \, 	c_2=-i\sinh{\alpha}$, according to eq.\eqref{eq:c1c2def}. The constants $c_1,c_2$ can be traded for the one-parameter family (denoted by $\alpha$) of the initial condition at $\eta \to -\infty$. A consistency check of eq.\eqref{eq:gravpropstep} is to realize that it satisfies the expected relation between the bulk-to-bulk (given by $G_k(\eta,\eta')$) and the bulk-to-boundary propagator (given by $\mathcal{K}_k(\eta) \equiv f_k(\eta)/f_k(\eta_0)$) as given below \footnote{Eq.\eqref{eq:proprel} can be derived from the relation explained in eq.(A.10) of \cite{Goon:2018fyu}
\begin{equation*}\label{eq:ogrel}
	\lim_{\eta' \to \eta_0} \sqrt{\bar{g}} \, n^{\mu} \nabla^{(\eta',\mathbf{y})}_{\mu} \mathcal{G}(\eta,\mathbf{x};\eta',\mathbf{y}) = \mathcal{K} (\eta,\mathbf{x}:\mathbf{y}) \, ,
\end{equation*}
where, $\mathcal{G}$ is the bulk to bulk propagator and $\mathcal{K}$ is the bulk to boundary propagator. To be precise, $\mathcal{K}(\eta,\mathbf{x}:\mathbf{y})$ is the Fourier transform of $\mathcal{K}_k(\eta)$, appearing in eq.\eqref{eq:proprel}. Similarly, $\mathcal{G}(\eta,\mathbf{x};\eta',\mathbf{y})$ is the Fourier transform of $G_k (\eta,\eta')$. Also, $\bar{g}$ is the determinant of the induced metric on the slice $\eta = \eta_0$ and $n^{\mu}$ is the outward pointing normal vector to this surface. Using this one can obtain eq.\eqref{eq:proprel} by substituting
\begin{equation*}
		\bar{g} = \dfrac{1}{H^6\eta^6_0} ~~ \text{as} ~~ \bar{g}_{ab} = \dfrac{1}{H^2\eta^2_0} d \mathbf{x}^2 \, , \quad \text{and} \quad
		n^{\mu} = (H\eta,0,0,0),  ~~ \text{such that $n^{\mu}n_{\mu} = -1$} \, .
\end{equation*}
It should be noted that our bulk-to-boundary propagator $\mathcal{K}_k(\eta)$ for the scalar modes in $\alpha$-vacua satisfies the following boundary condition $\text{lim}_{\eta \to\eta_0}\mathcal{K}_k(\eta)=1$, same as in \cite{Goon:2018fyu, Goodhew:2020hob, Goodhew:2021oqg, Bonifacio:2022vwa}.}
\begin{equation}\label{eq:proprel}
	\lim_{\eta' \to \eta_0}\partial_\eta G_k (\eta,\eta') = H^2\eta_0^2 \, \mathcal{K}_k(\eta) = H^2\eta_0^2 \, f_k(\eta)/f_k(\eta_0) \, ,
\end{equation}
with $\eta_0 \to 0^-$. We have checked that eq.\eqref{f_def_c1_c2} and eq.\eqref{eq:gravpropstep} satisfies eq.\eqref{eq:proprel}. 

Using eq.\eqref{eq:c1c2def}, one can further simplify eq.\eqref{eq:gravpropstep}, to obtain $G_k(\eta,\eta')$ for $\alpha$-vacua as 
\begin{equation}\label{grav_prop_alpha_simple}
	\begin{split}
	G_k^\alpha(\eta,\eta')&=\theta(\eta-\eta')\frac{H^2}{k^3}\frac{\cosh{\alpha}  \,U^{BD}_k(\eta')-i \,\sinh{\alpha} \, U_k^{BD*}(\eta')}{\cosh{\alpha}-i \,\sinh{\alpha}} \,\left[\sin{k\eta}-k\eta\cos{k\eta}\right]\\
	&\hspace{1cm}+\{\eta \leftrightarrow\eta'\} \, .
\end{split}
\end{equation}
Now, using the Bessel function formulae
\begin{equation}
	\begin{split}
		K_{3/2}(x) = \sqrt{\dfrac{\pi}{2}} x^{-\frac{3}{2}} e^{-x} (1+ x) \, , \quad 		I_{3/2}(x) = \sqrt{\dfrac{2}{\pi}} x^{-\frac{3}{2}} (x \cosh(x) - \sinh(x)) \, ,
	\end{split}
\end{equation}
and analytically continuing eq.\eqref{grav_prop_alpha_simple} through $\eta = i z$, one obtains the scalar bulk-to-bulk propagator in $EAdS_4$. Dressing with the projectors as shown in eq.\eqref{eq:gravGreenScal}, one finally obtains the transverse bulk-to-bulk graviton propagator in $EAdS_4$ as 
\begin{equation}\label{eq:gravalphatranprop}
	\begin{split}
		\widetilde{G}_{l k, j i}({\vect{k}}, z_1, z_2)&  =\dfrac{1}{c_1+c_2} {1 \over 2} \left(\pi^{\vect{k}}_{l j} \,\pi^{\vect{k}}_{k i} + \pi^{\vect{k}}_{l i} \,\pi^{\vect{k}}_{k j} - 
		\pi^{\vect{k}}_{l k} \,\pi^{\vect{k}}_{j i} \right) \times \\
		(z_1 z_2)^{\frac{3}{2}} & \bigg(\Theta(z_1 - z_2) \left[c_1 K_{3/2}(k z_1) I_{3/2}(k z_2) + c_2 K_{3/2}(-k z_1) I_{3/2}(-k z_2) \right] \\
		&+  \Theta(z_2-z_1) \left[ c_1 K_{3/2}(k z_2) I_{3/2}(k z_1) + c_2  K_{3/2}(-k z_2) I_{3/2}(-k z_1) \right] \bigg)  \, .
	\end{split}
\end{equation}

\subsection{Four point scalar correlator in alpha vacua} \label{ssec:4ptfullans}

In this sub-section, we compute the final expression for the scalar four point function in $\alpha$-vacua. 

\subsubsection*{The on-shell action:} 
\begin{figure}[htb] 
\begin{center}
\includegraphics[scale=.6]{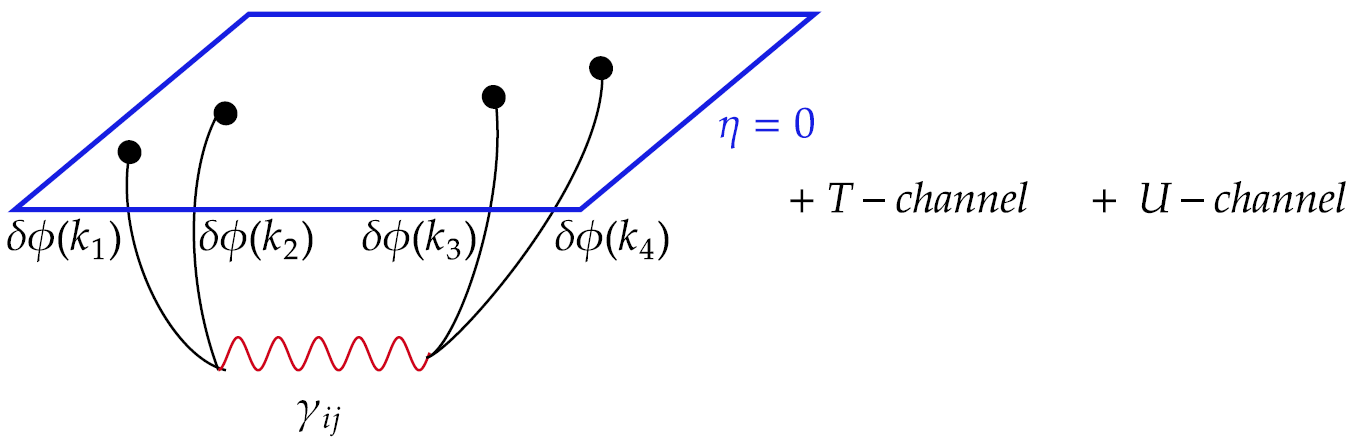}
\caption{Feynman-Witten diagram for $\langle \delta\phi\delta\phi\delta\phi\delta\phi \rangle$.}\label{4pointfig}
\end{center}
\end{figure}

We use the mode expansion of $\delta\phi$ with superposition of four external scalar modes from eq.\eqref{eq:supsoleads}, and the modified bulk-to-bulk graviton propagator in eq.\eqref{eq:gravalphatranprop} suitable for initial conditions with $\alpha$-vacuum in the on-shell action eq.\eqref{eq:onshellact1}, which is represented by the Feynman-Witten diagram in figure-\ref{4pointfig}.
. 
With these ingredients, following the steps outlined in \cite{Ghosh:2014kba} we find the on-shell action given by 
\begin{align}\label{eq:sAdS_onshell}
	S^{\text{AdS}}_{\text{on-shell}} ={M_{Pl}^2 R_{\text{AdS}}^2 \over 2} {1 \over 4}[\widetilde{W}_{\alpha} + 2 R_{\alpha}],
\end{align}
where $\widetilde{W}_{\alpha}$ is the transverse graviton contribution and $R_{\alpha}$ is the longitudinal graviton contribution. Details leading up to the calculation of eq.\eqref{eq:sAdS_onshell} is given in Appendix\,\ref{ap:onshellact}. The subscript $\alpha$ is used to denote the full $\alpha$ vacua contribution.

We explicitly write here the s-channel expression for $\widetilde{W}_{\alpha}$,
\begin{equation}\label{eq:wnew}
	\widetilde{W}_{\alpha}^S(\vect{k}_1,\vect{k}_2,\vect{k}_3,\vect{k}_4)= 16(2\pi)^3\delta^3\left(\sum_i\vect{k}_i\right) \left(\prod_{a=1}^4 \phi(\vect{k}_a)\right) \widehat{W}^S_{\alpha}(\vk, \vkk, \vkkk, \vkfour) 
\end{equation}
where,
\begin{equation}
	\widehat{W}^S_{\alpha}(\vk, \vkk, \vkkk, \vkfour)=  k_1^l\,  k_2^m\,  k_3^j\,  k_4^i \,  \mathcal{T}_{lmji}^{\vect{K}_s} \,  S_{\alpha}(k_1,k_2,k_3,k_4) \, ,
\end{equation}
where $\mathcal{T}^{\vect{k}}_{ijlm}$ is the transverse projector and is defined as follows,
\begin{equation}\label{eq:pol_sum}
	\mathcal{T}_{lkji}^{\vect{k}}\equiv \left(\pi^{\vect{k}}_{l j} \pi^{\vect{k}}_{k i} + \pi^{\vect{k}}_{l i} \pi^{\vect{k}}_{k j} - 
	\pi^{\vect{k}}_{l k} \pi^{\vect{k}}_{j i} \right) \, , \quad \pi_{ij}^{\vect{k}}\equiv\delta_{ij}-\frac{k_ik_j}{k^2} \, . 
\end{equation}
One can obtain 
\begin{equation}
	S_{\alpha}(k_1,k_2,k_3,k_4)= \frac{1}{c_1+c_2}\left[c_1\, S_{(1)}(k_1,k_2,k_3,k_4,K_s)+c_2 \, S_{(1)}(k_1,k_2,k_3,k_4,-K_s)\right] \, ,
\end{equation}
where $S_{(1)}(k_1,k_2,k_3,k_4,\pm K_s)$ can be written as,
\begin{equation}
	\begin{split}\label{eq:snew}
		S_{(1)}(k_{1},k_{2},k_{3}&,k_{4},\pm K_s) = c^4_1 S(k_1,k_2,k_3,k_4,\pm K_s)+ c^4_2 S(-k_1,-k_2,-k_3,-k_4,\pm K_s) \\
		&+ c^3_1 c_2 \big( S(-k_1,k_2,k_3,k_4,\pm K_s) + S(k_1,-k_2,k_3,k_4,\pm K_s) \\
		& \quad\quad + S(k_1,k_2,-k_3,k_4,\pm K_s) + S(k_1,k_2,k_3,-k_4,\pm K_s) \big) \\
		& + c^2_1 c^2_2 \big( S(-k_1,-k_2,k_3,k_4,\pm K_s) + S(-k_1,k_2,-k_3,k_4,\pm K_s) \\
		&\quad\quad+ S(-k_1,k_2,k_3,-k_4,\pm K_s) + S(k_1,-k_2,-k_3,k_4,\pm K_s) \\
		&\quad\quad+S(k_1,-k_2,k_3,-k_4,\pm K_s) + S(k_1,k_2,-k_3,-k_4,\pm K_s) \big) \\
		& + c_1 c^3_2 \big( S(-k_1,-k_2,-k_3,k_4,\pm K_s) + S(-k_1,k_2,-k_3,-k_4,\pm K_s) \\
		&\quad\quad+ S(-k_1,-k_2,k_3,-k_4,\pm K_s) + S(k_1,-k_2,-k_3,-k_4,\pm K_s) \big)
		 \, ,
	\end{split}
\end{equation}
such that $S(k_1,k_2,k_3,k_4,K_s)$ is the corresponding Bunch-Davies expression explicitly given in eq.\eqref{eq:sbd}.

Similarly, we can write the s-channel of the longitudinal graviton contribution as, 
\begin{equation}\label{eq:rnew}
	R^{S}_{\alpha}(\vk,\vkk,\vkkk,\vkfour) =16 (2\pi)^3 \ \delta^3\big(\sum_{J=1}^4 {\vect{k}}_J\big) \bigg[ \prod_{I=1}^4 \phi({\vect{k}}_I)\bigg]  
	\widehat{R}^S_{\alpha}(\vk,\vkk,\vkkk,\vkfour),
\end{equation}
where $\widehat{R}^S_{\alpha}$ is given in terms of the corresponding Bunch-Davies expressions 
\begin{equation}\label{eq:long_cont_wf}
	\begin{split}
		\widehat{R}^S_{\alpha}(k_{1},k_{2}&,k_{3},k_{4})= (c_1^4-c_2^4)\widehat{R}^S(k_{1},k_{2},k_{3},k_{4}) \\ 
		&+(c_1^3c_2-c_1c_2^3)\bigg[\widehat{R}^S(-k_{1},k_{2},k_{3},k_{4}) +\widehat{R}^S(k_{1},-k_{2},k_{3},k_{4})\\ 
		&\quad \quad +\widehat{R}^S(k_{1},k_{2},-k_{3},k_{4})
		+\widehat{R}^S(k_{1},k_{2},k_{3},-k_{4})\bigg] \, , 
	\end{split}
\end{equation}
and $\widehat{R}^S(k_{1},k_{2},k_{3},k_{4})$ is the Bunch-Davies expression for longitudinal graviton contribution. Some details of the calculation is given in Appendix\,\ref{ap:OOOO_calc}; see steps following eq.\eqref{defrs}. 

Using eq.\eqref{eq:wnew} and eq.\eqref{eq:rnew}, we can add all the three (s,t,u) channels and write the full on-shell action as, 
\begin{equation}\label{eq:on_shell_action_full}
	\begin{split}
		S^{\text{AdS}}_{\text{on-shell}} = {M_{\text{pl}}^2 R_{\text{\text{AdS}}}^2 \over 4}  \bigg[&\left(\frac{{\widetilde W}^S_{\alpha}(\vk, \vkk, \vkkk, \vkfour)}{2} + R^S_{\alpha}(\vk, \vkk, \vkkk, \vkfour)\right) \\ 
		&+ \big({\vkk \leftrightarrow \vkkk} \big)+\big({\vkk \leftrightarrow \vkfour} \big) \bigg] \, ,
	\end{split}
\end{equation}

\subsubsection*{The wave-function coefficient $\langle O(\vk) O(\vkk) O(\vkkk) O(\vkfour)\rangle_{\alpha}$:}
Next, we differentiate the on-shell action given in eq.\eqref{eq:on_shell_action_full} by the boundary value of the perturbations given in eq.\eqref{eq:scalsource}. Analytically continuing the $EAdS_4$ expression using eq.\eqref{anaz}, we obtain the $dS_4$ wave-function coefficient $\langle O(\vk) O(\vkk) O(\vkkk) O(\vkfour)\rangle_{\alpha}$ written as,
\begin{equation}\label{eq:wf_coff_dS}
	\begin{split}
		\langle & O(\vk)O(\vkk) O(\vkkk) O(\vkfour)\rangle_\alpha =
		-4(2 \pi)^3 \delta^3(\sum_{i=1}^3
		{\vect{k}}_i)  \frac{1}{(c_1+c_2)^4} \times \\ \bigg[&\left(\frac{{\widehat W}^S_{\alpha}(\vk, \vkk, \vkkk, \vkfour)}{2} + \widehat{R}^S_{\alpha}(\vk, \vkk, \vkkk, \vkfour)\right) + \big({\vkk \leftrightarrow \vkkk} \big)+\big({\vkk \leftrightarrow \vkfour} \big) \bigg] \, .
	\end{split}
\end{equation}

\subsubsection*{The wave-function coefficient $\langle O(\vk) O(\vkk) T_{ij}(\vkkk) \rangle_\alpha$:} Next, we need the wave-function coefficient $\langle OOT_{ij}\rangle_\alpha$, which has already been calculated in\,\cite{Jain:2022uja} for the $\alpha$-vacua. We write it explicitly here again for completeness 
\begin{equation} \label{OOT_ij_alpha}
	\langle O(\vk) O(\vkk) T_{ij}(\vkkk)\rangle_{\alpha} \epsilon^{s,ij}=(2\pi)^3 \delta^3\Bigg(\sum_{a=1}^{3}{\vect{k_a}}\Bigg)\epsilon^{s,ij}G_{\alpha}(k_{1},k_{2},k_{3})k^i_{1}k^j_{2} \, ,
\end{equation}
where $\epsilon^{ij}$ is the polarisation tensor for the exchange graviton and $G_{\alpha}(k_{1},k_{2},k_{3})$ is defined below as follows\,\cite{Jain:2022uja},
\begin{equation}\label{eq:gnew}
	\begin{split}
		G_{\alpha}(k_{1},k_{2},k_{3})&= Re \Bigg[\dfrac{1}{(c_1+c_2)^3} \left[ (c^3_1-c^3_2) G(k_1,k_2,k_3)\right. \\& \left. + (c^2_1 c_2- c_2 c^2_1)( G(-k_1,k_2,k_3)+ G(k_1,-k_2,k_3) + G(k_1,k_2,-k_3)) \right] \Bigg]\, ,
	\end{split}
\end{equation}
where $c_1$ and $c_2$ have already been defined in eq.\eqref{eq:c1c2def} and $G( k_1,k_2,k_3)$ is just the Bunch-Davies expression which has the following form,
\begin{equation}
	G( k_1,k_2,k_3) = (k_1+k_2+k_3) - {\sum_{i>j}k_i k_j \over (k_1+k_2+k_3)} - {k_1k_2k_3\over (k_1+k_2+k_3)^2} \, .
\end{equation}
Note that $Re$ in eq.\eqref{eq:gnew} implies imposing reality condition on the coefficients that appear.

\subsubsection*{From wave-function coefficients to $\langle \delta\phi(\vk) \delta\phi(\vkk) \delta\phi(\vkkk) \delta\phi(\vkfour)\rangle_\alpha$:}

Knowing the coefficient functions, we can now directly obtain the scalar four-point function $\langle \delta\phi \delta\phi \delta\phi \delta\phi \rangle_\alpha$. From eq.\eqref{eq:sc_corr_wf_coff_rel}, we see that there are two types of contribution to $\langle \delta\phi \delta\phi \delta\phi \delta\phi \rangle_\alpha$. The first one comes from the ``Coefficient Function" $\langle OOOO \rangle_\alpha$, hence, we call it $\langle \delta\phi \delta\phi \delta\phi \delta\phi \rangle^{\text{CF}}_\alpha$ with the superscript \textit{`CF'}. The other contribution comes from two $\langle OOT_{ij} \rangle_\alpha$ coefficients, hence, named the ``Extra Term" and denoted as $\langle \delta\phi \delta\phi \delta\phi \delta\phi \rangle^{\text{ET}}_\alpha$ with the suffix \textit{`ET'} \footnote{The nomenclature is derived from \cite{Ghosh:2014kba}.}. 

Using eq.\eqref{eq:wf_coff_dS}, it is now straightforward to obtain 
\begin{equation}\label{eq:scal_corr_CF}
\begin{split}
	\langle& \delta\phi(\vect{k}_1)\delta\phi(\vect{k}_2)\delta\phi(\vect{k}_3)\delta\phi(\vect{k}_4)\rangle^{\text{CF}}_\alpha=-\frac{H^6}{M_{\text{pl}}^6}\frac{2\Re{\langle O(\vect{k}_1)O(\vect{k}_2)O(\vect{k}_3)O(\vect{k}_4)\rangle_\alpha}}{\prod_{i=1}^4(2\Re{\langle O(\vect{k}_i)O(-\vect{k}_i)\rangle'_\alpha})} \\
	&= - 8 (2 \pi)^3 \delta^3\big(\sum_{J=1}^4 {\vect{k}}_J\big) {H^6 \over M_{Pl}^6} ~~ Re \Big\{ \frac{|c_{1}+c_{2}|^8}{(c_1+c_2)^4}{1 \over \prod_{J=1}^4 (2 k_J^3)} \times \\
	 &  \quad \bigg[\left(\frac{{\widehat W}^S_{\alpha}(\vk, \vkk, \vkkk, \vkfour)}{2} + \widehat{R}^S_{\alpha}(\vk, \vkk, \vkkk, \vkfour)\right) + \big({\vkk \leftrightarrow \vkkk} \big)+\big({\vkk \leftrightarrow \vkfour} \big) \bigg] \Big\} \, .
	\end{split}
\end{equation}
To arrive at eq.\eqref{eq:scal_corr_CF} we have used,
\begin{equation}
	\langle O(\vect{k})O(-\vect{k})\rangle_\alpha=\frac{1}{2 \langle\delta\phi(\vect{k})\delta\phi(\vect{-k})\rangle}=\frac{k^3}{\abs{c_1+c_2}^2} \, .
\end{equation}
The reality condition $Re$ in eq.\eqref{eq:scal_corr_CF} is a consequence of the comments made in footnote-\ref{footnote26}. In fact, we will see that this reality condition on the coefficients is crucial for the four point and three point correlators computed to match with the in-in result.

Similarly, using eq.\eqref{OOT_ij_alpha} in eq.\eqref{eq:sc_corr_wf_coff_rel} we get the `ET" contribution as follows,
\begin{equation}\label{eq:scal_corr_ET}
	\begin{split}
		\langle \delta \phi(\vk)&\delta\phi(\vkk) \delta \phi(\vkkk) \delta\phi(\vkfour) \rangle^{\text{ET}}_\alpha= 4 (2 \pi)^3 \delta^3\big(\sum_{J=1}^4 {\vect{k}}_J\big) {H^6 \over M_{Pl}^6}{|c_{1}+c_{2}|^{10} \over \prod_{J=1}^4 (2 k_J^3)} \times \\ & \bigg[\widehat{G}^S_{\alpha}(\vk, \vkk, \vkkk, \vkfour) +  \big({\vkk \leftrightarrow \vkkk} \big)+\big({\vkk \leftrightarrow \vkfour} \big)\bigg] \, ,
	\end{split}
\end{equation}
with $\widehat{G}^S_{\alpha}(\vk, \vkk, \vkkk, \vkfour)$ defined as,
\begin{equation}\label{eq:ghatgg}
	\begin{split}
		\widehat{G}^S_{\alpha}(\vk, \vkk, \vkkk, \vkfour)
		 = { G_{\alpha}(k_1,k_2,k_s) \times G_{\alpha}(k_s,k_3,k_4)\}  \over |\vk+\vkk|^3} \,k^i_1 k^j_2 k^l_3 k^m_4 (\mathcal{T}_{ijlm}^{\vect{k}_s})  \, .
	\end{split}
\end{equation}
Here $\widehat{G}^S_{\alpha}(\vk, \vkk, \vkkk, \vkfour)$ is given in eq.\eqref{eq:gnew}, $k_s\equiv\abs{\vect{k}_1+\vect{k}_2}$ and the other terms can be worked out from the following identity,
\begin{equation}\label{eq:tranpropcont}
	\begin{split}
		& k^i_1 k^j_2 k^l_3 k^m_4 (\mathcal{T}_{ijlm}^{\vect{k}_s})= \\
		& \bigg\{\vk.\vkkk+\frac{\{(\vkk+\vk).\vk\} \{(\vkfour+\vkkk).\vkkk\}}{|\vk+\vkk|^2}\bigg\} \bigg\{\vkk.\vkfour+  \frac{\{(\vk+\vkk).\vkk\} \{(\vkkk+\vkfour).\vkfour\}}{|\vk+\vkk|^2}\bigg\} \\ 
		& +  \bigg\{\vk.\vkfour+\frac{\{(\vkk+\vk).\vk\} \{(\vkfour+\vkkk).\vkfour\}}{|\vk+\vkk|^2}\bigg\}\bigg\{\vkk.\vkkk+\frac{\{(\vkk+\vk).\vkk\} \{(\vkfour+\vkkk).\vkkk\}} {|\vk+\vkk|^2}\bigg\}\\ 
		& - \bigg\{\vk.\vkk-\frac{\{(\vkk+\vk).\vk\} \{(\vk+\vkk).\vkk\}} {|\vk+\vkk|^2}\bigg\} \bigg\{\vkkk.\vkfour-\frac{\{(\vkkk+\vkfour).\vkfour\} \{(\vkfour+\vkkk).\vkkk\}}{|\vk+\vkk|^2}\bigg\} \, .
	\end{split}
\end{equation}
We note here that the structure of eq.\eqref{eq:gnew} is completely fixed by conformal invariance and permutation symmetry. This shows that conformal symmetry completely fixes the ``ET" part of the scalar four-point function. Now we can combine eq.\eqref{eq:scal_corr_CF} and \eqref{eq:scal_corr_ET} to write the full scalar four-point correlator formally as,
\begin{equation}\label{eq:sca_corr_full_wf}
	\begin{split}
		\langle \delta \phi(\vk)\delta\phi(\vkk) \delta \phi(\vkkk) \delta\phi(\vkfour) \rangle_\alpha=&\langle \delta \phi(\vk)\delta\phi(\vkk) \delta \phi(\vkkk) \delta\phi(\vkfour) \rangle^{\text{CF}}_\alpha \\ &+\langle \delta \phi(\vk)\delta\phi(\vkk) \delta \phi(\vkkk) \delta\phi(\vkfour) \rangle^{\text{ET}}_\alpha.
	\end{split}
\end{equation}

Finally, we can now move to gauge-$2$ by using eq.\eqref{infper} to obtain the scalar four-point correlator in terms of the variable $\zeta$,  
\begin{equation}\label{eq:zzzz_wf}
	\begin{split}
		\langle \zeta(\vk) \zeta(\vkk) \zeta(\vkkk) \zeta(\vkfour) \rangle_\alpha= \frac{1}{4\epsilon^2}\langle \delta \phi(\vk)\delta\phi(\vkk) \delta \phi(\vkkk) \delta\phi(\vkfour) \rangle_\alpha.
	\end{split}
\end{equation}

\subsubsection*{Consistency with the in-in result and some comments:} We have verified in Mathematica that the scalar four-point function calculated using wavefunctional formalism (eq.\eqref{eq:sca_corr_full_wf}) and the same calculated using the in-in formalism (eq.\eqref{eq:scalarfourpfinal_inin}) match with each other.

The contribution from the longitudinal part of the graviton propagator (denoted by $\widehat{R}_\alpha= \widehat{R}^S_{\alpha}(\vk, \vkk, \vkkk, \vkfour) + \big({\vkk \leftrightarrow \vkkk} \big)+\big({\vkk \leftrightarrow \vkfour} \big)$) to $\langle \delta \phi(\vk)\delta\phi(\vkk) \delta \phi(\vkkk) \delta\phi(\vkfour) \rangle^{\text{CF}}_\alpha$ in eq.\eqref{eq:sca_corr_full_wf} (which is given in eq.\eqref{eq:scal_corr_CF}) matches with the CI term of eq.\eqref{eq:scalarfourpfinal_inin} (given in eq.\eqref{eq:alpha_CI}). Note that $\widehat{R}_\alpha$ is given in terms of the Bunch-Davies rexpression $\widehat{R}_{\alpha=0}$ in eq.\eqref{eq:long_cont_wf}. Upon inspection, the coefficients of $\widehat{R}_{\alpha}$ in eq.\eqref{eq:scal_corr_CF} match with the coefficients of eq.\eqref{eq:alphaci2} because
\begin{equation}
	(c^*_1+c^*_2)^4 = \dfrac{|c_1+c_2|^8}{(c_1+c_2)^4} \, ,
\end{equation}
once $c_1, c_2$ are substituted from eq.\eqref{eq:c1c2def}.

It is useful to note that the structure of eq.\eqref{eq:long_cont_wf} matches the CFT predictions of \cite{Jain:2022uja} even though their analysis was restricted to three point functions. This is not surprising, since, only the longitudinal part of the CF contribution of the wavefunctional result contributes to the LHS and RHS of eq.\eqref{eq:scal_ward_3} and eq.\eqref{eq:sct_ward_3}.

Now the transverse contribution of eq.\eqref{eq:scal_corr_CF} $\widehat{W}_{\alpha}$ and the ET term contribution of eq.\eqref{eq:scal_corr_ET} combine together to match the GE contribution of the in-in result given by eq.\eqref{eq:ininfourptfinal}. Again, one must account for the coefficients $c_1, c_2$ from eq.\eqref{eq:c1c2def} in both the expressions. For the $\alpha$ generalization, the matching of the wave-function and in-in result was done at the level of individual coefficients of the form $\cosh[k](\alpha)\sinh[l](\alpha)$ in Mathematica \footnote{The Mathematica file can be accessed here: \url{https://github.com/arhumansari007/alpha_vacua_ward_identity}.}. 

It should be noted that the transverse part of eq.\eqref{eq:scal_corr_CF} and the ET term in eq.\eqref{eq:scal_corr_ET} do not have the same structure as the longitudinal part eq.\eqref{eq:long_cont_wf}. Hence, they are not fixed in terms of the corresponding Bunch-Davies result through a sign flip of the modulus of some momenta. One way to quickly see this is from eq.\eqref{eq:ghatgg}: the product of $G_{\alpha}$ (i.e. eq.\eqref{eq:gnew}) will involve terms that cannot be derived from the Bunch-Davies result. For instance, a term like $G(k_1,k_2,-k_s)G(k_s,-k_3,k_4)$ cannot be obtained from the corresponding Bunch-Davies expression $G(k_1,k_2,k_s)G(k_s,k_3,k_4)$ by replacing $k_s \to  -k_s$ because that will introduce a sign flip in both of the terms of the product. This is related to the fact that these terms do not contribute to the Ward identities, they drop out from eq.\eqref{eq:scal_ward_3} and eq.\eqref{eq:sct_ward_3}. Hence, the Ward identities do not entirely fix some parts of the correlator, which is expected since we are looking at a four point function. 

From our explicit results, we learn that $\langle \delta\phi \delta\phi \delta\phi \delta\phi \rangle \sim \mathcal{O}(\epsilon^0)$ in the leading slow-roll approximation. Therefore, it has been actually calculated in exact $dS_4$ geometry. Through the changing of gauge using eq.\eqref{infper}, we have retrieved the slow-roll dependence in $\langle \zeta\zeta\zeta\zeta \rangle \sim \mathcal{O}(1/\epsilon^2)$.

\subsection{Three point scalar correlator in alpha vacua}
\label{sec:3ptwfalpha}

We now focus on the computation of the three-point scalar correlator in inflation using wave-function formalism. 
Previously, in \S\,6 of \cite{Shukla:2016bnu}, the same method was used for the three-point scalar correlator but restricted to only Bunch-Davies initial condition. 

One has to compute the same Feynman-Witten diagram used for the four-point scalar correlator, albeit with the following modifications. In one of the four external legs of the four-point diagram, one has to replace the inflaton fluctuation $\delta\phi$ with the time derivative of the homogeneous inflaton background, i.e., $\dot{\bar{\phi}}$.

It is important to note that the wave-function method for computing the three-point scalar correlator is actually motivated by the Ward identities between the three-point and the four-point scalar correlator \footnote{See \S\,(6.3.1) in \cite{Ghosh:2014kba} and also \cite{Arkani-Hamed:2015bza, Lee:2016vti} for a discussion justifying this.}. The relevant terms in the wave-function eq.\eqref{eq:wfgauge2},  can be schematically written as 
\begin{equation} \label{wf_schematic}
\begin{split}
	\Psi[\delta\phi, \, \gamma_{ij}] = \text{exp}\bigg[-&\int \delta\phi \delta\phi \, \langle OO \rangle +\int \delta\phi \delta\phi \delta\phi \, \langle OOO \rangle \\ & + \int \delta\phi \delta\phi \gamma_{ij} \, \langle OOT_{ij}\rangle+ \int \delta\phi \delta\phi \delta\phi \delta\phi \, \langle OOOO \rangle+ \cdots \bigg] \, .
	\end{split}
\end{equation}
For $\langle \delta\phi \delta\phi \delta\phi \rangle$ we need to know the coefficient of a cubic term $\delta\phi^3$ in the wave function, see eq.\eqref{def_cor_fn}. Naively, there is only one such term in eq.\eqref{wf_schematic} involving the coefficient function $\langle OOO \rangle$. But the Ward identity in eq.\eqref{WI_O3_O4} relates $\langle OOO \rangle$ to the term $(\dot{\bar{\phi}}/H) \, \langle O(\vect{k}_1)O(\vect{k}_{2})O(\vect{k}_{3})O(\vect{k}_{4})\rangle \big|_{\vect{k}_{4}\rightarrow 0}$. This is effectively achieved, in the language of computing Feynman-Witten diagram as follows: substitute $\dot{\bar{\phi}}/H$ for one of the four $\delta\phi$'s in the $\delta\phi \delta\phi \delta\phi \delta\phi \, \langle OOOO \rangle$ term on the RHS in eq.\eqref{wf_schematic}. Consequently, the on-shell action, $S^{\text{AdS}}_{\text{on-shell}}$ receives a contribution cubic in $\delta\phi$ but involving $\langle OOOO \rangle$, leading us to $\langle OOO \rangle$. Note that the order of slow-roll parameters plays a significant role in this.

Previously, in \S\,\ref{secWI:check}, we have checked that the scalar three and four-point functions computed using the in-in formalism in $\alpha$-vacua satisfy the scaling and special conformal Ward identities. In this section, we will explicitly show that the final answer for $\langle \zeta \zeta \zeta \rangle^\alpha$ computed using the wave-function method matches the same correlator worked out in \cite{Shukla:2016bnu} using in-in formalism. Given the discussion above, this will serve as a further justification for the validity of Ward identities being satisfied by the $\langle \zeta \zeta \zeta \rangle^\alpha$ and $\langle \zeta \zeta \zeta \zeta \rangle^\alpha$.

Hence, we want to highlight that the computation of $\langle \zeta \zeta \zeta \rangle^\alpha$ through the wave-function formalism is not only a technical advancement that confirms the consistency of the corresponding in-in calculation. Instead, it justifies something novel about the Ward identities. Even without the knowledge of the four-point function, just the matching of the final results for $\langle \zeta \zeta \zeta \rangle^\alpha$ between the wave-function method and the in-in calculation is powerful enough to validate the Ward-identity between $\langle \zeta \zeta \zeta \rangle^\alpha$ and $\langle \zeta \zeta \zeta \zeta \rangle^\alpha$. 

\subsubsection*{Computing the coefficient function $\langle OOO \rangle_\alpha$ and obtaining $\langle \zeta \zeta \zeta \rangle^\alpha$:}

\begin{figure}[htb] 
\begin{center}
\includegraphics[scale=.6]{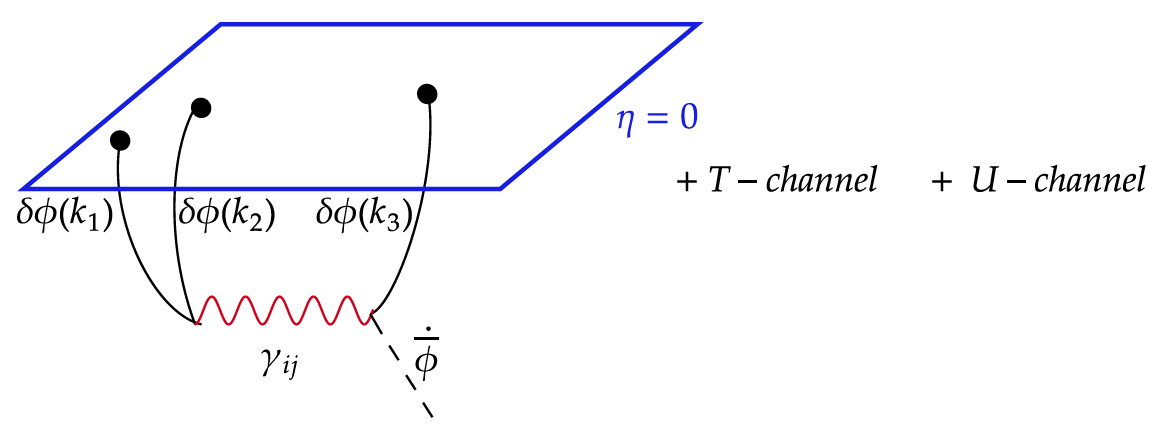}
\caption{Feynman-Witten diagram for $\langle \delta\phi\delta\phi\delta\phi \rangle$.}\label{3pointfig}
\end{center}
\end{figure}

We evaluate the partition function in $EAdS_4$ by computing the on-shell action eq.\eqref{eq:expact}. We look at the Feynman-Witten diagram involving a graviton exchange between four scalar modes with one of the scalar modes set to the time derivative of the background, see figure-\ref{3pointfig}.
We will follow \S\,5 of \cite{Shukla:2016bnu}, but keeping in mind that we are looking at $\alpha$-vacuum and eq.\eqref{eq:supsoleads} and eq.\eqref{eq:gravalphatranprop} should be used in appropriate places. 

In \S\,\ref{ssec:4ptfullans}, while computing the on-shell action relevant for the four-point scalar correlator, we have noticed that the Feynman-Witten diagrams with graviton exchange (summing up the $s,\, t, \, u$ channels) lead us to the total contribution, see eq.\eqref{eq:sAdS_onshell},
\begin{equation}\label{eq:onshellads3pt}
	S^{EAdS}_{on-shell} = \frac{1}{2}\, M_{\text{Pl}}^2 \, \text{R}_{\text{AdS}}^2 \,(\mathcal{W} + 2 \mathcal{R}) \, ,
\end{equation}
where $\mathcal{W}$ is the transverse graviton contribution,
\begin{equation}
	\mathcal{W} = \int dz_1 dz_2\, d^3x_1 d^3x_2 \,  T_{lk}(z_1,\vect{x_1}) \tilde{\mathcal{G}}_{lk,ji}(z_1,\vect{x_1};z_2,\vect{x_2}) T_{ji}(z_2,\vect{x_2}) \, ,
\end{equation}
and the contribution from the longitudinal part of the exchanged graviton is denoted by the ``remainder'' term $\mathcal{R}$. This $\mathcal{R}$ can further be written with the following form
\be
\label{remainder}
\mathcal{R} = \mathcal{R}_1 + \mathcal{R}_2 + \mathcal{R}_3 ,
\ee
such that $\mathcal{R}_1, \mathcal{R}_2, \mathcal{R}_3$ are given by
\begin{equation}
	\begin{split}
		\label{r123pos}
		&\mathcal{R}_1 = -\int \frac{dz}{z^2} \, d^3x\,  T_{zj}(z,\vect{x}) \, \frac{1}{\partial^{\,2}} \, T_{zj}(z,\vect{x})\, ,\\
		&\mathcal{R}_2 = - \frac{1}{2} \int \frac{dz}{z} \, d^3x\,  \partial_j T_{zj}(z,\vect{x}) \, \frac{1}{\partial^{\,2}} \, T_{zz}(z,\vect{x})\, ,\\
		&\mathcal{R}_3 = - \frac{1}{4} \int \frac{dz}{z^2} \, d^3x\,  \partial_j T_{zj}(z,\vect{x}) \, \bigg(\frac{1}{\partial^{\,2}}\bigg)^2 \, \partial_i T_{zi}(z,\vect{x})\, .
	\end{split}
\end{equation}

Next, we should identify where to substitute one $\delta\phi$ with $\dot{\bar{\phi}}/H$ in the external legs. Let us start with a term involving the stress-tensor components: $T_{zj}$ and $T_{zz}$. For the four-point function calculation in \S\,\ref{ssec:4ptfullans}, we had the following structure
\be
\label{tzj2dp}
\begin{split}
T_{zj}(z,&\vect{x}) = \partial_z\delta\phi \, \partial_j \delta\phi\, , \quad T_{zz}(z,\vect{x}) = \half \Big( (\partial_z\delta\phi)^2 - (\partial_i \delta\phi)^2 \Big)\, .
\end{split}
\ee
For the three-point function calculation, we need to modify eq.\eqref{tzj2dp}. When computing the $\mathcal{R}$ terms, out of the two factors of $T_{\mu\nu}$ in each of the remainder terms in eq.\eqref{r123pos}, one must contribute one factor of $\bar\phi$ and one factor of $\delta\phi$. With this in mind, we obtain 
\be
\label{tzj1dp}
T_{zj}(z,\vect{x}) = \partial_z\bar\phi \, \partial_j \delta\phi. 
\ee
By substituting eq.\eqref{eq:supsoleads} for $\delta\phi$ in eq.\eqref{tzj1dp}, and moving to momentum space, we obtain
\be
\label{tzj1dpm}
T_{zj}(z,\vect{k}) = i \,(2\pi)^3 \delta^3(\vect{p}-\vect{k}) \,\phi_{0}(\vect{p})\,\partial_z\bar\phi \, p_j \big[c_{1}(1+pz)\text{e}^{-pz}+c_{2}(1-pz)\text{e}^{pz}\big],
\ee
where $\vect{p}$ is the momentum carried by the external leg $\delta\phi$. Similarly, for $T_{zz}$ one should accordingly modify as the following \footnote{We have incorporated a potential $V(\phi)$ for the scalar field in the action eq.\eqref{eq:eadsaction}. We use the notation $V'(\bar\phi) \equiv {dV(\bar\phi)}/{d\bar\phi}$. Also, note that we have an additional factor of $\text{R}_{\text{AdS}}^2$ multiplying the $V'(\bar\phi)$ term in eq. \eqref{tzz1dp} as compared to eq.\eqref{eq:defint}. It is because in writing the $EAdS_4$ on-shell action eq.\eqref{eq:eadsaction}, we extracted out an overall factor of $\text{R}_{\text{AdS}}^2$.}
\be
\label{tzz1dp}
T_{zz}(z,\vect{x}) = \partial_z\bar\phi \, \partial_z \delta\phi - \frac{\text{R}_{\text{AdS}}^2}{z^2} \, V'(\bar\phi)\, \delta\phi \, ,
\ee
which upon using eq.\eqref{eq:supsoleads}, leads us to
\be
\label{tzz1dpm}
\begin{split}
	T_{zz}(z,\vect{k}) = - (2\pi)^3 \delta^3(\vect{p}-\vect{k}) \bigg[&p^2 z \partial_z\bar\phi[c_{1}(1+pz)\text{e}^{-pz}+c_{2}(1-pz)\text{e}^{pz}] \\& ~~ + \frac{\text{R}_{\text{AdS}}^2}{z^2} \, V'(\bar\phi) [c_{1}\text{e}^{-pz}+c_{2}\text{e}^{pz}]\bigg] \phi_{0}(\vect{p}) \, .
\end{split}
\ee

One can now compute $\mathcal{R}$ in eq.\eqref{remainder}, which can be expressed as 
\be
\begin{split}\label{eq:rs3ptfinal}
	&\mathcal{R}(k_{1},k_{2},k_{3})=(c_1^3-c_2^3)\mathcal{R}_{BD}\\
	&+(c_1^2c_2-c_1c_2^2)[\mathcal{R}_{BD}(-k_{1},k_{2},k_{3})+\mathcal{R}_{BD}(k_{1},-k_{2},k_{3})+\mathcal{R}_{BD}(k_{1},k_{2},-k_{3})] \, ,
\end{split}
\ee
where 
\begin{equation}
	\begin{split}
		\label{rpfin}
		\mathcal{R}_{BD}(k_{1},k_{2},k_{3}) = \half \, \frac{\dot{\bar\phi}}{H}& \, (2\pi)^3\delta^3\Big(\sum_{a=1}^3 \vect{k_a}\Big) \Big( \prod_{a=1}^3 \phi_0(\vect{k_a})\Big) \times \\
		&\Bigg[- \half \sum_{a=1}^3 k_a^{\,3} + \half \sum_{a \neq b} k_a k_b^{\,2} + \frac{4}{K}\sum_{a<b} k_a^{\,2} k_b^{\,2}  \Bigg].
	\end{split}
\end{equation}
Note that $\mathcal{R}$, in eq.\eqref{eq:rs3ptfinal}, is entirely determined by the answer in Bunch-Davies vacua, i.e., $\mathcal{R}_{BD}$, through the permutation of the signs of the modulus value of the external momenta precisely the same way as predicted by \cite{Jain:2022uja}. 

Next, we want to calculate the contribution to the $EAdS_4$ on-shell action coming from the transverse part of the exchanged graviton. This contribution is denoted by $\mathcal{W}$ defined in eq.\eqref{eq:onshellads3pt}. Out of the two $T_{lk}$ terms in $\mathcal{W}$, one must contribute a piece proportional to $\delta\phi \delta\phi$, and the other must contribute a piece proportional to $\partial_z\bar\phi \, \delta\phi$. So, we get
\be
\label{tijd}
T_{lk}(z,\vect{x}) = -\bigg( \partial_z\bar\phi \, \partial_z\delta\phi + \frac{\text{R}_{\text{AdS}}^2}{z^2} \, V'(\bar\phi) \, \delta\phi\bigg) \delta_{lk}\, .
\ee
The key point to note in eq.\eqref{tijd} is that $T_{lk}$ is proportional to $\delta_{lk}$. Thus, while calculating $\mathcal{W}$, the contraction of $T_{lk}$ with the transverse graviton propagator $\tilde{\mathcal{G}}_{lk,ji}$ gives the trace $\tilde{\mathcal{G}}_{ll,ji}$, which vanishes. Therefore, there is no contribution from the transverse graviton exchange to the $EAdS_4$ on-shell action, i.e.
\be
\label{wv}
\mathcal{W} = 0\, .
\ee
Now, the $EAdS_4$ on-shell action is obtained to be
\begin{equation}
	\begin{split}
		&S^{EAdS}_{on-shell} = \frac{1}{4}\, M_{\text{Pl}}^2 \, \text{R}_{\text{AdS}}^2 \frac{\dot{\bar\phi}}{H} \, (2\pi)^3\delta^3\Big(\sum_{a=1}^3\vect{k_{a}}\Big) \Big( \prod_{a=1}^3 \phi_0(\vect{k_{a}})\Big) \times \\~~&  \Bigg[(c_{1}^3-c_{2}^3)\Bigg\{- \sum_{a=1}^3 k_a^{\,3} + \sum_{a\neq b}k_a k_b^{\,2} + \frac{8}{k_{1}+k_{2}+k_{3}} \sum_{a<b}k_a^{\,2} k_b^{\,2}\Bigg\} \\
		& + (c_{1}^2c_2-c_1c_2^2)\Bigg\{ - \sum_{a=1}^3 k_a^{\,3} + \sum_{a\neq b}k_a k_b^{\,2} \\
		&~~~+ 8 \bigg(\sum_{a<b}k_a^{\,2} k_b^{\,2}\bigg) \bigg(\frac{1}{k_1+k_2-k_3} + \frac{1}{k_1-k_2+k_3} + \frac{1}{-k_1+k_2+k_3}\bigg) \Bigg\}\Bigg] \, .
	\end{split}
\end{equation}
Next, by taking derivatives of the on-shell action  with respect to the boundary value of the field $\delta\phi$, i.e., $\phi_{0}(c_1+c_2)$, we obtain the three-point coefficient function to be
\begin{equation}
	\begin{split}
		\label{dsoc}
		&\langle O(\vect{k_1}) O(\vect{k_2}) O(\vect{k_3}) \rangle = -\frac{1}{4}\,  \frac{\dot{\bar\phi}}{H} \, (2\pi)^3\delta^3\Big(\sum_{a=1}^3\vect{k_{a}}\Big) \times \\
		&\Bigg[\frac{(c_{1}^3-c_{2}^3)}{(c_1+c_2)^3}\Bigg\{- \sum_{a=1}^3 k_a^{\,3} + \sum_{a\neq b}k_a k_b^{\,2} + \frac{8}{k_{1}+k_{2}+k_{3}} \sum_{a<b}k_a^{\,2} k_b^{\,2}\Bigg\}  \\
		&+\frac{(c_{1}^2c_2-c_1c_2^2)}{(c_1+c_2)^3}\Bigg\{ - \sum_{a=1}^3 k_a^{\,3} + \sum_{a\neq b}k_a k_b^{\,2} \\
		&~~+ 8 \bigg(\sum_{a<b}k_a^{\,2} k_b^{\,2}\bigg) 
		\bigg(\frac{1}{k_1+k_2-k_3} + \frac{1}{k_1-k_2+k_3} + \frac{1}{-k_1+k_2+k_3}\bigg) \Bigg\}\Bigg] \, .
	\end{split}
\end{equation}
From eq.\eqref{wf_schematic}, we can express
\be
\label{dporel}
\langle \delta\phi(\vect{k_1})  \delta\phi(\vect{k_2})  \delta\phi(\vect{k_3}) \rangle = \frac{1}{4} \, \frac{H^4}{M_{\text{Pl}}^4} \, \frac{Re \,\langle O(\vect{k_1}) O(\vect{k_2}) O(\vect{k_3}) \rangle}{\prod_{a=1}^3 Re \, \langle O(\vect{k_a}) O(-\vect{k_a}) \rangle'}\, ,
\ee
where one has a similar reality condition as eq.\eqref{eq:scal_corr_CF}. We substitute the result eq.\eqref{dsoc} in eq.\eqref{dporel}, and use \footnote{This follows from 
\begin{equation*}
	\langle \delta\phi(\vect{k_1})\delta\phi(\vect{k_2})\rangle = (2\pi)^3 \delta^3\Big(\sum_{a=1}^2 \vect{k_a}\Big) \frac{|c_1+c_2|^2 }{2k_1^3}=(2\pi)^3 \delta^3\Big(\sum_{a=1}^2 \vect{k_a}\Big) \frac{\cosh(2\alpha) }{2k_1^3} \, .
\end{equation*}}
\be \label{eq:twopointcoeff}
\langle O(\vect{k_1})O(\vect{k_2})\rangle = (2\pi)^3 \delta^3\Big(\sum_{a=1}^2 \vect{k_a}\Big) \frac{k_1^{\,3}}{|c_1+c_2|^2 } =(2\pi)^3 \delta^3\Big(\sum_{a=1}^2 \vect{k_a}\Big) \frac{k_1^{\,3}}{\cosh(2\alpha)} \, ,
\ee 
to finally get 
\begin{equation}
	\begin{split}
		\label{dp3}
		&\langle \delta\phi(\vect{k_1})  \delta\phi(\vect{k_2})  \delta\phi(\vect{k_3}) \rangle \\
		&= -\frac{1}{8} \frac{H^4}{M_{\text{Pl}}^4}  \frac{\dot{\bar\phi}}{H}(2\pi)^3\delta^3\Big(\sum_{a=1}^3\vect{k_{a}}\Big) \Big( \prod_{a=1}^3 \frac{1}{k_a^3}\Big)|c_{1}+c_{2}|^6 \\ &~~ \times \Bigg[Re\frac{(c_{1}^3-c_{2}^3)}{(c_1+c_2)^3}\Bigg\{- \sum_{a=1}^3 k_a^{\,3} + \sum_{a\neq b}k_a k_b^{\,2} + \frac{8}{k_{1}+k_{2}+k_{3}} \sum_{a<b}k_a^{\,2} k_b^{\,2}\Bigg\} \\
		&  +Re\frac{(c_{1}^2c_2-c_1c_2^2)}{(c_1+c_2)^3}\Bigg\{ - \sum_{a=1}^3 k_a^{\,3} + \sum_{a\neq b}k_a k_b^{\,2}
		\\&+ 8 \bigg(\sum_{a<b}k_a^{\,2} k_b^{\,2}\bigg)\bigg(\frac{1}{k_1+k_2-k_3} + \frac{1}{k_1-k_2+k_3} + \frac{1}{-k_1+k_2+k_3}\bigg) \Bigg\}\Bigg] \, .
	\end{split}
\end{equation}
Now, we want to find the correlation functions of the metric perturbations $\zeta$ by a change of gauge. A gauge relating $\zeta$ and $\delta\phi$, which is up to second order given by see \cite{Ghosh:2014kba}
\be
\label{change}
\zeta = - \frac{H}{\dot{\bar\phi}} \delta\phi + \half \bigg( \half + \frac{\ddot{\bar\phi} H}{\dot{\bar\phi}^3}\bigg) \delta\phi^2 .
\ee
One can invert this relation eq.\eqref{change} and express $\delta\phi$ in terms of $\zeta$. Using that, we can compute the final result for the three-point function for scalar fluctuations to be 
\begin{equation}
	\begin{split}
		\label{ptall2}
		&{}_{\mathrm{I}}\langle \alpha | \zeta(\vect{k_1}) \zeta(\vect{k_2}) \zeta(\vect{k_3})|\alpha\rangle_{\mathrm{I}} = (2\pi)^3 \delta^3\Big(\sum_{a=1}^3 \vect{k_a}\Big) \Big[\prod_{a=1}^3 \frac{1}{(2k_a^{\,3})}\Big]|c_{1}+c_{2}|^6 \frac{H^4}{4\epsilon} \times\\
		&\hspace{2mm}\Bigg[Re\frac{(c_{1}^3-c_{2}^3)}{(c_1+c_2)^3}\Bigg\{- \sum_{a=1}^3 k_a^{\,3} + \sum_{a\neq b}k_a k_b^{\,2} + \frac{8}{k_{1}+k_{2}+k_{3}} \sum_{a<b}k_a^{\,2} k_b^{\,2}\Bigg\} \\& ~~~ + \frac{2(\epsilon + \delta)}{\epsilon} \, |c_{1}+c_{2}|^{-2} \Big(\sum_{a=1}^3 k_a^{\,3} \Big)
		+ Re\frac{(c_{1}^2c_2-c_1c_2^2)}{(c_1+c_2)^3}\Bigg\{ - \sum_{a=1}^3 k_a^{\,3} + \sum_{a\neq b}k_a k_b^{\,2}
		\\&+ 8 \bigg(\sum_{a<b}k_a^{\,2} k_b^{\,2}\bigg) \bigg(\frac{1}{k_1+k_2-k_3} + \frac{1}{k_1-k_2+k_3} + \frac{1}{-k_1+k_2+k_3}\bigg) \Bigg\}\Bigg],
	\end{split}
\end{equation}
Setting $c_1=\cosh(\alpha),c_2=-i\sinh(\alpha)$ from eq.\eqref{eq:c1c2def}, we can calculate the coefficients to be
\begin{equation}
	|c_1+c_2|^2=\cosh(2\alpha)\, , \quad 
	Re \frac{c_1^3-c_2^3}{(c_1+c_2)^3}=\frac{1}{\cosh^3(2\alpha)}\, , \quad 
	Re \frac{c_1^2c_2-c_2^2c_1}{(c_1+c_2)^3}=\frac{\sinh^2(2\alpha)}{\cosh^3(2\alpha)}\, .
\end{equation}
Thus expressed in terms of $\alpha$, we get
\begin{equation}
	\begin{split} \label{eq:threepointalpha}
		&{}_{\mathrm{I}}\langle \alpha | \zeta(\vect{k_1}) \zeta(\vect{k_2}) \zeta(\vect{k_3})|\alpha\rangle_{\mathrm{I}} = (2\pi)^3 \delta^3\Big(\sum_{a=1}^3 \vect{k_a}\Big) \Big[\prod_{a=1}^3 \frac{1}{(2k_a^{\,3})}\Big] \frac{H^4}{4\epsilon} \times\\
		&\hspace{2mm}\Bigg[\Bigg\{- \sum_{a=1}^3 k_a^{\,3} + \sum_{a\neq b}k_a k_b^{\,2} + \frac{8}{k_{1}+k_{2}+k_{3}} \sum_{a<b}k_a^{\,2} k_b^{\,2}\Bigg\} \\
		& + \frac{2(\epsilon + \delta)}{\epsilon} \, \cosh^2(2\alpha)\Big(\sum_{a=1}^3 k_a^{\,3} \Big) + \sinh^2(2\alpha)\Bigg\{ - \sum_{a=1}^3 k_a^{\,3} + \sum_{a\neq b}k_a k_b^{\,2}
		\\&+ 8 \bigg(\sum_{a<b}k_a^{\,2} k_b^{\,2}\bigg) \bigg(\frac{1}{k_1+k_2-k_3} + \frac{1}{k_1-k_2+k_3} + \frac{1}{-k_1+k_2+k_3}\bigg) \Bigg\}\Bigg] \, ,
	\end{split}
\end{equation}
This calculation from wave-function formalism matches the result obtained from the in-in calculation done in \cite{Shukla:2016bnu}, see eq.(5.17) of that paper.


\section{Discussions} \label{sec:disco}

This section will summarize our main results and discuss and comment on their implications and future directions. We start by mentioning our results first. 
\begin{itemize}
\item We have computed the inflationary scalar four-point function in $\alpha$-vacuum using the in-in formalism to the leading order in slow-roll expansion. The expression for $\langle \zeta\zeta\zeta\zeta \rangle_\alpha$ is written in eq.\eqref{eq:zeta_phi_rel}, eq.\eqref{eq:scalarfourpfinal_inin} (sum of eq.\eqref{eq:alpha_CI} and eq.\eqref{eq:ininfourptfinal}).  
\item After working out $\langle \zeta\zeta\zeta\zeta \rangle_\alpha$, and using $\langle \zeta\zeta\zeta \rangle_\alpha$ from eq.(5.17) in \cite{Shukla:2016bnu} we checked that the Ward identities of conformal transformations (eq.\eqref{eq:scal_ward_3} for the scaling, and eq.\eqref{eq:sct_ward_3} for the special conformal transformation) are satisfied between them. This is one of the main results of our paper and is very intriguing, given that the Ward identities between $\langle \zeta\zeta \rangle_\alpha$ and $\langle \zeta\zeta\zeta \rangle_\alpha$ do not hold, as reported in \cite{Shukla:2016bnu}.
 
\item There are two distinctive parts in $\langle \zeta\zeta\zeta\zeta \rangle_\alpha$ eq.\eqref{eq:scalarfourpfinal_inin}. The contact interaction part $\langle \zeta\zeta\zeta\zeta \rangle^{\text{CI}}_\alpha$ is completely fixed by $\langle \zeta\zeta\zeta\zeta \rangle^{\text{CI}}_{BD}$ (i.e. $\langle \zeta\zeta\zeta\zeta \rangle^{\text{CI}}_{\alpha=0}$) as shown in eq.\eqref{eq:finalformfactorci}. This is consistent with the results derived in \cite{Jain:2022uja}, where the general solution of conformal Ward identities for three-point CFT correlators was studied in momentum space. Only the $\langle \zeta\zeta\zeta\zeta \rangle^{\text{CI}}_\alpha$ part of the full $\langle \zeta\zeta\zeta\zeta \rangle_\alpha$ contributes to the RHS in the Ward identities, eq.\eqref{eq:scal_ward_3} and eq.\eqref{eq:sct_ward_3}. Hence, $\langle \zeta\zeta\zeta\zeta \rangle^{\text{CI}}_\alpha$ gets fixed by the conformal Ward identities.

\item The other contribution to $\langle \zeta\zeta\zeta\zeta \rangle_\alpha$ comes from a graviton exchange involving two scalar-scalar-graviton three-point vertices, denoted as $\langle \zeta\zeta\zeta\zeta \rangle^{\text{GE}}_\alpha$. This is also related to the corresponding part of the Bunch-Davies result $\langle \zeta\zeta\zeta\zeta \rangle^{\text{GE}}_{BD}$, see eq.\eqref{eq:ininfourptfinal}. However, this relation is not the same as eq.\eqref{eq:finalformfactorci}, related to the fact that $\langle \zeta\zeta\zeta\zeta \rangle^{\text{GE}}_\alpha$ does not contribute to the RHS of the Ward identities in eq.\eqref{eq:scal_ward_3} and eq.\eqref{eq:sct_ward_3}.

\item We also generalized the wave-function method of computing inflationary scalar correlators in $\alpha$-vacua. We adopted the formalism used in \cite{Ghosh:2014kba} to evaluate the Feynman-Witten diagram with modifications needed for modes in $\alpha$-vacua. A modified scalar mode expansion, see eq.\eqref{eq:supsoleads}, was required to compute the on-shell action. Furthermore, the graviton bulk-to-bulk propagator was also changed as in eq.\eqref{eq:gravalphatranprop} to make it consistent with the different boundary conditions implied by the use of $\alpha$-vacua. 

\item The resulting expressions for $\langle \zeta\zeta\zeta\zeta \rangle_\alpha$ in the wave-function method are given in eq.\eqref{eq:zzzz_wf} and eq.\eqref{eq:sca_corr_full_wf}, precisely matching with the in-in result, written in eq.\eqref{eq:scalarfourpfinal_inin} in \S\,\ref{sec:4pininalpha}. This matching between the two methods works similarly to the Bunch-Davies case considered in \cite{Ghosh:2014kba}. 

The transverse graviton part denoted by ${\widetilde W}^S_{\alpha}$ (see eq.\eqref{eq:wnew} and eq.\eqref{eq:scal_corr_CF}) in $\langle \zeta\zeta\zeta\zeta \rangle^{\text{CF}}_\alpha$ together with the $\langle \zeta\zeta\zeta\zeta \rangle^{\text{ET}}_\alpha$ contribution (see eq.\eqref{eq:scal_corr_ET}), matches with $\langle \zeta\zeta\zeta\zeta \rangle^{\text{GE}}_\alpha$ piece in the in-in result. The longitudinal graviton contribution denoted by $R^S_{\alpha}$ in the wave-function result (see eq.\eqref{eq:scal_corr_CF} and eq.\eqref{eq:rnew}) matches with the $\langle \zeta\zeta\zeta\zeta \rangle^{\text{CI}}_\alpha$ part in the in-in result. 

\item We also computed $\langle \zeta\zeta\zeta \rangle_\alpha$ by computing the same Feynman-Witten diagram used for the four-point function but with one external leg substituted by the time derivative of the background inflaton. Our result in eq.\eqref{eq:threepointalpha} matches $\langle \zeta\zeta\zeta \rangle_\alpha$ computed using the in-in formalism. This matching of $\langle \zeta\zeta\zeta \rangle_\alpha$ between the computation using both the methods is an independent justification of the conformal Ward identities being satisfied between $\langle \zeta\zeta\zeta \rangle_\alpha$ and $\langle \zeta\zeta\zeta\zeta \rangle_\alpha$, without even knowing $\langle \zeta\zeta\zeta\zeta \rangle_\alpha$, see the discussions in \S\,\ref{sec:3ptwfalpha}. 
\end{itemize}

Following the summary of our results, let us now comment on certain subtleties we encountered in the technical part of our calculations. This has to do with the implementation of the analytic continuation $\eta \to \eta(1 \pm i \epsilon)$ at the far past $\eta \to - \infty$. This is necessary for the calculation of correlators via the in-in method (e.g., to perform integrations as in eq.\eqref{eq:genalphaint}) to ensure that the correct boundary conditions for the $\alpha$-vacua state are imposed. To project onto the correct ground state and obtain convergent $\eta$-integrals, one has to choose an appropriate sign for the $i\epsilon$ prescription depending upon the sign of the exponent within the integrand, as was also previously observed in \cite{Shukla:2016bnu}. We leave the possibility of looking for a better prescription to regulate these integrals for future work. 
 
We encountered a similar situation in the wave-function method while computing the on-shell action in $EAdS_4$. As discussed in \S\,\ref{sec:analyticcont}, we had to impose a non-regular boundary condition for $\delta\phi$ at $z \to \infty$ in the deep interior of $EAdS_4$ - due to the growing mode $e^{kz}$ as $z \to \infty$ in eq.\eqref{eq:solscalads}. This is not natural in standard calculations in $AdS/CFT$ literature. But in the context relevant to this paper, the growing mode at $z \to \infty$ in $EAdS_4$ is inherited via analytic condition from $dS_4$ and bears the tell-tale signature of the fact that $\alpha$-vacua is an excited state at $\eta \to - \infty$.  

It is also instructive to note that in our calculation of the four-point scalar correlator, there was an implicit assumption to consider both the scalar and tensor perturbations with the same choice of $\alpha$-vacuum modes. The same set of coefficients $c_1, \, c_2$ has been used in eq.\eqref{eq:modefunctions} and eq.\eqref{eq:gammamode}. This choice has simplified the final expression of $\langle \zeta \zeta \zeta \zeta \rangle$. However, it is important to note that choosing a different alpha vacuum for the graviton mode (i.e., selecting different values for $c_1$ and $c_2$ in eq.\eqref{eq:gammamode}) will not affect our result that the conformal Ward identities described in eq.\eqref{eq:scal_ward_3} and eq.\eqref{eq:sct_ward_3} of \S\ref{secWI:check} will still hold true. The reason for this is as follows. A different choice of $\alpha$-vacuum for the graviton mode will not change the ``CI" part of $\langle \delta\phi \delta\phi\delta\phi\delta\phi \rangle$ in eq.\eqref{eq:scalarfourpfinal_inin}, since this entirely arises from contact interactions involving only the scalar mode. The ``GE" part in eq.\eqref{eq:scalarfourpfinal_inin} will change resulting in more complicated coefficients in eq.\eqref{eq:ininfourptfinal}. However, the ``GE" will continue vanishing independently in eq.\eqref{eq:scal_ward_3} and eq.\eqref{eq:sct_ward_3} and thus they will not contribute to the Ward identities. This is due to the fact that the differential operators in the RHS of the Ward identities are at most linear in the derivatives, and the ``GE" term will continue being some linear combination of the pure Bunch-Davies answers. 

Finally, let us highlight certain implications of our results and mention possible future directions. 
\begin{enumerate}
\item \textit{Two versus three-point Ward identity for $\alpha$-vacua correlators:} Our results in this paper deepen the question of whether the inflationary cosmological correlators computed in a conformally invariant initial state, namely, in the $\alpha$-vacuum, are indeed consistent with conformal invariance. That the Ward identities are satisfied by $\langle \zeta\zeta\zeta \rangle_\alpha$ and $\langle \zeta\zeta\zeta\zeta \rangle_\alpha$ is in stark contrast with the findings in \cite{Shukla:2016bnu}, where the Ward identities were shown to be invalid between $\langle \zeta\zeta \rangle_\alpha$ and $\langle \zeta\zeta\zeta \rangle_\alpha$. Although we have not been able to resolve this, we believe our results should be convincing enough to revisit the case of \cite{Shukla:2016bnu}. Since $\alpha$-vacua is conformally invariant, we certainly expect that the conformal invariance should be maintained with correlators in $\alpha$-vacua. 

We note certain subtle points and a few speculative remarks that may be crucial in resolving this anomaly. So far, the cosmological correlators have been computed only for the leading slow-roll order. Hence, the Ward identities are also examined in that leading order. Let us look at the scaling  Ward identity between the two and three-point correlator eq.\eqref{eq:scal_ward1} and try to understand how different powers of $\epsilon$ in $\langle \zeta\zeta \rangle_\alpha$ and $\langle \zeta\zeta\zeta \rangle_\alpha$ are contributing in it. The LHS of eq.\eqref{eq:scal_ward1} gives us the change in $\langle \zeta\zeta \rangle_\alpha$ under scaling which is of order unity, $\mathcal{O}(\epsilon^{0})$. One can check that both $\langle \zeta\zeta \rangle_\alpha$ and $\langle \zeta\zeta\zeta \rangle_\alpha$ are of order $\mathcal{O}(\epsilon^{-1})$, making the RHS of eq.\eqref{eq:scal_ward1} $\sim \mathcal{O}(\epsilon^{0})$. This is also clear from eq.\eqref{maldcon}, with $n_s \sim \mathcal{O}(\epsilon)$. However, as was shown in \cite{Shukla:2016bnu}, see eq.(5.21) there, in the limit $\vect{k_3} \to 0$,  
$\langle \zeta\zeta\zeta \rangle_\alpha \sim \epsilon^{-1} k_3^{-1} \sinh 2\alpha$ blows up, thus failing the Ward identity. 

On the other hand, in this paper, we examined the Ward identity (let's focus on the scaling one) between $3$-point and $4$-point correlator, eq.\eqref{eq:scal_ward2}. It is crucial that $\langle \zeta\zeta\zeta \zeta\rangle_\alpha$ goes as $\mathcal{O}(\epsilon^{-2})$ to the leading order, but does not blow up as $\vect{k_4} \to 0$. Therefore, both sides of eq.\eqref{eq:scal_ward2} cancels each other at $\mathcal{O}(\epsilon^{-1})$. Now, coming back to the case of $\langle \zeta\zeta \rangle_\alpha$ and $\langle \zeta\zeta\zeta \rangle_\alpha$, it might be possible that there are other effects due to $\alpha$-vacua that enhances the breaking of scaling symmetry in the $2$-point $\langle \zeta\zeta \rangle_\alpha$. When considered on the LHS of eq.\eqref{eq:scal_ward1}, this may counter-balance the RHS of the same. 


Consideration of loop effects may offer an alternative explanation for this puzzle. It is plausible that modes in the $\alpha$-vacuum decaying in the loop could result in the dynamical renormalization of $\alpha$ to zero at later times \footnote{We thank the anonymous referee for suggesting this to us.}. It would be interesting to investigate if such a phenomenon can produce an effect at the appropriate order in the slow-roll parameter mentioned above for both sides of equation \eqref{eq:scal_ward1}. However, we have not yet found a solution to this puzzle, and we leave this for future exploration.

\item \textit{Check of Ward identities with other non-Gaussianities:} To the question of checking conformal Ward identities, we have found results in affirmative but contradictory to the negative result in \cite{Shukla:2016bnu}. To take steps towards settling this issue, It will be essential to perform further checks of this principle in other examples. Since any $n$-point correlator has Ward identities relating it to $(n+1)$-point correlator, we can perform this checking with other inflationary correlators computed in $\alpha$-vacuum, even involving the graviton fluctuations. Recently, $\alpha$-vacua correlators signifying non-Gaussianities involving primordial gravitational waves in the CMB spectrum has been computed, see for example in \cite{Kundu:2013gha, Akama:2020jko, Kanno:2022mkx, Gong:2023kpe, Akama:2023jsb}. It will be interesting to check if they also satisfy appropriate Ward identities.

\item \textit{Comments on possible observational significance:} Let us comment on possible observational connections of our results. The three-point correlator (also known as the bispectrum) and the four-point correlator (also known as the trispectrum) are reflected in the measurement of the non-Gaussianities in the CMB spectrum, denoted by the parameters $f_{NL}$ for the three-point, and $\tau_{NL}$ and $g_{NL}$ for the four-point correlator. For situations where these parameters are independent of momenta (also known as the local form \cite{Seery:2008ax}), one gets
\begin{equation} 
\begin{split}
\zeta &= \zeta_g - \frac{3}{5} \,  f^{local}_{NL}  \, (\zeta_g^2 - \langle\zeta_g^2\rangle)\, ,\\ 
\zeta &= \zeta_g + \frac{1}{2} \,  (\tau^{local}_{NL})^{1/2} \,  (\zeta_g^2 - \langle\zeta_g^2\rangle)+\frac{9}{25} g^{local}_{NL} \, \zeta_g^3 \, ,
\end{split}
\end{equation}
with $\zeta_g$ being the Gaussian field. In principle, we should expect to see signatures of $\alpha$-vacuum in these observational parameters. It has been noted in the literature that the three-point correlator in $\alpha$ vacua can enhance the $f_{NL}$ parameter by a factor $\sim e^{2\alpha}$, see \cite{ Kanno:2022mkx, Gong:2023kpe, Akama:2023jsb}. To be precise, we know that in the Bunch-Davies vacuum, it is small since $f_{NL} \sim \mathcal{O}(\epsilon)$. In $\alpha$-vacuum it is still $\sim \mathcal{O}(\epsilon)$, but obtains an enhancement on top of it proportional to $e^{2\alpha}$. 

For the scalar four-point function $\langle \zeta\zeta \zeta\zeta \rangle_{\alpha=0}$ in the Bunch-Davies vacuum, possible observational connections have been estimated in \cite{Seery:2008ax}. Our final expression for $\langle \zeta\zeta \zeta\zeta \rangle_{\alpha}$ is pretty complicated, and seemingly no immediate connection can be established with a local form of the non-Gassianity parameter $\tau^{local}_{NL}$. However, following similar arguments in \cite{Seery:2008ax}, we can at least get a rough estimate that $\tau_{NL}\sim \epsilon$ even for the trispectrum in $\alpha$-vacuum. Furthermore, we also expect a possibility of an enhancement in $\tau_{NL}$ due to the effect of $\alpha$ parameter, similar as in $f_{NL}$ with $\alpha$.

However, we must admit that these are speculative theoretical predictions now and certainly beyond the scope of immediate experimental verifications, given an inadequate resolution of present-day observations of the non-Gaussianity.

\end{enumerate}

\section*{Acknowledgements}
We would like to thank Ashish Shukla for collaboration in the initial stage, for numerous insightful discussions, and for useful comments on our manuscript. We greatly appreciate stimulating conversations with Sunil Kumar Sake and Sandip Trivedi. We are also thankful to Dionysios Anninos, Arjun Bagchi, Debtosh Chowdhury, Diptarka Das, Shibam Das, Apratim Kaviraj, Carlos Duaso Pueyo, Alok Laddha, Subhendra Mohanty, Sarah Shandera and Nemani Suryanarayana for useful discussions. AA acknowledges the support from a Senior Research Fellowship, granted by the Human Resource Development Group, Council of Scientific and Industrial Research, Government of India. PB would like to acknowledge the warm hospitality of Tata Institute of Fundamental Research Mumbai during the later stages of this work. PD acknowledges the warm hospitality of University of Cambridge, University of Southampton, Durham University, King's College London, University of Edinburgh, IIT Bombay, Institute of Mathematical Sciences Chennai, Chennai Mathematical Institute, and BITS Goa during the course of this work. PD would like to duly acknowledge the Council of Scientific and Industrial Research (CSIR), New Delhi, for financial assistance through the Senior Research Fellowship (SRF) scheme and the partial support from the Royal Society of London international exchange grant with the University of Edinburgh. NK acknowledges support from a MATRICS research grant (MTR/2022/000794) from the Science and Engineering Research Board (SERB), India.
%
\appendix
\section{Details of the in-in calculation in \S\,\ref{sec:4pininalpha}}
\label{ap:I1234_calc}

\subsection*{Details for eq.\eqref{eq:actionfourth_mb}}
Here we write down the detailed expressions for the quantities used in the quartic action written in eq.\eqref{eq:actionfourth_mb}. Derivations of them are outlined in \cite{Seery:2006vu}. Rewriting the quartic action $S_4$ as 
\begin{equation}\label{eq:actionfourth}
	S_4 \simeq \int dt \, d^3x \, \left( - \dfrac{1}{4 a} \beta_{2j} \partial^2 \beta_{2j} + a \vartheta_2 \Sigma + \dfrac{3}{4} a^3 \partial^{-2} \Sigma \partial^{-2}\Sigma - a \delta \dot{\phi} \beta_{2j} \partial_j \delta \phi \right) \, ,
\end{equation}
we note that $\beta_{2j}$, $\vartheta_2$ and $\Sigma$ can be obtained as,
\begin{equation}\label{eq:beta2j}
	\dfrac{1}{2a^2} \beta_{2j} \simeq \partial^{-4} \left( \partial_m \partial_j \delta \dot{\phi} \partial_m \delta \phi + \partial_j \delta \dot{\phi} \partial^2 \delta \phi - \partial^2 \delta \dot{\phi} \partial_j \delta \phi - \partial_m \delta \dot{\phi} \partial_m \partial_j \delta \phi \right) \, ,
\end{equation}
\begin{equation}\label{eq:theta2}
	\begin{split}
		\dfrac{4H}{a^2} &\partial^2 \vartheta_2 = - \dfrac{1}{a^2} \partial_i \delta \phi \partial_i \delta \phi - V''(\phi) \delta \phi^2 +\dfrac{1}{a^4} \partial^2 \vartheta_1 \partial^2 \vartheta_1 - \dfrac{1}{a^4} \partial_i \partial_j \vartheta_1 \partial_i \partial_j \vartheta_1 \\
		&- \delta \dot{\phi} \delta \dot{\phi} + 2 \dfrac{\dot{\phi}}{a^2} \partial_i \vartheta_1 \partial_i \delta \phi + 2 \sigma_1 \left( \dfrac{4H}{a^2} \partial^2 \vartheta_1 + 2 \dot{\phi} \delta \dot{\phi} \right) - (3\sigma^2_1 -2 \sigma_2 )(-6 H^2 + \dot{\phi}^2) \, ,
	\end{split}
\end{equation}
and
\begin{equation}\label{eq:sigma2}
	\sigma_2 = \dfrac{\sigma^2_1}{2}+\dfrac{1}{2H} \partial^{-2} \left( \Sigma + \dfrac{1}{a^2}\partial^2 \sigma_1 \partial^2 \vartheta_1 - \dfrac{1}{a^2} \partial_i \partial_j \sigma_1 \partial_i \partial_j \vartheta_1 \right) \, ,
\end{equation}
with $\Sigma$ defined as,
\begin{equation}
	\Sigma \equiv \partial_j \delta \dot{\phi} \partial_j \delta \phi + \delta\dot{\phi} \partial^2 \delta \phi \, .
\end{equation}
We should remember that various objects here are related to the shift and lapse functions defined in eq.\eqref{NNizero}, as $N = 1+ \sigma , \, N_j = \nabla_j \vartheta + \beta_j$. Also, the variables $\sigma, \vartheta, \beta_j$ are expanded as a power series in the field perturbations $\sigma = \sum_{m=1}^{\infty} \sigma_m  , \, \vartheta = \sum_{m=1}^{\infty} \vartheta_m  , \, \beta_j = \sum_{m=1}^{\infty} \beta_{mj}$, such that the subscript $m$ denotes the number of $\delta \phi$ present.
\subsection*{Details for eq.\eqref{eq:alphaci2}}

eq.\eqref{eq:alphaci2} is derived from eq.\eqref{eq:alpha_CI} after incorporating the sign flips of the magnitude of the momenta as in eq.\eqref{eq:wijci_0}. Thus, the integrals in eq.\eqref{eq:wijci} ultimately get modified and one can sum over all such $\alpha$ terms to get a new form factor given by
\begin{equation}\label{eq:newformfactor}
	\begin{split}
		\cM^{\text{new}}_4&(k_1,k_2,k_3,k_4) \\
		&= \text{Re} \bigg\{ (c^*_1 +c^*_2)^4 \bigg[ c^4_1 \cM_4(k_1,k_2,k_3,k_4) + c^4_2 \cM_4(-k_1,-k_2,-k_3,-k_4) \\
		&+c^3_1 c_2 \bigg(\cM_4(-k_1,k_2,k_3,k_4) + \cM_4(k_1,-k_2,k_3,k_4) \\
		&+  \cM_4(k_1,k_2,-k_3,k_4) + \cM_4(k_1,k_2,k_3,-k_4) \bigg) \\
		& + c_1 c^3_2 \bigg( \cM_4(k_1,-k_2,-k_3,-k_4) + \cM_4(-k_1,k_2,-k_3,-k_4) \\
		& +  \cM_4(-k_1,-k_2,k_3,-k_4) + \cM_4(-k_1,-k_2,-k_3,k_4) \bigg) \\
		&+ 2 c^2_1 c^2_2 \bigg( \cM_4(-k_1,-k_2,k_3,k_4) + \cM_4(-k_1,k_2,-k_3,k_4)  \\
		&+ \cM_4(-k_1,k_2,k_3,-k_4) + \cM_4(k_1,-k_2,-k_3,k_4)  \\
		&+  \cM_4(k_1,-k_2,k_3,-k_4) + \cM_4(k_1,k_2,-k_3,-k_4) \bigg) \bigg] \bigg\} \, ,
	\end{split}
\end{equation}
where the reality condition comes as we need to add the complex conjugate of each integral to itself. We have explicitly checked that,
\begin{equation}
	\begin{split}
		\cM_4(-k_1,k_2,k_3,k_4) &= - \cM_4(k_1,-k_2,-k_3,-k_4) \, , \\
		\cM_4(-k_1,-k_2,k_3,k_4) &= - \cM_4(k_1,k_2,-k_3,-k_4) \, , \\
		\cM_4(-k_1,-k_2,-k_3,-k_4) &= - \cM_4(k_1,k_2,k_3,k_4) \, .
	\end{split}
\end{equation}
The same can be checked for other permutations straightforwardly. Thus, eq.\eqref{eq:newformfactor} can be simplified as eq.\eqref{eq:alphaci2} which leads to eq.\eqref{eq:finalformfactorci}.

\subsection*{Calculation of $\mathcal{I}_{1234}^\alpha$}

In this appendix we layout the details of the calculation leading to $\mathcal{I}_{1234}^\alpha$ given in eq.\eqref{eq:ininfourptfinal}. For the sake of completeness, we are writing the explicit form of  $G_{k_1}^{>\alpha}(\eta_*,\eta)G_{k_2}^{>\alpha} (\eta_*,\eta)$, which is given as,
\begin{equation}\label{eq:GG}
\begin{split}
   &G_{k_1}^{>\alpha}(\eta_*,\eta)G_{k_2}^{>\alpha} (\eta_*,\eta)= U_{k_1}^\alpha(\eta_*)U_{k_1}^{\alpha*}(\eta)U_{k_2}^\alpha(\eta_*)U_{k_2}^{\alpha*}(\eta)\\
   &=[c_1U_{k_1}(\eta_*)+c_2U_{-k_1}(\eta_*)][c_1U_{-k_1}(\eta)-c_2U_{k_1}(\eta)]\\
   &[c_1U_{k_2}(\eta_*)+c_2U_{-k_2}(\eta_*)][c_1U_{-k_2}(\eta)-c_2U_{k_2}(\eta)]\\
   &=c_1^4\Bigg\{U_{k_1}(\eta_*)U_{-k_1}(\eta)U_{k_2}(\eta_*)U_{-k_2}(\eta)\Bigg\}\\
   &+c_1^3c_2\Bigg\{-U_{k_1}(\eta_*)U_{-k_1}(\eta)U_{k_2}(\eta_*)U_{k_2}(\eta)+U_{k_1}(\eta_*)U_{-k_1}(\eta)U_{-k_2}(\eta_*)U_{-k_2}(\eta)\\
   &-U_{k_1}(\eta_*)U_{k_1}(\eta)U_{k_2}(\eta_*)U_{-k_2}(\eta)+U_{-k_1}(\eta_*)U_{-k_1}(\eta)U_{k_2}(\eta_*)U_{-k_2}(\eta)\Bigg\}\\
   &+c_1^2c_2^2\Bigg\{-U_{-k_1}(\eta_*)U_{k_1}(\eta)U_{k_2}(\eta_*)U_{-k_2}(\eta)+U_{-k_1}(\eta_*)U_{-k_1}(\eta)U_{-k_2}(\eta_*)U_{-k_2}(\eta)\\
   &-U_{-k_1}(\eta_*)U_{-k_1}(\eta)U_{k_2}(\eta_*)U_{k_2}(\eta)-U_{k_1}(\eta_*)U_{k_1}(\eta)U_{-k_2}(\eta_*)U_{-k_2}(\eta)\\
   &+U_{k_1}(\eta_*)U_{k_1}(\eta)U_{k_2}(\eta_*)U_{k_2}(\eta)-U_{k_1}(\eta_*)U_{-k_1}(\eta)U_{-k_2}(\eta_*)U_{k_2}(\eta)\Bigg\}\\
   &+c_1c_2^3\Bigg\{-U_{-k_1}(\eta_*)U_{k_1}(\eta)U_{-k_2}(\eta_*)U_{-k_2}(\eta)+U_{-k_1}(\eta_*)U_{k_1}(\eta)U_{k_2}(\eta_*)U_{k_2}(\eta)\\
   &-U_{-k_1}(\eta_*)U_{-k_1}(\eta)U_{-k_2}(\eta_*)U_{k_2}(\eta)+U_{k_1}(\eta_*)U_{k_1}(\eta)U_{-k_2}(\eta_*)U_{k_2}(\eta)\Bigg\}\\
   &+c_2^4\Bigg\{U_{-k_1}(\eta_*)U_{k_1}(\eta)U_{-k_2}(\eta_*)U_{k_2}(\eta)\Bigg\} \, .
   \end{split}
\end{equation}
We can evaluate $F^>_{k_{12}}$ in a similar manner as well.

We now rewrite eq.\eqref{eq:GG} by substituting the values of $c_1$ and $c_2$ using eq.\eqref{eq:c1c2def}, then we get,
\begin{equation}\label{k1k2block}
	\begin{split}
		G_{k_1}^{>\alpha}(\eta_*,\eta)&G_{k_2}^{>\alpha} (\eta_*,\eta)=\cosh^4{\alpha}\Bigg(\sum^{k_1,k_2}_{0}\Bigg)-i\cosh^3{\alpha}\sinh{\alpha}\Bigg(\sum^{k_1,k_2}_{1}\Bigg) \\
		&-\cosh^2{\alpha}\sinh^2{\alpha}\Bigg(\sum^{k_1,k_2}_{2}\Bigg)+i\cosh{\alpha}\sinh^3{\alpha}\Bigg(\sum^{k_1,k_2}_{3}\Bigg)+\sinh^4{\alpha}\Bigg(\sum^{k_1,k_2}_{4}\Bigg) \, ,
	\end{split}
\end{equation}
where we have defined,
\begin{equation}\label{eq:sumdef}
	\begin{split}
		\sum^{k_1,k_2}_{0}&\equiv U_{k_1}(\eta_*)U_{-k_1}(\eta)U_{k_2}(\eta_*)U_{-k_2}(\eta) \\
		\sum^{k_1,k_2}_{1}&\equiv -U_{k_1}(\eta_*)U_{-k_1}(\eta)U_{k_2}(\eta_*)U_{k_2}(\eta)+U_{k_1}(\eta_*)U_{-k_1}(\eta)U_{-k_2}(\eta_*)U_{-k_2}(\eta) \\
		&-U_{k_1}(\eta_*)U_{k_1}(\eta)U_{k_2}(\eta_*)U_{-k_2}(\eta)+U_{-k_1}(\eta_*)U_{-k_1}(\eta)U_{k_2}(\eta_*)U_{-k_2}(\eta) \\
		\sum^{k_1,k_2}_{2}&\equiv -U_{-k_1}(\eta_*)U_{k_1}(\eta)U_{k_2}(\eta_*)U_{-k_2}(\eta)+U_{-k_1}(\eta_*)U_{-k_1}(\eta)U_{-k_2}(\eta_*)U_{-k_2}(\eta) \\
		&-U_{-k_1}(\eta_*)U_{-k_1}(\eta)U_{k_2}(\eta_*)U_{k_2}(\eta)-U_{k_1}(\eta_*)U_{k_1}(\eta)U_{-k_2}(\eta_*)U_{-k_2}(\eta) \\
		&+U_{k_1}(\eta_*)U_{k_1}(\eta)U_{k_2}(\eta_*)U_{k_2}(\eta)-U_{k_1}(\eta_*)U_{-k_1}(\eta)U_{-k_2}(\eta_*)U_{k_2}(\eta) \\ 
		\sum^{k_1,k_2}_{3}&\equiv -U_{-k_1}(\eta_*)U_{k_1}(\eta)U_{-k_2}(\eta_*)U_{-k_2}(\eta)+U_{-k_1}(\eta_*)U_{k_1}(\eta)U_{k_2}(\eta_*)U_{k_2}(\eta) \\
		&-U_{-k_1}(\eta_*)U_{-k_1}(\eta)U_{-k_2}(\eta_*)U_{k_2}(\eta)+U_{k_1}(\eta_*)U_{k_1}(\eta)U_{-k_2}(\eta_*)U_{k_2}(\eta) \\
		\sum^{k_1,k_2}_{4}&\equiv U_{-k_1}(\eta_*)U_{k_1}(\eta)U_{-k_2}(\eta_*)U_{k_2}(\eta) \, .
	\end{split}
\end{equation}

Similarly we can write,
\begin{equation}\label{grav_cont}
	\begin{split}
		F^>_{k_{12}}&(\eta,\eta')=[c_1\gamma_{k_{12}}(\eta)+c_2\gamma_{-k_{12}}(\eta)][c_1\gamma_{-k_{12}}(\eta')-c_2\gamma_{k_{12}}(\eta')] \\
		&=\cosh^2{\alpha}\Bigg(\sum^{k_{12}}_{0}\Bigg)-i\cosh{\alpha}\sinh{\alpha}\Bigg(\sum^{k_{12}}_{1}\Bigg)-\sinh^2{\alpha} \Bigg(\sum^{k_{12}}_{2}\Bigg) \, ,
	\end{split}
\end{equation}
where the definition of the sum is apparent from the above eq.\eqref{eq:sumdef}. The $k_3,k_4$ block will have the same structure as eq.\eqref{k1k2block}, hence we are not writing it explicitly. Now we have to put all this back in eq.\eqref{eq:ininfourptfirst} and take the imaginary part. For convenience we have divided the full answer into 3 blocks corresponding to the 3 blocks of the graviton contribution as in eq.\eqref{grav_cont}. We have written the 3 blocks explicitly below,

\underline{\textbf{$\cosh^2{\alpha}$ block}:}
\begin{equation}\label{eq:c12_block}
	\begin{split}
		&\Bigg\{\cosh^4{\alpha}\Im\Bigg(\sum^{k_1,k_2}_{0}\Bigg)-\cosh^3{\alpha}\sinh{\alpha}\Re\Bigg(\sum^{k_1,k_2}_{1}\Bigg)-\cosh^2{\alpha}\sinh^2{\alpha}\Im\Bigg(\sum^{k_1,k_2}_{2}\Bigg) \\
		&+\cosh{\alpha}\sinh^3{\alpha}\Re\Bigg(\sum^{k_1,k_2}_{3}\Bigg)+\sinh^4{\alpha}\Im\Bigg(\sum^{k_1,k_2}_{4}\Bigg)\Bigg\} \\
		&\times\Bigg\{\cosh^6{\alpha}\Im\Bigg(\sum^{k_3,k_4}_{0}\sum^{k_{12}}_{0}\Bigg)-\cosh^5{\alpha}\sinh{\alpha}\Re\Bigg(\sum^{k_3,k_4}_{1}\sum^{k_{12}}_{0}\Bigg)-\cosh^4{\alpha}\sinh^2{\alpha}\Im\Bigg(\sum^{k_3,k_4}_{2}\sum^{k_{12}}_{0}\Bigg) \\
		&+\cosh^3{\alpha}\sinh^3{\alpha}\Re\Bigg(\sum^{k_3,k_4}_{3}\sum^{k_{12}}_{0}\Bigg)+\cosh^2{\alpha}\sinh^4{\alpha}\Im\Bigg(\sum^{k_3,k_4}_{4}\sum^{k_{12}}_{0}\Bigg)\Bigg\}
	\end{split}
\end{equation}

\underline{\textbf{$\cosh{\alpha}\sinh{\alpha}$ block}:}
\begin{equation}\label{eq:c1c2_block}
	\begin{split}
		&\Bigg\{\cosh^4{\alpha}\Im\Bigg(\sum^{k_1,k_2}_{0}\Bigg)-\cosh^3{\alpha}\sinh{\alpha}\Re\Bigg(\sum^{k_1,k_2}_{1}\Bigg)-\cosh^2{\alpha}\sinh^2{\alpha}\Im\Bigg(\sum^{k_1,k_2}_{2}\Bigg)  \\
		&+\cosh{\alpha}\sinh^3{\alpha}\Re\Bigg(\sum^{k_1,k_2}_{3}\Bigg)+\sinh^4{\alpha}\Im\Bigg(\sum^{k_1,k_2}_{4}\Bigg)\Bigg\} \\
		&\times\Bigg\{-\cosh^5{\alpha}\sinh{\alpha}\Re\Bigg(\sum^{k_3,k_4}_{0}\sum^{k_{12}}_{1}\Bigg)-\cosh^4{\alpha}\sinh^2{\alpha}\Im\Bigg(\sum^{k_3,k_4}_{1}\sum^{k_{12}}_{1}\Bigg) \\
		&+\cosh^3{\alpha}\sinh^3{\alpha}\Re\Bigg(\sum^{k_3,k_4}_{2}\sum^{k_{12}}_{1}\Bigg)+\cosh^2{\alpha}\sinh^4{\alpha}\Im\Bigg(\sum^{k_3,k_4}_{3}\sum^{k_{12}}_{1}\Bigg)  \\
		&-\cosh{\alpha}\sinh^5{\alpha}\Re\Bigg(\sum^{k_3,k_4}_{4}\sum^{k_{12}}_{1}\Bigg)\Bigg\}
	\end{split}
\end{equation}

\underline{\textbf{$\sinh^2{\alpha}$ block}:}
\begin{equation}\label{eq:c22_block}
	\begin{split}
		&\Bigg\{\cosh^4{\alpha}\Im\Bigg(\sum^{k_1,k_2}_{0}\Bigg)-\cosh^3{\alpha}\sinh{\alpha}\Re\Bigg(\sum^{k_1,k_2}_{1}\Bigg)-\cosh^2{\alpha}\sinh^2{\alpha}\Im\Bigg(\sum^{k_1,k_2}_{2}\Bigg) \\
		&+\cosh{\alpha}\sinh^3{\alpha}\Re\Bigg(\sum^{k_1,k_2}_{3}\Bigg)+\sinh^4{\alpha}\Im\Bigg(\sum^{k_1,k_2}_{4}\Bigg)\Bigg\} \\
		&\times\Bigg\{-\cosh^4{\alpha}\sinh^2{\alpha}\Im\Bigg(\sum^{k_3,k_4}_{0}\sum^{k_{12}}_{2}\Bigg)+\cosh^3{\alpha}\sinh^3{\alpha}\Re\Bigg(\sum^{k_3,k_4}_{1}\sum^{k_{12}}_{2}\Bigg) \\
		&+\cosh^2{\alpha}\sinh^4{\alpha}\Im\Bigg(\sum^{k_3,k_4}_{2}\sum^{k_{12}}_{2}\Bigg)-\cosh{\alpha}\sinh^5{\alpha}\Re\Bigg(\sum^{k_3,k_4}_{3}\sum^{k_{12}}_{2}\Bigg)-\sinh^6{\alpha}\Im\Bigg(\sum^{k_3,k_4}_{4}\sum^{k_{12}}_{2}\Bigg)\Bigg\}
	\end{split}
\end{equation}
Now we evaluate this for each term in eq.\eqref{eq:c12_block}, \eqref{eq:c1c2_block}, and \eqref{eq:c22_block} and add up everything to get the full answer. If we look at the structure of these equations, we see it has 4 types of terms: $\Im{}\Re{}$, $\Re{}\Im{}$, $\Im{}\Im{}$ and $\Re{}\Re{}$. We have checked explicitly in mathematica, that all the terms which have structure like $\Im{}\Re{}$ and $\Re{}\Im{}$ vanish. This is expected as such a contribution will be purely imaginary and thus it adds to its complex conjugate to vanish. So we are only left with contributions which have like $\Im{}\Im{}$ and $\Re{}\Re{}$. We will now look at how to compute them.

Before moving ahead, let us define
\begin{equation}
	\begin{split}
		\mathcal{I}_{1234}^1&=\int_{-\infty}^{\eta_*}\frac{d\eta}{\eta^2}\int_{-\infty}^{\eta}\frac{d\eta'}{\eta'^2}\text{Im}[X]\text{Im}[YZ] \, , \\
		\mathcal{I}_{1234}^2&=\int_{-\infty}^{\eta_*}\frac{d\eta}{\eta^2}\int_{-\infty}^{\eta}\frac{d\eta'}{\eta'^2}\text{Re}[X]\text{Re}[YZ] \, ,
	\end{split}
\end{equation}
which can be calculated to be \cite{Seery:2008ax},
\begin{equation}\label{eq:i1234_iaib}
	\begin{split}
		\mathcal{I}_{1234}^1= -\frac{1}{4}[I_A-I_B-I_C+I_D] \, , \\
		\mathcal{I}_{1234}^2=\frac{1}{4}[I_A+I_B+I_C+I_D] \, ,
	\end{split}
\end{equation}
where
\begin{equation}
	I_A=\int_{-\infty}^{\eta_*}\frac{d\eta}{\eta^2}\int_{-\infty}^{\eta}\frac{d\eta'}{\eta'^2} X Y Z \, ,
\end{equation}
and
\begin{equation}
	\begin{split}
		&I_B=I_A/.\{k_1\rightarrow -k_1,k_2\rightarrow -k_2\} \, ,\nonumber\\
		&I_C=I_A/.\{k_3\rightarrow -k_3,k_4\rightarrow -k_4,k_{12}\rightarrow -k_{12}\} \, ,\nonumber\\
		&I_D=I_A/.\{k_1\rightarrow -k_1,k_2\rightarrow -k_2,k_3\rightarrow -k_3,k_4\rightarrow -k_4,k_{12}\rightarrow -k_{12}\} \, .
	\end{split}
\end{equation}
Here $/.$ means that we replace the selected $k_i$ by $-k_i$ in the explicit expressions. $I_A$ can be calculated straight forwardly and it is given as,
\begin{equation}\label{eq:Ia}
	\begin{split}
		I_A=&-\frac{2\gamma_5}{\mu^3}+\frac{\gamma_4}{\mu^2}+\frac{\gamma_3}{\mu}+\frac{1}{2}\delta\mu-\gamma_1\delta-\frac{1}{4}\delta^2 \\
		&+\frac{1}{2\eta_*^2}+\frac{1}{4}\sum_ak_a^2+\frac{i}{\eta_*}\Bigg(\gamma_1+\frac{\delta}{2}-\frac{\mu}{2}\Bigg)-\Bigg(\gamma_2-\gamma_1\mu+\frac{\mu^2}{2}\Bigg)\Gamma[0,i\mu\eta_*] \, ,
	\end{split}
\end{equation}
where
\begin{align}
	&\alpha_1=(\epsilon _3k_1+\epsilon _4k_2+\epsilon _5k_{12}) & \beta_1&=(\epsilon _8k_3+\epsilon _9k_4+\epsilon _{10}k_{12})\nonumber\nonumber\\
	&\alpha_2=-(\epsilon _3\epsilon _4k_1k_2+\epsilon _5k_{12}(\epsilon _3k_1+\epsilon _4k_2)) & \beta_2&=-(\epsilon _8\epsilon _9k_3k_4+\epsilon _{10}k_{12}(\epsilon _8k_3+\epsilon _9k_4))\nonumber\\
	&\alpha_3=-\epsilon _3\epsilon _4\epsilon _5k_1k_2k_{12} & \beta_3&=-\epsilon _8\epsilon _9\epsilon _{10}k_3k_4k_{12}\nonumber\\
	&\gamma_1=\alpha_1-\left(\frac{\beta_1\beta_2+\beta_3}{\beta_1^2}\right) & \delta&=\epsilon _1k_1+\epsilon _2k_2+\epsilon _6k_3+\epsilon _7k_4\nonumber\\
	&\gamma_2=-\alpha_2-\alpha_1\left(\frac{\beta_1\beta_2+\beta_3}{\beta_1^2}\right)-\frac{\beta_3}{\beta_1} & \mu&=\alpha_1+\beta_1\nonumber\\
	&\gamma_3=-\alpha_2\left(\frac{\beta_1\beta_2+\beta_3}{\beta_1^2}\right)+\frac{\alpha_1\beta_3}{\beta_1}+\alpha_3 & \gamma_5&=\frac{\alpha_3\beta_3}{\beta_1}\nonumber\\
	&\gamma_4=-\frac{\alpha_2\beta_3}{\beta_1}-\alpha_3\left(\frac{\beta_1\beta_2+\beta_3}{\beta_1^2}\right)
\end{align}
We have checked that the divergent pieces in eq.\eqref{eq:Ia} cancel when combined as in eq.\eqref{eq:i1234_iaib} and all channels and flips are summed together. Also let us define a notation which will be useful later as,
\begin{equation}\label{eq:i121234}
	\begin{split}
		\mathcal{I}_{1234}^1(i,j,k)&=\int_{-\infty}^{\eta_*}\frac{d\eta}{\eta^2}\int_{-\infty}^{\eta}\frac{d\eta'}{\eta'^2}\Im\Bigg[\sum^{k_1,k_2}_{i}\Bigg]\Im\Bigg[\sum^{k_3,k_4}_{j}\sum^{k_{12}}_{k}\Bigg] \, ,\\
		\mathcal{I}_{1234}^2(i,j,k)&=\int_{-\infty}^{\eta_*}\frac{d\eta}{\eta^2}\int_{-\infty}^{\eta}\frac{d\eta'}{\eta'^2}\Re\Bigg[\sum^{k_1,k_2}_{i}\Bigg]\Re\Bigg[\sum^{k_3,k_4}_{j}\sum^{k_{12}}_{k}\Bigg]  \, .
	\end{split}
\end{equation}

Once we have this, we can write the full $\mathcal{I}_{1234}^{\alpha}$ using the eq.\eqref{eq:i1234_iaib} as,
\begin{equation}\label{eq:ialpha1234full}
\begin{split}
	\mathcal{I}&_{1234}^\alpha=\cosh^{10}{\alpha}\Bigg[\mathcal{I}_{1234}^1(0,0,0)\Bigg] \\
	& -\cosh^8{\alpha}\sinh^{2}{\alpha}\Bigg[\mathcal{I}_{1234}^1(0,2,0)+\mathcal{I}_{1234}^1(2,0,0)+\mathcal{I}_{1234}^1(0,1,1)+\mathcal{I}_{1234}^1(0,0,2) \\
	&\hspace{3cm}-\mathcal{I}_{1234}^2(1,1,0)-\mathcal{I}_{1234}^2(1,0,1)\Bigg] \\
	&+\cosh^6{\alpha}\sinh^{4}{\alpha}\Bigg[\mathcal{I}_{1234}^1(0,4,0)+\mathcal{I}_{1234}^1(4,0,0)+\mathcal{I}_{1234}^1(2,2,0)+\mathcal{I}_{1234}^1(0,3,1) \\
	&\hspace{3cm}+\mathcal{I}_{1234}^1(2,1,1)+\mathcal{I}_{1234}^1(0,2,2)+\mathcal{I}_{1234}^1(2,0,2)-\mathcal{I}_{1234}^2(1,3,0) \\
	&\hspace{3cm}-\mathcal{I}_{1234}^2(3,1,0)-\mathcal{I}_{1234}^2(1,2,1)-\mathcal{I}_{1234}^2(3,0,1)-\mathcal{I}_{1234}^2(1,1,2)\Bigg] \\
	&-\cosh^4{\alpha}\sinh^{6}{\alpha}\Bigg[\mathcal{I}_{1234}^1(2,4,0)+\mathcal{I}_{1234}^1(4,2,0)+\mathcal{I}_{1234}^1(2,3,1)+\mathcal{I}_{1234}^1(4,1,1) \\
	&\hspace{3cm}+\mathcal{I}_{1234}^1(0,4,2)+\mathcal{I}_{1234}^1(2,2,2)+\mathcal{I}_{1234}^1(4,0,2)-\mathcal{I}_{1234}^2(3,3,0) \\
	&\hspace{3cm}-\mathcal{I}_{1234}^2(1,4,1)-\mathcal{I}_{1234}^2(3,2,1)-\mathcal{I}_{1234}^2(1,3,2)-\mathcal{I}_{1234}^2(3,1,2)\Bigg] \\
	&+\cosh^2{\alpha}\sinh^{8}{\alpha}\Bigg[\mathcal{I}_{1234}^1(4,4,0)+\mathcal{I}_{1234}^1(4,3,1)+\mathcal{I}_{1234}^1(2,4,2)+\mathcal{I}_{1234}^1(4,2,2) \\
	&\hspace{3cm}-\mathcal{I}_{1234}^2(3,4,1)-\mathcal{I}_{1234}^2(3,3,2)\Bigg] \\
	&-\sinh^{10}{\alpha}\Bigg[\mathcal{I}_{1234}^1(4,4,2)\Bigg] \, .
	\end{split}
\end{equation}

\section{Details of the wave-function formalism for computing correlators}

In this Appendix, we will provide the details of the computations in section \ref{sec:4pwfalpha}.
\subsection{Details of the on-shell action}
\label{ap:onshellact}

The wave-function coefficient $\langle O(\vect{k}_1)O(\vect{k}_2)O(\vect{k}_3)O(\vect{k}_4)\rangle $ can be calculated by computing Feynmann-Witten diagram in EAdS. For this, we first need to compute the on-shell action eq.\eqref{eq:sAdS_onshell}. This is detailed in section 4.1 of \cite{Ghosh:2014kba} and we review the main steps here. On-shell fluctuations of $\delta \phi$ sources the metric fluctuations $\delta g_{\mu\nu}$ through the matter stress tensor. These metric perturbations are physical only after fixing the gauge. We choose the axial gauge to fix the gauge as
\begin{equation}\label{eq:gmetp}
	\delta g_{z z}=0 \, , ~~~~ \delta g_{z k}=0 \, , 
\end{equation}
where $k=1,2,3$. Substituting for $\delta \phi$ and $\delta g_{\mu\nu}$ in eq.\eqref{eq:expact}, we get the on-shell action eq.\eqref{eq:onshellact1}. The on-shell action eq.\eqref{eq:onshellact1} allows us to obtain the partition function as a functional of the source eq.\eqref{eq:scalsource}. Here $T_{\mu\nu}$ is the matter stress tensor that is given by eq.\eqref{eq:defint} and $G^{{\text{grav}}}_{l_2 k_2, j_2 i_2} (x_1, z_1, x_2, z_2)$ is the Graviton propagator given by:
\begin{equation}\label{eq:gravfullprop}
	G^{{\text{grav}}}_{l k, j i} =  \int {d^3 {\vect{k}} \over (2\pi)^3} e^{i{\vect{k}}\cdot
		({\vect{x}}_1-{\vect{x}}_2)}\int_0^{\infty} {d p^2\over2} \bigg[{ J_{{3 \over 2}}(p z_1) J_{{3 \over 2}} (p z_2) \over  
		\sqrt{z_1 z_2}\left({\vect{k}}^2 + p^2 \right)}   {1 \over 2}\left({{\cal T}}_{l j} {{\cal T}}_{k i} + {{\cal T}}_{l i} {{\cal T}}_{k j} - 
	{{\cal T}}_{l k} {{\cal T}}_{j i} \right) \bigg] \, ,
\end{equation}
where
\begin{equation}
	{\cal T}_{i j} = \delta_{i j} + {k_{i} k_{j} \over p^2} \, .
\end{equation}

We see that the projector which appears in the graviton propagator is not transverse (i.e $k_i\mathcal{T}_{ij}\neq 0$) and traceless. This hints at the fact that the axial gauge propagator  also has a longitudinal component.  For the purpose of this calculation, it is convenient to 
break up our answer into the contribution from the transverse graviton propagator, and the longitudinal propagator. It is convenient to break up the graviton propagator into a transverse contribution and a longitudinal contribution given by
\begin{equation}\label{eq:onshelleadsfinal4pt}
	S^{\text{AdS}}_{\text{on-shell}} ={M_{Pl}^2 R_{\text{AdS}}^2 \over 2} {1 \over 4}[\widetilde{W}_{\alpha} + 2 R_{\alpha}] \, , 
\end{equation}
where the transverse contribution $\widetilde{W}_{\alpha}$ after fourier transformation from position space $\vect{x}$ to $\vect{k}$ \footnote{The convention for fourier transform is $T_{ij}({\vect{k}},z)=\int d^3 {\vect{x}} \ T_{ij}({\vect{x}},z) \ e^{-i \ {\vect{k}}\cdot {\vect{x}}}.$} is given by:
\begin{equation}\label{eq:trancont}
	\widetilde{W}_{\alpha} =\int dz_1 dz_2 {d^3 {\vect{k}} \over (2 \pi)^3} T_{l_1 k_1}(-{\vect{k}}, z_1) \delta^{l_1 l_2} \delta^{k_1 k_2} \widetilde{G}_{l_2 k_2, j_2 i_2}({\vect{k}}, z_1, z_2) \delta^{j_1 j_2} \delta^{i_1 i_2} T_{j_1 i_1}({\vect{k}}, z_2) \, .
\end{equation}
Here $\widetilde{G}_{l_2 k_2, j_2 i_2}({\vect{k}}, z_1, z_2)$ is just the transverse part of the graviton propagator eq.\eqref{eq:gravfullprop} given by
\begin{equation}\label{eq:gravtranprop}
	\begin{split}
		\widetilde{G}_{l k j i} (z_1, {\vect{x}}_1; z_2,{\vect{x}}_2) = & \int {d^3 {\vect{k}} \over (2\pi)^3} e^{i{\vect{k}}\cdot
			({\vect{x}}_1-{\vect{x}}_2)}\\ &~~~~~\times \int_0^{\infty} {d p^2\over2} \bigg[{ J_{{3 \over 2}}(p z_1) J_{{3 \over 2}} (p z_2) \over  
			\sqrt{z_1 z_2}\left({\vect{k}}^2 + p^2 \right)}   {1 \over 2}  \left(\pi^{\vect{k}}_{l j} \pi^{\vect{k}}_{k i} + \pi^{\vect{k}}_{l i} \pi^{\vect{k}}_{k j} - 
		\pi^{\vect{k}}_{l k} \pi^{\vect{k}}_{j i} \right) \bigg] \, ,
	\end{split}
\end{equation}
where $\pi^{\vect{k}}_{ij}$ is the transverse projector which projects all momentas perpendicular to ${\vect{k}}$:
\begin{equation}
	\pi^{\vect{k}}_{ij} = \delta_{ij} - \dfrac{k_i k_j}{k^2} \, .
\end{equation}
Clearly $\widetilde{G}_{l k, j i} k^i = 0$ and $\widetilde{G}_{l k j j} = 0$ which shows that the propagator is transverse. The longitudinal contribution is given by
\begin{equation}\label{eq:longcont}
	\begin{split}
		R_{\alpha} = \int \dfrac{d z_1}{z^2_1} \dfrac{d^3 \vect{k}}{(2\pi)^3} \left[ T_{zj}(\vect{k},z_1)\dfrac{1}{k^2}T_{zj}(-\vect{k},z_1) \right. &+ \dfrac{1}{2}i k_j T_{zj}(\vect{k},z_1)\dfrac{1}{k^2}T_{zz}(-\vect{k},z_1)  \\  & \left. - \dfrac{1}{4}k_j T_{zj}(\vect{k},z_1)\dfrac{1}{k^4} k_iT_{zi}(-\vect{k},z_1) \right] \, .
	\end{split}  
\end{equation}
\subsection{wave-function coefficient in alpha vacua}\label{ap:OOOO_calc}

Once we have the on-shell action as eq.\eqref{eq:onshelleadsfinal4pt}, we have the partition function as a functional of the sources eq.\eqref{eq:scalsource}. Thus, we only need to compute $\widetilde{W}_{\alpha}$ and $R_{\alpha}$ to find the wave-function coefficient $\langle OOOO \rangle$. Below we highlight the details of the computation for the transverse and longitudinal graviton contributions. This is detailed in Appendix C of \cite{Ghosh:2014kba}.

\subsubsection*{Transverse Graviton contribution}

If we integrate eq.\eqref{eq:gravtranprop} over $p$ with Bunch-Davies boundary condition, we get
\begin{equation}\label{eq:bdvacuaprop}
	\begin{split}
		\widetilde{G}_{l k, j i}({\vect{k}}, z_1, z_2) = (z_1 z_2)^{\frac{3}{2}} &\left(\Theta(z_1 - z_2) K_{3/2}(k z_1) I_{3/2}(k z_2) + \Theta(z_2-z_1) K_{3/2}(k z_2) I_{3/2}(k z_1) \right) \\
		& {1 \over 2} \left(\pi^{\vect{k}}_{l j} \pi^{\vect{k}}_{k i} + \pi^{\vect{k}}_{l i} \pi^{\vect{k}}_{k j} - 
		\pi^{\vect{k}}_{l k} \pi^{\vect{k}}_{j i} \right) \, .
	\end{split}
\end{equation}
The generalization of this for the case of alpha vacua is given by eq.\eqref{eq:gravalphatranprop}. eq.\eqref{eq:gravalphatranprop} can be written compactly as
\begin{equation}\label{eq:alphapropcompact}
	\widetilde{G}_{l k, j i}({\vect{k}}, z_1, z_2) = \dfrac{c_1}{c_1 + c_2} G^{(1)}_{lk,ji}(k,z_1,z_2) + \dfrac{c_2}{c_1+c_2} G^{(1)}_{lk,ji}(-k,z_1,z_2) \, .
\end{equation}
The effects of the alpha vacua can be incoporated by first calculating all the expressions with respect to $G^{(1)}_{lk,ji}$ and then interchanging the sign of the magnitude of the exchange momenta.

The $z$ components of the stress tensor do not appear because of our choice of gauge, eq.\eqref{eq:gmetp}. In eq.\eqref{eq:trancont}, we see that the graviton propagator $\widetilde{G}_{lk,ji}({\vect{k}}, z_1, z_2)$ is contracted against two factors of the stress tensors $T_{ij}({\vect{k}},z)$, inserted at two different values of the radial coordinate $z_{1}, z_{2}$. From eq.\eqref{eq:defint}, we get 
\begin{align}
	\label{eq:Tij}
	T_{ij}(z,{\vect{x}})&=2(\partial_i\delta \phi)(\partial_j\delta\phi)-\delta_{ij}\left[(\partial_z\delta\phi)^2+{\eta^{mn}}(\partial_m\delta\phi)(\partial_{n}\delta\phi)\right].
\end{align}
For the exchange process, we have three channels namely $S,T,U$. To compute the S channel result, we insert the $\phi(\vk)\phi(\vkk)$ at $z=z_1$ and similarly $\phi(\vkkk)\phi(\vkfour)$ at $z=z_2$. For the $T$- channel and $U$- channel contributions one just needs to perform exchange operations on two of the external momenta like $\vkk \leftrightarrow \vkkk$ and $\vkk \leftrightarrow \vkfour$ respectively. In momentum space, the stress tensors take the expressions of the form
\begin{align}
	\label{bnexplct}
	\begin{split}
		T_{lk}[\phi_1(z_1),\phi_2(z_1)]&=-4\left\{k_{1l}k_{2k}\phi_1\phi_2+\frac{1}{2}\eta_{lk}\left[(\partial_{z_1}\phi_1)(\partial_{z_1}\phi_2)-\vect{k_1}\cdot\vect{k_2}\,\phi_1\phi_2\right]\right\}\,,\\
		T_{ji}[\phi_3(z_2),\phi_4(z_2)]&=-4\left\{k_{3j}k_{4i}\phi_3\phi_4+\frac{1}{2}\eta_{ji}\left[(\partial_{z_2}\phi_3)(\partial_{z_2}\phi_4)-\vect{k_3}\cdot\vect{k_4}\,\phi_3\phi_4\right]\right\}\, ,
	\end{split}
\end{align}
where
$\partial_m\phi=-ik_m\phi$ and $g^{ij}=z^2\eta^{ij}$ 
and the abbreviations stand for the on-shell values of the scalar fields eq.\eqref{eq:supsoleads}:
\begin{equation}\label{eq:phivals}
	\phi_1\equiv \phi(\vk)[c_{1}(1+ k_1 z_1)e^{-k_1 z_1} +  c_{2} (1-k_1z_1)e^{k_1z_1}] e^{-i \vect{k} \cdot \vect{x}} \, ,
\end{equation}
and similarly for the others. 
\par
It is evident that only the first term on the RHS of eq.\eqref{bnexplct} in both $T_{lk}$ and $T_{ji}$ contributes to $\widetilde{W}$ in eq.\eqref{eq:trancont}. The second term in $T_{lk}$ on the RHS of eq.\eqref{bnexplct} carries $\eta_{lk}$ which when contracted with the transverse projector  $\left(\pi^{\vect{k}}_{l j} \pi^{\vect{k}}_{k i} + \pi^{\vect{k}}_{l i} \pi^{\vect{k}}_{k j} - 
\pi^{\vect{k}}_{l k} \pi^{\vect{k}}_{j i} \right)$ of the graviton propagator $\widetilde{G}_{l_2 k_2, j_2 i_2}({\vect{k}}, z_1, z_2)$ in eq.\eqref{eq:gravtranprop}, gives zero, as can be checked explicitly
\be
\eta_{lk} \left(\pi^{\vect{k}}_{l j} \pi^{\vect{k}}_{k i} + \pi^{\vect{k}}_{l i} \pi^{\vect{k}}_{k j} - 
\pi^{\vect{k}}_{l k} \pi^{\vect{k}}_{j i} \right) =0 \, .
\ee
Therefore, the relevant terms for our calculation in the stress tensors are,
\begin{align}
	\label{bnexplct1}
	\begin{split}
		T_{lk}(z_1)=-4k_{1l}k_{2k}\phi_1\phi_2 \, , ~~~~
		T_{ji}(z_2) =-4k_{3j}k_{4i}\phi_3\phi_4 \, .
	\end{split}
\end{align}

Finally substituting for $T_{lk}$ from eq.\eqref{bnexplct1}, $\phi_i$'s from eq.\eqref{eq:phivals} and the graviton propagator of eq.\eqref{eq:alphapropcompact} in eq.\eqref{eq:trancont}, we obtain $\widetilde{W}^S(\vk,\vkk,\vkkk,\vkfour)$. To keep track of the expressions, we first analyse the contribution to eq.\eqref{eq:trancont} just coming from $G^{(1)}_{lk,ji}$ of eq.\eqref{eq:alphapropcompact}. Thus for the $S-$channel, we first have
\be
\label{defwtsdet}
\begin{split}
	\widetilde{W}^S_{(1)}(\vk,\vkk,\vkkk,\vkfour) =&16 (2\pi)^3 \delta^3(\sum_i {\vect{k}}_i)\phi(\vk) \phi(\vkk) \phi(\vkkk) 
	\phi(\vkfour) \\ & k^l_1 k^k_2 k^j_3 k^i_4\left(\pi^{\vect{k}}_{l j} \pi^{\vect{k}}_{k i} + \pi^{\vect{k}}_{l i} \pi^{\vect{k}}_{k j} - 
	\pi^{\vect{k}}_{l k} \pi^{\vect{k}}_{j i} \right)  S_{(1)}(k_1, k_2, k_3, k_4,K_s) \, ,
\end{split}
\ee
with $S_{(1)}(k_{1},k_{2},k_{3},k_{4},K_s)$ given by eq.\eqref{eq:snew}. The subscript 1 is used to denote the fact that we are looking at the contribution only from $G^{(1)}_{ij,kl}$ of eq.\eqref{eq:alphapropcompact}, where $S_{(1)}(k_1, k_2, k_3, k_4,K_s)$ is explicitly written in eq.\eqref{eq:snew} with the Bunch-Davies expression $S(k_1, k_2, k_3, k_4,K_s)$ given by
\begin{equation}\label{eq:sbd}
	\begin{split}
		S(k_1, k_2, k_3, k_4,K_s) &= \int_0^{\infty} {d p^2 \over 2(p^2 + K_s^2)} \times \\ &\int_0^{\infty} {d z_1 \over z_1^2} (1 + k_1 z_1) (1 + k_2 z_1) (z_1)^{3 \over 2} J_{3 \over 2}(p z_1) e^{-(k_1 + k_2) z_1} \\ & \times \int_0^{\infty} {d z_2 \over z_2^2} (1 + k_3 z_2) (1 + k_4 z_2) (z_2)^{3 \over 2} J_{3 \over 2}(p z_2) e^{-(k_3 + k_4) z_1} \\
		= \int_{0}^{\infty} \dfrac{dz_1}{z^2_1} \dfrac{dz_2}{z^2_2} (1+ k_1 z_1) (1+ &k_2 z_1) (1+ k_3 z_2) (1+ k_4 z_2) e^{-(k_1+k_2)z_1} e^{-(k_3 + k_4)z_1} \\
		(z_1 z_2)^{3/2}\left(\Theta(z_1 - z_2) \right.& \left. K_{3/2}(k z_1) I_{3/2}(k z_2) + \Theta(z_2-z_1) K_{3/2}(k z_2) I_{3/2}(k z_1) \right) \, .
	\end{split}
\end{equation}
$K_s$ is the norm of momentum $\vect{k}_s$ of the graviton exchanged in the $S$ channel given by:
\be
\label{defks}
{\vect{k}}_s=\vk+\vkk=-(\vkkk+\vkfour) \, .
\ee
The explicit form of $S(k_1,k_2,k_3,k_4)$ is given by eq.(C.17) of \cite{Ghosh:2014kba}. Also, the transverse projector involving $\pi^{\vect{k}}_{l j}$'s contracted with $\vect{k}_i$'s is given by eq.\eqref{eq:tranpropcont} (eq.(C.18) of \cite{Ghosh:2014kba}).
Now these can be put back into the equation for the transverse graviton propagator eq.\eqref{defwtsdet} which finally yields the result for the $S$ channel contribution for $\widetilde {W}$ in eq.\eqref{defwtsdet}:
\be
\label{defwts}
\widetilde{W}^S_{(1)}(\vk,\vkk,\vkkk,\vkfour) =16 (2\pi)^3 \ \delta^3\bigg(\sum_{J=1}^4 {\vect{k}}_J\bigg) \bigg( \prod_{I=1}^4 \phi({\vect{k}}_I)\bigg) \widehat{W}^S_{(1)}(\vk,\vkk,\vkkk,\vkfour) \, .
\ee
where
\begin{equation}
	\widehat{W}^S_{(1)}(\vk,\vkk,\vkkk,\vkfour) =  k^l_1 k^k_2 k^j_3 k^i_4\left(\pi^{\vect{k}}_{l j} \pi^{\vect{k}}_{k i} + \pi^{\vect{k}}_{l i} \pi^{\vect{k}}_{k j} - 
	\pi^{\vect{k}}_{l k} \pi^{\vect{k}}_{j i} \right)  S_{(1)}(k_1, k_2, k_3, k_4, K_s) \, .
\end{equation}
The interchange of the signs of $k_i$'s in eq.\eqref{eq:snew} comprise the effect of alpha vacua in the modes eq.\eqref{eq:phivals}. In order to incorporate the full propagator of eq.\eqref{eq:alphapropcompact}, we simply have to interchange the sign of the magnitude of the exchanged momenta. Thus, we get
\begin{equation}\label{eq:wsfinal}
	\widehat{W}^{S}_{\alpha}(k_{1},k_{2},k_{3},k_{4}) = \dfrac{c_1}{c_1+c_2}\widehat{W}^{S}_{(1)}(k_{1},k_{2},k_{3},k_{4},K_s) + \dfrac{c_2}{c_1+c_2} \widehat{W}^{S}_{(1)}(k_{1},k_{2},k_{3},k_{4},-K_s) \, .
\end{equation}
This leads to the final $S-$channel contribution
\begin{equation}\label{eq:schanwnew}
	\begin{split}
		\widetilde{W}^S_{\alpha}(\vk,\vkk,\vkkk,\vkfour) =16 (2\pi)^3 \ \delta^3\bigg(\sum_{J=1}^4 {\vect{k}}_J\bigg) \bigg( \prod_{I=1}^4 \phi({\vect{k}}_I)\bigg) \widehat{W}^S_{\alpha}(\vk,\vkk,\vkkk,\vkfour) \, .
	\end{split}
\end{equation}
After summing over $S,T,U$ channels, we can get the entire transverse graviton contribution of eq.\eqref{eq:trancont}.

\subsubsection*{Longitudinal Graviton contribution}

Now, we turn our attention to the longitudinal graviton contribution of eq.\eqref{eq:longcont}. As done in eq.\eqref{bnexplct}, we first find the $S-$channel contribution.  The ``remainder" terms are given in momentum space, and the relevant expressions for the energy momentum tensor components in $S-$ channel are given by:
\begin{equation}
	\begin{split}
		T_{zi}(\phi_{1},\phi_{2})= iz_{1}[(c_{1}(1+k_{1}z_{1})e^{-k_{1}z_{1}}+c_{2}(1-k_{1}z_{1})e^{k_{1}z_{1}})(c_{1}e^{-k_{2}z_{1}}+c_{2}e^{k_{2}z_{1}})k_{1i}k_{2}^{2}\\+(c_{1}(1+k_{2}z_{1})e^{-k_{2}z_{1}}+c_{2}(1-k_{2}z_{1})e^{k_{2}z_{1}})(c_{1}e^{-k_{1}z_{1}}+c_{2}e^{k_{1}z_{1}})k_{2i}k_{1}^{2}] \, ,
	\end{split}
\end{equation}
\begin{equation}
	\begin{split}
		k_iT_{zi}=iz_1 [(c_1(1+k_1z_1)e^{-k_1z_1}+c_2(1-k_1z_1)e^{k_1z_1})(c_1e^{-k_2z_1}+c_2e^{k_2z_1})k_{1i}(k_{1i}+k_{2i})k_{2}^{2}+\\(c_1(1+k_2z_1)e^{-k_2z_1}+c_2(1-k_2z_1)e^{k_2z_1})(c_1e^{-k_1z_1}+c_2e^{k_1z_1})k_{2i}(k_{1i}+k_{2i})k_{1}^{2}] \, ,
	\end{split}
\end{equation}
\begin{equation}
	\begin{split}
		&T_{zz}(\phi_1,\phi_2)=k_{1}^{2}k_{2}^{2}z^2(c_1e^{-k_1z_1}+c_2e^{k_1z_1})(c_1e^{-k_2z_1}+c_2e^{k_2z_1})+\\&(\vect{k_1}.\vect{k_2})[(c_1(1+k_1z_1)e^{-k_1z_1}+c_2(1-k_1z_1)e^{k_1z_1})(c_1(1+k_2z_1)e^{-k_2z_1}+c_2(1-k_2z_1)e^{k_2z_1})] \, .
	\end{split}
\end{equation}
The $S$ channel contribution for $R$ is denoted by $R^S(\vk, \vkk, \vkkk, \vkfour)$ and is given by
\be
\label{defrs}
R^S_{\alpha}(\vk,\vkk,\vkkk,\vkfour) =16 (2\pi)^3 \ \delta^3\big(\sum_{J=1}^4 {\vect{k}}_J\big) \bigg[ \prod_{I=1}^4 \phi({\vect{k}}_I)\bigg]  
\widehat{R}^S_{\alpha}(\vk,\vkk,\vkkk,\vkfour),
\ee
with
\begin{equation}\label{eq:rsnewhat}
	\begin{split}
		\widehat{R}^S_{\alpha}(k_{1},k_{2},k_{3},k_{4})= (c_1^4-c_2^4)\widehat{R}^S(k_{1},k_{2},k_{3},k_{4})+(c_1^3c_2-c_1c_2^3)\bigg[\widehat{R}^S(-k_{1},k_{2},k_{3},k_{4})\\+\widehat{R}^S(k_{1},-k_{2},k_{3},k_{4})+\widehat{R}^S(k_{1},k_{2},-k_{3},k_{4})
		+\widehat{R}^S(k_{1},k_{2},k_{3},-k_{4})\bigg] \, ,
	\end{split}
\end{equation}
where $\widehat{R}^S(k_{1},k_{2},k_{3},k_{4})$ is the expression for eq.\eqref{eq:longcont} in the Bunch-Davies vacua given by eq.(4.27) of \cite{Ghosh:2014kba}. Due to eq.\eqref{eq:phivals}, we just need to reverse the signs of the norm of the momenta of external particles. In particular, one shouldn't reverse the signs of vector momenta. We have chosen to suppress the explicit dependence of $\widehat{R}^S_{\text{new}}(k_{1},k_{2},k_{3},k_{4})$ on $\vk,\vkk,\vkkk,\vkfour$ for the sake of clarity.
\par 
The structure of $\widehat{R}^S_{\alpha}(k_{1},k_{2},k_{3},k_{4})$ is identical to eq.\eqref{eq:snew} and thus we don't write it again. However, we now have the following additional relations due to the explicit expression of $\widehat{R}^S$ in the Bunch-Davies vacua:
\begin{equation}\label{eq:rsbdrel}
	\begin{split}
		\widehat{R}^S(k_{1},k_{2},k_{3},k_{4}) &=- \widehat{R}^S(-k_{1},-k_{2},-k_{3},-k_{4}) \, ,\\
		\widehat{R}^S(-k_{1},k_{2},k_{3},k_{4}) &=- \widehat{R}^S(k_{1},-k_{2},-k_{3},-k_{4}) \, ,\\
		\widehat{R}^S(k_{1},-k_{2},k_{3},k_{4}) &=- \widehat{R}^S(-k_{1},k_{2},-k_{3},-k_{4}) \, ,\\
		\widehat{R}^S(k_{1},k_{2},-k_{3},k_{4}) &=- \widehat{R}^S(-k_{1},-k_{2},k_{3},-k_{4}) \, ,\\
		\widehat{R}^S(k_{1},k_{2},k_{3},-k_{4}) &=- \widehat{R}^S(-k_{1},-k_{2},-k_{3},k_{4})\, ,\\
		\widehat{R}^S(k_{1},k_{2},-k_{3},-k_{4}) &=- \widehat{R}^S(-k_{1},-k_{2},k_{3},k_{4}) \, ,\\
		\widehat{R}^S(-k_{1},k_{2},-k_{3},k_{4}) &=- \widehat{R}^S(k_{1},-k_{2},k_{3},-k_{4}) \, , \\
		\widehat{R}^S(-k_{1},k_{2},k_{3},-k_{4}) &=- \widehat{R}^S(k_{1},-k_{2},-k_{3},k_{4}) \, .
	\end{split}
\end{equation}
eq.\eqref{eq:rsbdrel} results in eq.\eqref{eq:rsnewhat}. eq.\eqref{eq:rsbdrel} helps us to get the final result for $R^S_{\alpha}(k_{1},k_{2},k_{3},k_{4})$ of eq.\eqref{defrs} as
\begin{equation}\label{eq:rsnew}
	\begin{split}
		R^S_{\alpha}(k_{1},k_{2},k_{3},k_{4}) &= (c^4_1 - c^4_2) R^S(k_1,k_2,k_3,k_4) + (c^3_1 c_2 - c_1 c^3_2) \left[R^S(-k_1,k_2,k_3,k_4) \right. \\
		&~~ \left. + R^S(k_1,-k_2,k_3,k_4) + R^S(k_1,k_2,-k_3,k_4) + R^S(k_1,k_2,k_3,-k_4) \right] \, . \\
	\end{split}
\end{equation}
Here, we have again chosen to suppress the dependence on $\vect{k}_i$ for the sake of clarity. $R^S(k_1,k_2,k_3,k_4)$ is the standard Bunch-Davies vacua result given by eq.(4.26) of \cite{Ghosh:2014kba}.
\par
The  full answer for $S_{\text{on-shell}}^{\text{AdS}}$ is obtained by adding the contributions from all the three $S, T, U$ channels. Summing eq.\eqref{eq:schanwnew} and eq.\eqref{eq:rsnew} leads to eq.\eqref{eq:on_shell_action_full}:
\begin{align}
	\label{fullanswerS}
	\begin{split}
		S^{\text{AdS}}_{\text{on-shell}} = {M_{\text{pl}}^2 R_{\text{\text{AdS}}}^2 \over 4}  \bigg[&\left(\frac{{\widetilde W}^S_{\alpha}(\vk, \vkk, \vkkk, \vkfour)}{2} + R^S_{\alpha}(\vk, \vkk, \vkkk, \vkfour)\right) \\ 
		&+ \big({\vkk \leftrightarrow \vkkk} \big)+\big({\vkk \leftrightarrow \vkfour} \big) \bigg] \, .
	\end{split}
\end{align}





\bibliographystyle{JHEP}
\bibliography{References}

\end{document}